\documentclass[12pt]{article}
\usepackage{amssymb,amsmath}
\makeatletter

\usepackage{arydshln}

\newcount\comment@nesting

\begingroup
\xdef\comment@begincomment{\string\\begin\string\{comment\string\}}
\xdef\comment@endcomment{\string\\end\string\{comment\string\}}
\endgroup

\begingroup
\catcode`\^^M=\active
\def\@temp{\endgroup\def\comment@processline##1^^M}%
\@temp{%
    \def\comment@curline{#1}%
    \let\@next=\comment@processline
    \ifx\comment@curline\comment@endcomment
        \ifnum\comment@nesting=0
            \def\@next{\end{comment}}%
        \else
            \global\advance\comment@nesting by -1
        \fi
    \else
        \ifx\comment@curline\comment@begincomment
            \global\advance\comment@nesting by 1
        \fi
    \fi
    \@next
}

\makeatother

\usepackage[dvipdfmx]{graphicx}
\usepackage{bm}
\usepackage{color}
\usepackage{here}
\usepackage{enumerate}
\usepackage{here}
\usepackage[vcentermath]{youngtab}
\usepackage{subfigure}
\usepackage{tikz}
\usepackage{hyperref}
\newcommand{\Zb}{\mathbb{Z}}

\newcommand{\Ncal}{\mathcal{N}}

\DeclareMathOperator*{\Tr}{{\rm Tr}}

\newcommand{\II}{\mathbb{II}}

\numberwithin{equation}{section}

\allowdisplaybreaks[1]

\definecolor{mygreen}{rgb}{0,0.714,0.286}

\begin{document}

%%%%%%%%%%%%%%%%%%%%%%%%%%%%%%%%%%%%%%%%%%%%
\thispagestyle{empty}
\begin{flushright}
%%%%%%%%%%%%%%%%%%%%%%%%%%%%%%%%%%%%%%%%%%%%%%%%%%%%%%%%%%%%%%%%%
%input \\
%%%%%%%%%%%%%%%%%%%%%%%%%%%%%%%%%%%%%%%%%%%%%%%%%%%%%%%%%%%%%%%%%

\end{flushright}
\vskip1.5cm
\begin{center}
{\Large \bf 
Boundary lines and Askey-Wilson type moments
}

\vskip1.5cm
Tadashi Okazaki\footnote{tokazaki@seu.edu.cn}

\bigskip
{\it School of Physics and Shing-Tung Yau Center, Southeast University,\\
Yifu Architecture Building, No.2 Sipailou, Xuanwu district, \\
Nanjing, Jiangsu, 210096, China
}

\bigskip
and
\\
\bigskip
Douglas J. Smith\footnote{douglas.smith@durham.ac.uk}

\bigskip
{\it Department of Mathematical Sciences, Durham University,\\
Upper Mountjoy, Stockton Road, Durham DH1 3LE, UK}

\end{center}

%%%%%%%%%%%%%%%%%%%%%%%%%%%%%%%%%%%%%%%%%%%%
\vskip1cm
\begin{abstract}
The Wilson line defect half-indices for 3d $\mathcal{N}=2$ gauge theories with boundary confining phases 
admit a formulation in terms of the Askey-Wilson type moments. 
In the dual Landau-Ginzburg description the dual line operators can be realized as vortex line defects 
which induce singular behavior of chiral multiplets associated with the minimal monopole operators, together with additional one-dimensional degrees of freedom. 
By incorporating such a singular structure as an effective spin shift into the index computation, 
we obtain exact closed-form expressions for the line defect half-indices which are Askey-Wilson type moments. 
\end{abstract}
%%%%%%%%%%%%%%%%%%%%%%%%%%%%%%%%%%%%%%%%%%
\newpage
\setcounter{tocdepth}{3}
\tableofcontents
%%%%%%%%%%%%%%%%%%%%%%%%%%

%%%%%%%%%%%%%%%%%%%%%%%%%%%%%%%%%%%%%%%%%%%%%%%%%%%
%%%%%%%%%%%%%%%%%%%%%%%%%%%%%%%%%%%%%%%%%%%%%%%%%%%
\section{Introduction and conclusion}
\label{sec_intro}
%%%%%%%%%%%%%%%%%%%%%%%%%%%%%%%%%%%%%%%%%%%%%%%%%%%
%%%%%%%%%%%%%%%%%%%%%%%%%%%%%%%%%%%%%%%%%%%%%%%%%%%

In the previous work \cite{Okazaki:2023hiv,Okazaki:2023kpq}, 
we proposed boundary confining dualities in 3d $\mathcal{N}=2$ supersymmetric gauge theories. 
These dualities arise in the presence of the $\mathcal{N}=(0,2)$ half-BPS boundary conditions \cite{Gadde:2013wq,Okazaki:2013kaa} 
in such a way that the bulk gauge degrees of freedom can effectively confine at the boundary, 
leading to an infrared description purely in terms of gauge invariant composite operators.
A central result of the work is based on the correspondence 
between the equalities of the half-indices \cite{Gadde:2013wq,Gadde:2013sca,Yoshida:2014ssa,Dimofte:2017tpi}
associated with the boundary confinement phenomena 
and the non-trivial $q$-hypergeometric integral identities of Askey-Wilson type $q$-beta integrals \cite{MR783216,MR772878,MR845667,MR1139492,MR1266569,MR2267266}. 

%What we do here
A central theme of this work is the incorporation of line defect operators into the above correspondence. 
In particular, we consider an insertion of Wilson line operators compatible with the half-BPS boundary conditions 
involving the Neumann boundary conditions for the vector and chiral multiplets 
and develop a framework for the exact computation of the resulting line defect half-indices. 
Remarkably, this physical generalization admits a natural interpretation in terms of \textit{Askey-Wilson type moments} \cite{MR2630104,MR2831874,MR2946941,MR3207468,MR3816505,MR3715708}. 
Motivated by this correspondence, we propose several explicit and conjecturally exact formulas for the line defect half-indices, 
which, in turn, provide new integral representations and evaluation formulas for Askey-Wilson type moments. 

%gauge theory interpretation
For the 3d $\mathcal{N}=2$ supersymmetric gauge theories with $s$-confining properties, 
the low-energy physics can be described by a dual Landau-Ginzburg (LG) theory of gauge invariant composites 
(see e.g. \cite{Amariti:2015kha,Nii:2016jzi,Pasquetti:2019uop,Pasquetti:2019tix,Nii:2019dwi,Benvenuti:2020gvy,Benvenuti:2021nwt,Bajeot:2022lah,Amariti:2022wae,Okazaki:2023hiv,Okazaki:2023kpq,Amariti:2023wts}). 
The LG theory typically includes mesonic operators $M$, baryonic operators $B$ and a singlet chiral multiplet $V$. 
While the mesonic and baryonic operators map to certain gauge invariant combinations of the matter fields, 
the singlet $V$ corresponds to the minimal monopole operator. 
While the monopole operator in the UV gauge theory is naturally labeled by a cocharacter $m \in \Lambda_{\mathrm{cochar}}(G)$, 
this labeling does not survive as independent data in the LG description but it is reorganized into the charge assignments of $V$ in the IR. 

%Wilson and vortex lines
On the other hand, the Wilson line operator can be re-expressed, upon restriction to the Coulomb branch, as a defect that sources a dual gauge field through a BF interaction, 
where the BF coupling enforces a vortex-like singularity associated with magnetic symmetry \cite{Kapustin:2012iw}. 
Since monopole operators carrying magnetic charge map to a singlet chiral multiplet $V$ in the LG description, 
the vortex line operator dual to a Wilson line in the gauge theory can induce a singular profile in $V$. 

%boundary lines
We test the above picture by computing the half-index in the presence of a boundary. 
Imposing Neumann boundary conditions for vector and chiral multiplets in the gauge theory, 
the half-index is given by the Askey-Wilson type integral and admits an exact factorized expression matching the dual LG description, 
as discussed in \cite{Okazaki:2023hiv,Okazaki:2023kpq}. 
When we introduce the Wilson line, the integral is promoted to the line defect half-index, whose exact evaluation is generically non-trivial. 
From the perspective of the dual LG description, however, the effect is naturally expected from the underlying physical picture:  
the corresponding vortex line induces a singular profile only in the chiral multiplet $V$ obeying the Dirichlet boundary condition, leading to an effective spin shift. 
Incorporating this shift, we obtain various exact results, expressed as the original factorized form together with a finite number of extra contributions, 
which admit a natural interpretation as 1d degrees of freedom localized on the defect. 
These contributions are naturally organized into characters of the flavor symmetries acting on the charged matter multiplets, 
corresponding to the Higgs branch isometry, and thus furnish Higgs branch data.
The normalized one-dimensional defect index can be interpreted as the Askey-Wilson type moment. 

%models studied in this work
For the rank-$1$ $SU(2)$ gauge theory with $N_f=4$, 
the line defect half-indices admit an exact representation in terms of Askey-Wilson moments \cite{MR2946941,MR3207468}. 
In this case, the above physical picture based on line operator insertions and their interpretation can be rigorously justified. 
Motivated by this structure, we extend the analysis to higher-rank gauge groups. 
In particular, for theories with gauge groups $SU(N)$, $USp(2n)$, $SO(N)$, and $G_2$, 
whose half-indices without Wilson line insertions are known to be expressible as Gustafson-type integrals, 
we propose exact closed-form expressions for line defect half-indices based on the same physical principles. 
In the case of $U(N)$ gauge theories with $N_f$ fundamental and $N_a$ antifundamental chirals where $N_f=N_a=N$, 
which are likewise $s$-confining, we propose analogous integral formulas based on the same prescription. 
In contrast to simple gauge groups, the Wilson lines in $U(N)$ gauge theory carry an additional $U(1)$ electric charge. 
On the dual side, there exist two chiral multiplets $Y$ and $Z$ associated with the monopole operators, which carry distinct topological $U(1)$ charges. 
The Wilson line charge selects one of these chiral multiplets, which accordingly undergoes a spin shift.
To the best of our knowledge, these formulas have not appeared in the existing literature and thus constitute new exact results. 
A particularly illuminating example is provided by the $USp(2n)$ gauge theory with a rank-two antisymmetric chiral multiplet and four fundamental chirals. 
Based on its expected LG description, we apply the above prescription to derive a closed-form expression for the line defect half-index, which gives an exact result for a \textit{Macdonald-Koornwinder moment} which we define. 
While we are not aware of any results in the mathematical literature evaluating such Macdonald-Koornwinder moment, remarkably, upon an appropriate specialization of fugacities, 
this expression reduces to a Koornwinder moment and reproduces the known results studied in the mathematical literature \cite{MR3816505}.

%%%%%%%%%%%%%%%%%%%%%%%%%%%%%%%%%%%%%%%%%%%%%%%%%%%
\subsection{Structure}
%%%%%%%%%%%%%%%%%%%%%%%%%%%%%%%%%%%%%%%%%%%%%%%%%%%
The paper is organized as follows. 
We begin in section~\ref{sec_ind} with a review of half-indices and the inclusion of Wilson line defects, with vortex line defects in a dual confining theory discussed in section~\ref{sec_VortexLG}.
This is followed in section \ref{sec_AWmoment} with an exact evaluation of the Wilson line defect half-index for the rank-one $SU(2)$ gauge theory, 
expressed in terms of Askey-Wilson moments. 
In section \ref{sec_Gmoment}, we extend the analysis to higher-rank gauge theories. 
In the absence of the Wilson line operators, the relevant integrals realize Gustafson-type integrals. 
Motivated by this structure, we introduce higher-rank generalizations of Askey-Wilson moments and propose a variety of exact closed-form expressions.
In section \ref{sec_MK_moments}, we study the Wilson line defect half-index for $USp(2n)$ gauge theories with a rank-$2$ antisymmetric chiral multiplet and four fundamental chirals. 
After proposing the relevant duality for this class of theories, 
we formulate a general expression for the first Macdonald-Koornwinder moment and show that, upon a suitable specialization of fugacities, 
the line defect index reduces to the Koornwinder moment, thereby providing a natural generalization. 

%%%%%%%%%%%%%%%%%%%%%%%%%%%%%%%%%%%%%%%%%%%%%%%%%%%
\subsection{Future works}
%%%%%%%%%%%%%%%%%%%%%%%%%%%%%%%%%%%%%%%%%%%%%%%%%%%

\begin{itemize}

\item 
There are other types of theories with boundary confining descriptions 
associated with different Askey-Wilson type integrals, including the Nassrallah-Rahman integral \cite{MR772878,MR845667} and Gustafson-Rakha integrals \cite{MR1266569}. 
In these cases, monopole operators are not dynamical operators in the IR chiral ring, as they are lifted by the monopole superpotential.
In particular, the singlet chiral multiplet $V$ is absent in these cases, and hence the above prescription cannot be directly applied. 
It would be interesting to understand the dual line defects in the corresponding LG descriptions.

\item 
It would be of considerable interest to establish rigorous proofs of the closed-form expressions for 
the line defect half-indices and the Askey-Wilson type moments obtained in this work. 
A salient feature of half-indices is that they are often constrained by 
$q$-difference equations generated by operators acting multiplicatively on fugacities, such as $x\rightarrow qx$. 
This structure furnishes a systematic method for establishing identities between half-indices. 
If two such quantities satisfy the same first-order $q$-difference equation and coincide at a suitable set of fugacity values, their equality follows. 

\item 
An intriguing problem is to explore combinatorial underpinnings and the 1d quantum mechanical descriptions of the line defect half-indices. 
As discussed in this work, the Wilson line defect half-indices for the $s$-confining 3d $\mathcal{N}=2$ gauge theories can be identified with the Askey-Wilson type moments. 
Motivated by a close connection \cite{MR2065218} between the Askey-Wilson polynomial and the asymmetric exclusion process (ASEP) partition function, 
it was argued in \cite{MR2946941} that 
the Askey-Wilson moments can be viewed as a specialized ASEP partition function 
and there is a combinatorial interpretation of the Askey-Wilson moment in terms of the \textit{staircase tableaux} \cite{MR2831874,MR2630104}. 
Furthermore, there is a natural extension to the Koornwinder moments in terms of \textit{rhombic staircase tableaux} associated with the two-species ASEP, 
as discussed in \cite{MR3715708}. 
These works naturally raise the question of whether more general line defect half-indices admit a formulation within the framework of the ASEP, 
and, if so, whether such a correspondence can be endowed with a clear physical interpretation. 

\item 
A natural extension of the present analysis is to consider the $s$-confining theories that include an adjoint chiral multiplet. 
In such cases, the Wilson line half-index exhibits a structure identical to the line defect half-index of 4d $\mathcal{N}=4$ super Yang-Mills theory in a half-space. 
For classical gauge groups of type A, B, C, and D, closed-form expressions have been obtained in \cite{Hatsuda:2025yzp}. 
These results can be rigorously derived by exploiting the orthogonality properties of Macdonald polynomials. 
It is therefore expected to extend the results to exceptional gauge groups of type E,F and G as well.

\end{itemize}

%%%%%%%%%%%%%%%%%%%%%%%%%%%%%%%%%%%%%%%%%%%%%%%%%%%
%%%%%%%%%%%%%%%%%%%%%%%%%%%%%%%%%%%%%%%%%%%%%%%%%%%
\section{Line defect half-indices}
\label{sec_ind}
%%%%%%%%%%%%%%%%%%%%%%%%%%%%%%%%%%%%%%%%%%%%%%%%%%%
%%%%%%%%%%%%%%%%%%%%%%%%%%%%%%%%%%%%%%%%%%%%%%%%%%%

%%%%%%%%%%%%%%%%%%%%%%%%%%%%%%%%%%%%%%%%%%%%%%%%%%%
\subsection{Wilson line defect half-indices}
%%%%%%%%%%%%%%%%%%%%%%%%%%%%%%%%%%%%%%%%%%%%%%%%%%%
The half-index of the class of theories under consideration admits 
a description in terms of the UV boundary conditions and multiplet data, following \cite{Gadde:2013wq,Gadde:2013sca,Yoshida:2014ssa,Dimofte:2017tpi}. 
We begin with the contribution of a vector multiplet of gauge group $G$ subject to Neumann boundary conditions, 
which furnishes the universal measure factor
\begin{align}
    \frac{(q)_\infty^{\mathrm{rank}(G)}}{|\mathrm{Wey}(G)|} 
    \oint \frac{ds}{2\pi i s}
    \prod_{\alpha\in \mathrm{root}(G)} (s^{\alpha}; q)_\infty ,
\end{align}
where the $q$-Pochhammer symbol is defined by
\begin{align}
    (x; q)_{\infty} = \prod_{n =0}^{\infty} (1 - xq^n)
\end{align}
and later we also use notation
\begin{align}
    (x^{\pm}; q)_{\infty} = (x; q)_{\infty} (x^{-1}; q)_{\infty} \; .
\end{align}

Matter contributions are incorporated multiplicatively in the integrand, with their form dictated by boundary conditions. 
The 3d chiral multiplets of R-charge $r$ transforming as the representation $\mathbf{R}$ of $G$ 
with Neumann boundary conditions contributes
\begin{align}
\prod_{\lambda \in \mathrm{wt}(\mathbf{R})} \frac{1}{(q^{\frac{r}{2}} s^{\lambda}; q)_\infty} ,
\end{align}
In contrast, imposing Dirichlet boundary conditions reverses the role of holomorphic and anti-holomorphic modes, leading to
\begin{align}
\prod_{\lambda\in \mathrm{wt}(\mathbf{R})} (q^{1 - \frac{r}{2}} s^{-\lambda} ; q)_\infty .
\end{align}

Boundary degrees of freedom in two dimensions admit a parallel description. 
In particular, the 2d $\mathcal{N}=(0,2)$ Fermi multiplet of vanishing R-charge transforming as the representation $\mathbf{R}$ of $G$ produces the contribution
\begin{align}
\prod_{\lambda \in \mathrm{wt}(\mathbf{R})} (q^{\frac{1}{2}} s^{\lambda} ; q)_\infty (q^{\frac{1}{2}} s^{-\lambda} ; q)_\infty,
\end{align}

When the matter fields transform non-trivially under such a global symmetry, 
their charges can be tracked by introducing additional fugacity parameters associated with the global symmetry group. 
Accordingly, the integrand, which originally depends on the gauge fugacities $s$, 
is naturally extended to a function of both $s$ and the global fugacities.

Let us then incorporate line defects ending on the boundary. 
A Wilson line in a representation $\mathcal{R}$ modifies the half-index 
by inserting the corresponding character, thereby projecting onto gauge invariant states in the conjugate representation $\overline{\mathcal{R}}$. 
Operationally, this amounts to the insertion of $\mathrm{Tr}_\mathcal{R}(s)$ into the integrand. 

%%%%%%%%%%%%%%%%%%%%%%%%%%%%%%%%%%%%%%%%%%%%%%%%%%%
\subsection{Vortex lines in LG theory}
\label{sec_VortexLG}
%%%%%%%%%%%%%%%%%%%%%%%%%%%%%%%%%%%%%%%%%%%%%%%%%%%

An electric Wilson line in the 3d $\mathcal{N}=2$ gauge theory is expected to admit 
a dual magnetic description as the vortex line defect associated with a singular profile of fields along its support in the path integral. 

A semiclassical realization of this correspondence can be obtained on the Coulomb branch, 
where the gauge group $G$ is broken to its maximal torus and the low-energy dynamics becomes Abelian. 
In this regime, each Cartan component of the gauge field, valued in the Cartan subalgebra $\mathfrak{t} \subset \mathfrak{g}=\mathrm{Lie}(G)$, 
can be dualized to a periodic scalar, i.e. the dual photon
\begin{align}
\gamma \in \mathfrak{t}/(2\pi \Lambda_{\mathrm{cochar}}), 
\end{align}
defined by $d\gamma = *F$. 
Together with the real scalar $\sigma$ in the vector multiplet, these fields provide a convenient description of the Coulomb branch. 
Monopole operators labeled by magnetic charges $m \in \Lambda_{\mathrm{cochar}}(G)$ admit a semiclassical description as
\begin{align}
V_m&\sim 
\exp\left(
-\langle m,\sigma\rangle+i\langle m,\gamma\rangle
\right), 
\end{align}
where $\langle \cdot,\cdot \rangle$ denotes the canonical pairing. 
These operators parametrize the Coulomb branch.

In an $s$-confining phase, 
the low-energy theory is described by the LG model of gauge invariant chiral multiplets. 
The Coulomb branch is effectively reduced to a one-dimensional direction, 
and the corresponding monopole operator is mapped to a single chiral multiplet
\begin{align}
V&\sim \exp\left(-\sigma_{\textrm{eff}}+i\gamma_{\textrm{eff}}\right), \qquad 
\gamma_{\textrm{eff}}=\langle v_{\textrm{eff}},\gamma\rangle, 
\end{align}
with $v_{\mathrm{eff}} \in \mathfrak{t}$ specifying the effective Coulomb branch direction. 

We now consider a Wilson line in a representation $\mathcal{R}$ of $G$. 
On the Coulomb branch, its effect can be captured in the Abelianized description 
by introducing a singular background connection that induces a localized magnetic flux. 
This can be represented as \cite{Kapustin:2012iw}
\begin{align}
A_{\omega}&=kd\theta, 
\end{align}
where $k \in \Lambda_{\mathrm{cochar}}(G)$ encodes the effective magnetic charge 
and $\theta$ is the angular coordinate on the plane transverse to the line operator. 
The resulting flux
\begin{align}
F&=dA_{\omega}=2\pi k\delta^{(2)}(z), 
\end{align}
induces, via the relation $d\gamma = *F$, a monodromy of the dual photon,
\begin{align}
\gamma&\sim k\theta. 
\end{align}
Projecting onto the effective Coulomb branch direction leads to
\begin{align}
\gamma_{\textrm{eff}}&\sim 
k_{\textrm{eff}} \theta, \qquad 
k_{\textrm{eff}}=\langle v_{\textrm{eff}},k\rangle. 
\end{align}
Substituting into the effective expression for $V$, we obtain
\begin{align}
V(z)&\sim \exp\left(ik_{\textrm{eff}} \theta \right). 
\end{align}
This behavior is equivalently represented as
\begin{align}
\label{singular_V}
V(z)&\sim z^{k_{\textrm{eff}}}. 
\end{align}
up to multiplication by a single-valued function. 
Thus, the Wilson line in the gauge theory will induce a singular profile for the chiral multiplet $V$ corresponding to the monopole operator in the LG description.

The singular behavior (\ref{singular_V}) of the chiral multiplet $V$, 
characterized by a zero or a pole, induces a non-trivial monodromy, which in turn results in an effective shift of the spin.
In particular, the parameter $k_{\mathrm{eff}}$ determines the shift in the $q$-grading of the contribution of the index for the chiral multiplet $V$.\footnote{The analogous effect of the vortex line on the full-index, as obtained via supersymmetric localization, was analyzed in \cite{Drukker:2012sr}.} 
In our analysis, we impose $\mathcal{N}=(0,2)$ Neumann boundary conditions for the vector and chiral multiplets on the gauge theory side. 
Under the duality, this is mapped to a boundary condition in the LG description that includes a Dirichlet boundary condition for the chiral multiplet $V$. 
Consequently, the contribution of $V$ to the half-index reflects the shifted $q$-grading dictated by $k_{\mathrm{eff}}$, 
in a manner compatible with the semiclassical analysis described above. 

However, this semiclassical analysis does not exhaust the full structure of the index. 
In particular, there exist additional contributions that cannot be accounted for solely by the singular configuration of $V$, 
and hence are not visible within the Abelianized Coulomb branch description above.
These contributions can be expressed in terms of the characters of flavor symmetries rotating the charged matter multiplets in the gauge theory, 
corresponding to the isometry of the Higgs branch.  

Accordingly, the full line defect half-index receives contributions both from the Coulomb branch sector, 
governed by the singular profile of the singlet chiral multiplet $V$ dual to the minimal monopole operator, 
and from additional contributions that incorporate, in particular, Higgs branch data. 
The normalized one-dimensional defect index 
\begin{align}
\label{normalized_hlineind}
\mathcal{I}_{\mathcal{R}_1;\mathcal{R}_2;\cdots;\mathcal{R}_k}^{\textrm{1d defect}}
=\frac{\langle W_{\mathcal{R}_1}W_{\mathcal{R}_2} \cdots W_{\mathcal{R}_k} \rangle_{\mathcal{B}}}{\mathbb{II}_{\mathcal{B}}}
\end{align}
defined as the ratio of the Wilson line defect half-index $\langle W_{\mathcal{R}_1}W_{\mathcal{R}_2} \cdots W_{\mathcal{R}_k} \rangle_{\mathcal{B}}$ 
to the half-index $\mathbb{II}_{\mathcal{B}}$ for a given boundary condition $\mathcal{B}$, 
admits a natural interpretation in terms of a moment $\mu_{*}$ of Askey-Wilson type \cite{MR2630104,MR2831874,MR2946941,MR3207468,MR3816505}. 
In the following sections, we illustrate this interpretation concretely by identifying instances of these Askey-Wilson type moments in various theories.

%Witten index
Furthermore, we consider the extraction of the Witten index for the line operator from the normalized line defect half-index (\ref{normalized_hlineind}). 
A priori, the superconformal R-charge should be determined e.g. via the extremization.
If one instead evaluates the index with a trial R-charge, the parameter $q$, which couples to R-charge, may introduce an explicit dependence on this unfixed R-charge. 
Consequently, naively setting all other fugacities to unity while keeping $q$ arbitrary, and then taking the limit $q \to 1$, can lead to an expression that retains an unphysical dependence on the trial R-charge. In such cases there is also an order of limits issue as taking $q \to 1$ before taking the flavor fugacities to one produces a different result.
On the other hand, the Witten index for the line operator should be independent of such ambiguities. 
We propose that one should first take the limit $q\to 1$, which removes the trial R-charge dependence 
and projects onto the supersymmetric ground states, thereby eliminating any residual sensitivity to the trial R-charge.
After this projection, the remaining fugacities can be safely turned off. 
Taking the limit $q\to 1$ first provides a prescription that is insensitive to the R-charge ambiguity and produces the Witten index
\begin{align}
\Tr (-1)^F&=\lim_{x \rightarrow 1} \lim_{q\rightarrow 1} \mathcal{I}_{\mathcal{R}_1;\mathcal{R}_2;\cdots;\mathcal{R}_k}^{\textrm{1d defect}}(x;q). 
\end{align}
For notational simplicity, we present expressions in which the fugacities are specialized. 
This should be understood with the prescription that the limit $q \to 1$ is taken prior to setting the remaining fugacities to one, 
in accordance with the non-commutativity of limits discussed above.
 
From a physical perspective, 
the Witten index obtained through the above construction is expected to be integer-valued 
whenever the associated one-dimensional quantum mechanics defined by the corresponding BPS line operator has a discrete spectrum. 
In this situation, the Witten index can be interpreted as a graded trace over a well-defined Hilbert space of supersymmetric ground states, 
and therefore should be protected against continuous deformations.
In all examples studied in the present work, we indeed find that the resulting Witten index takes integer values. 
However, from a purely mathematical standpoint, this integrality is not manifest from the ratios of matrix integrals, and appears to be a non-trivial property. 
This integrality property suggests the existence of an underlying combinatorial interpretation. 
In other words, the appearance of integer-valued Witten indices is expected to reflect an underlying counting problem associated with the relevant BPS sector. 
In fact, a combinatorial interpretation of the Askey-Wilson moments in terms of staircase tableaux has been proposed in the literature \cite{MR2831874,MR2630104}.
It is therefore natural to ask whether a similar combinatorial structure underlies the more general Witten indices or the Askey-Wilson type moments considered in this work. 
Exploring such a generalization appears to be an interesting direction for future research. 

%%%%%%%%%%%%%%%%%%%%%%%%%%%%%%%%%%%%%%%%%%%%%%%%%%%
%%%%%%%%%%%%%%%%%%%%%%%%%%%%%%%%%%%%%%%%%%%%%%%%%%%
\section{Askey-Wilson moments}
\label{sec_AWmoment}
%%%%%%%%%%%%%%%%%%%%%%%%%%%%%%%%%%%%%%%%%%%%%%%%%%%
%%%%%%%%%%%%%%%%%%%%%%%%%%%%%%%%%%%%%%%%%%%%%%%%%%%

%%%%%%%%%%%%%%%%%%%%%%%%%%%%%%%%%%%%%%%%%%%%%%%%%%%
\subsection{$SU(2)$ with $N_f=4$}
\label{sec_su2Nf4}
%%%%%%%%%%%%%%%%%%%%%%%%%%%%%%%%%%%%%%%%%%%%%%%%%%%

%%%%%%%%%%%%%%%%%%%%%%%%%%%%%%%%%%%%%%%%%%%%%%%%%%%
\subsubsection{Boundary conditions}
%%%%%%%%%%%%%%%%%%%%%%%%%%%%%%%%%%%%%%%%%%%%%%%%%%%
Consider theory A as a 3d $\mathcal{N}=2$ $SU(2)$ gauge theory with 
$N_f=4$ fundamental chirals $Q_{\alpha}$, $\alpha \in \{1,2,3,4\}$ with R-charge $r$.
This $SU(2)$ gauge theory is free from gauge anomaly without any 2d chiral or Fermi multiplets
when we impose the $\mathcal{N}=(0,2)$ Neumann boundary conditions for the $SU(2)$ vector multiplet and the chiral multiplets. 

The half-index is
\begin{align}
\label{su2_4a}
\mathbb{II}^{A}_{(\mathcal{N},N)}
&=\frac{(q)_{\infty}}{2}
\oint \frac{ds}{2\pi is}
(s^{\pm2};q)_{\infty}
\prod_{\alpha=1}^{4}
\frac{1}{(q^{\frac{r}{2}} s^{\pm} a x_{\alpha};q)_{\infty}}, 
\end{align}
where $\prod_{\alpha = 1}^4 x_{\alpha} = 1$.
Here the $x_{\alpha}$ are the fugacities for the global $SU(N_f = 4)$ flavor symmetry while $a$ is the fugacity for the $U(1)_a$ axial symmetry. Throughout $q$ is used for the fugacity combining the $U(1)_R$ R-symmetry and the spin, i.e.\ the power of $q$ in a $q$-series identifies $J + \frac{1}{2}R$ where $J$ is the spin and $R$ is the R-charge.
The integral (\ref{su2_4a}) is identified with the Askey-Wilson integral \cite{MR783216} 
which is the $q$-extension of the classical beta integral. 
It is equal to 
\begin{align}
\label{su2_4b}
\mathbb{II}^{B}_{N, D} & = 
\frac{(q^{2r} a^4;q)_{\infty}}
{\prod_{\alpha<\beta}^{4} (q^{r} a^2 x_{\alpha} x_{\beta};q)_{\infty}} \; . 
\end{align}
As discussed in \cite{Okazaki:2023hiv}, 
the expression (\ref{su2_4b}) can be interpreted as the half-index of Theory B which is a theory of chiral multiplets consisting of 
a chiral multiplet $M_{\alpha\beta}$ with Neumann boundary condition transforming 
as the rank-$2$ antisymmetric representation of the $SU(4)$ flavor symmetry 
and a chiral multiplet $V$ with $U(1)_a$ charge $-4$ with Dirichlet boundary conditions.
The field content of these dual theories is summarized as
\begin{align}
\label{SU2_4_charges}
\begin{array}{c|c|c|c|c|c}
& \textrm{bc} & SU(2) & 
SU(4)& U(1)_a & U(1)_R \\ \hline
\textrm{VM} & \mathcal{N} & {\bf Adj} & {\bf 1} & 0 & 0 \\
Q_{\alpha} & \textrm{N} & {\bf 2} & {\bf 4} & 1 & r \\
 \hline
M_{\alpha \beta} & \textrm{N} & {\bf 1} & {\bf 6} & 2 & 2r \\
V & \textrm{D} & {\bf 1} & {\bf 1} & -4 & 2 - 4r
\end{array}
\end{align}
Note that the dependence of the R-charge on the parameter $r$ corresponds to the $U(1)_a$ charge so for simplicity later we simply set $r = 0$ but it can always be restored by the redefinition of the $U(1)_a$ fugacity $a \to a q^{r}$ and the corresponding shift of the R-charge.

%%%%%%%%%%%%%%%%%%%%%%%%%%%%%%%%%%%%%%%%%%%%%%%%%%%
\subsubsection{One-point function}
%%%%%%%%%%%%%%%%%%%%%%%%%%%%%%%%%%%%%%%%%%%%%%%%%%%
%Wilson line
In the presence of boundary, we can further introduce Wilson line operators in the $SU(2)$ gauge theory. 
Then the Neumann half-index (\ref{su2_4a}) is decorated by inserting characters of the Wilson line representation into the integrand of \eqref{su2_4a}. 
For example the fundamental Wilson line defect half-index is 
\begin{align}
\label{su2_4a_FundWL}
\langle W_{\tiny \yng(1)}\rangle^{A}_{(\mathcal{N},N)}
&=\frac{(q)_{\infty}}{2}
\oint \frac{ds}{2\pi is}
(s^{\pm2};q)_{\infty}
\left( \prod_{\alpha=1}^{4}
\frac{1}{(q^{\frac{r}{2}} s^{\pm} a x_{\alpha};q)_{\infty}} \right)
(s+s^{-1}). 
\end{align}
We find that it agrees with 
\begin{align}
\label{su2_4b_FundWL}
&
\mathcal{I}^{\textrm{1d defect}}_{\tiny \yng(1)}(a,x_{\alpha};q)
\times 
\mathbb{II}^{B}_{N, D}
 \; , 
\end{align}
where
\begin{align}
\label{1d_su2Nf4_1ptQM}
\mathcal{I}^{\textrm{1d defect}}_{\tiny \yng(1)}(a,x_{\alpha};q)
&=\frac{q^{\frac{r}{2}}a
\left(
\sum_{\alpha=1}^{4}x_{\alpha}
-q^{r}a^2 \sum_{\alpha=1}^4x_{\alpha}^{-1}
\right)
}
{
1 - q^{2r} a^4
}
\end{align}
is the normalized Wilson line defect half-index that is attributed to one-dimensional quantum mechanical degrees of freedom localized on the line operator. 
We note that the denominator factor in (\ref{1d_su2Nf4_1ptQM}) arises from the contribution of the singlet field $V$. This does not introduce a pole in $\langle W_{\tiny \yng(1)}\rangle^{A}_{(\mathcal{N},N)}$ as it cancels a factor in the numerator of $\mathbb{II}^{B}_{N, D}$ through the following shift
\begin{align}
q^{2r}a^4\rightarrow q^{2r+1}a^4
\end{align}
in $(q^{2r} a^4;q)_{\infty}$. This is the spin shift effect alluded to in sections~\ref{sec_intro} and \ref{sec_ind}.

In the unflavored limit $a\rightarrow 1$, $x_{\alpha}\rightarrow 1$ the expression (\ref{1d_su2Nf4_1ptQM}) reduces to
\begin{align}
\mathcal{I}^{\textrm{1d defect}}_{\tiny \yng(1)}(1,1;q)
&=\frac{4q^{\frac{r}{2}}}{1+q^{r}}. 
\end{align}

%proof
A key observation is that 
the above result can be derived from general properties of the Askey-Wilson polynomials \cite{MR783216}. 
In particular, the structure of the expression follows directly from their defining orthogonality, thereby providing a conceptual explanation of the formula. 
The Askey-Wilson polynomials are defined by \cite{MR783216}
\begin{align}
\label{AskeyWilsonPoly_def}
    p_n(\cos \theta) & = a^{-n} x_1^{-n} \left( \prod_{\alpha = 2}^4 (a^2 x_1 x_{\alpha}; q)_n \right) {}_4\phi_3 \left[ \begin{array}{cccc} q^{-n} & a^4 q^{n-1} & a x_1 s & a x_1 s^{-1} \\ a^2 x_1 x_2 & a^2 x_1 x_3 & a^2 x_1 x_4 & \end{array} ; q, q \right], 
\end{align}
with $\cos \theta = \frac{s + s^{-1}}{2}$. 
They obey the orthogonality relation
\begin{align}
\label{AskeyWilsonPoly_orthog}
    \frac{(q)_{\infty}}{2}
\oint \frac{ds}{2\pi is}
(s^{\pm2};q)_{\infty}
\left( \prod_{\alpha=1}^{4}
\frac{1}{(q^{\frac{r}{2}} s^{\pm} a x_{\alpha};q)_{\infty}} \right)
p_m p_n
 & = \tilde{h}_n \delta_{mn}, 
\end{align}
where $\tilde{h}_n$ are the algebraic expressions
\begin{align}
\label{AW_h_n}
\tilde{h}_n & = 
\frac{(q; q)_n (q^{n - 1 + 2r} a^4;q)_{\infty}}
{(1 - q^{2n - 1}a^4) \prod_{\alpha<\beta}^{4} (q^{n + r} a^2 x_{\alpha} x_{\beta};q)_{\infty}} \; . 
\end{align}
In particular, the case of $m = n = 0$ gives the Askey-Wilson integral
which is the equality of \eqref{su2_4a} with \eqref{su2_4b}.

Above, ${}_4\phi_3$ is a unilateral $q$-hypergeometric (or basic hypergeometric) function which is defined in general by
\begin{align}
    {}_j\phi_k  \left[ \begin{array}{ccc} a_1 \cdots & a_j \\ b_1 \cdots & b_k \end{array} ; q, z \right] = & \sum_{m = 0}^{\infty} \frac{(a_1, \ldots , a_j; q)_m}{(b_1, \ldots , b_k, q; q)_m} \left( (-1)^m q^{\frac{1}{2}m(m-1)} \right)^{1 + k - j} z^m, 
\end{align}
where $(a_1, \ldots , a_j; q)_m = \prod_{I = 1}^j (a_I; q)_m$.
So, in the case of $j = k + 1$ we have the simpler expression
\begin{align}
    {}_{k+1}\phi_k  \left[ \begin{array}{ccc} a_1 \cdots & a_{k+1} \\ b_1 \cdots & b_k \end{array} ; q, z \right] = & \sum_{m = 0}^{\infty} \frac{(a_1, \ldots , a_j; q)_m}{(b_1, \ldots , b_k, q; q)_m} z^m, 
\end{align}

Note that $p_0 = 1$ while
\begin{align}
    p_1 = & a^{-1}x_1^{-1} (1 - a^2 x_1 x_2) (1 - a^2 x_1 x_3) (1 - a^2 x_1 x_4)
    \nonumber \\
    & + \frac{q (1 - q^{-1}) (1 - a^4)}{a x_1 (1 - q)}\left(1 - ax_1(s + s^{-1}) + a^2x_1^2 \right)
    \nonumber \\
    = & (1 - a^4)(s + s^{-1}) + \sum_{\alpha = 1}^4 (-ax_{\alpha} + a^3 x_{\alpha}^{-1}) \; ,
\end{align}
where for simplicity we have set $r = 0$ as the dependence can be restored simply by replacing $a \to q^{r/2} a$. It is then trivial to use the Askey Wilson integral along with the orthogonality of $p_1$ and $p_0$ in \eqref{AskeyWilsonPoly_orthog} to derive the equality of \eqref{su2_4a_FundWL} with \eqref{su2_4b_FundWL}.

%%%%%%%%%%%%%%%%%%%%%%%%%%%%%%%%%%%%%%%%%%%%%%%%%%%
\subsubsection{Multi-point functions}
%%%%%%%%%%%%%%%%%%%%%%%%%%%%%%%%%%%%%%%%%%%%%%%%%%%
Let us consider, more generally, multi-point functions of fundamental Wilson lines of the form
\begin{align}
\label{su2_4a_FundWLmult}
\langle \underbrace{W_{\tiny \yng(1)}\cdots W_{\tiny \yng(1)}}_{n} \rangle^{A}_{(\mathcal{N},N)}
&=\frac{(q)_{\infty}}{2}
\oint \frac{ds}{2\pi is}
(s^{\pm2};q)_{\infty}
\left( \prod_{\alpha=1}^{4}
\frac{1}{(q^{\frac{r}{2}} s^{\pm} a x_{\alpha};q)_{\infty}} \right)
(s+s^{-1})^n. 
\end{align}
We observe that these quantities can be exactly computed as the Askey-Wilson moments $\mu_n(a, x_{\alpha}; q)$ \cite{MR2946941}, 
thereby placing them within the well-established framework of Askey-Wilson orthogonal polynomials. 
To make this identification precise, we recall the definition of the Askey-Wilson moments.  
The Askey-Wilson moments $\mu_n(a, x_{\alpha}; q)$ are defined as \cite{MR2946941}\footnote{Our normalization conventions differ from \cite{MR2946941} by a factor of $2^n$.} 
\begin{align}
\label{AskeyWilson_moments}
    \mu_n(a, x_{\alpha}; q) \mathbb{II}^{A}_{(\mathcal{N},N)}
    & = 
    \frac{(q)_{\infty}}{2}
\oint \frac{ds}{2\pi is}
(s^{\pm2};q)_{\infty}
\left( \prod_{\alpha=1}^{4}
\frac{1}{(q^{\frac{r}{2}} s^{\pm} a x_{\alpha};q)_{\infty}} \right)
(s+s^{-1})^n \; .
\end{align}
In other words, they are identified with the normalized $n$-point Wilson line defect half-indices
\begin{align}
 \mu_n(a, x_{\alpha}; q)
&:=\frac{\langle \underbrace{W_{\tiny \yng(1)}\cdots W_{\tiny \yng(1)}}_{n} \rangle^{A}_{(\mathcal{N},N)}}
{\mathbb{II}^{A}_{(\mathcal{N},N)}}
=\mathcal{I}^{\textrm{1d defect}}_{\underbrace{\tiny\yng(1); \cdots ;\yng(1)}_{n}}(a,x_{\alpha};q), 
\end{align}
which should encode the one-dimensional quantum mechanical degrees of freedom localized on a collection of $n$ fundamental Wilson line operators. 

Any normalized Wilson line defect half-index can then be written as a linear combination of $\mu_n(a, x_{\alpha}; q)$ with constant coefficients. 
The Askey-Wilson moments can be evaluated using the orthogonal properties of the Askey-Wilson polynomials 
and this has been calculated in \cite{MR2946941} to give the closed form expression
\begin{align}
\label{AskeyWilson_moments_eval}
    \mu_n(a, x_{\alpha}; q)
    = & \sum_{k = 0}^n \frac{\prod_{\alpha = 2}^4 (a^2 x_1 x_{\alpha}; q)_{k}}{(a^4; q)_{k}} q^k \sum_{j = 0}^k q^{-j^2} a^{-2j} x_1^{-2j}
    \nonumber \\
    & \frac{(q^j a x_1 + q^{-j} a^{-1} x_1^{-1})^n}{(q; q)_{j} (q^{1 - 2j} a^{-2} x_1^{-2}; q)_{j} (q; q)_{k - j} (q^{1 + 2j} a^{2} x_1^{2}; q)_{k - j}}
    \; ,
\end{align}
which, like the Askey-Wilson polynomials, is symmetric in the $x_{\alpha}$ for all $\alpha$ although the symmetry is only manifest in this expression for $\alpha \in \{2, 3, 4\}$. 
In addition $(a^4; q)_n \mu_n$ is polynomial in $q$ but this is not manifest from the above expression. 
An explicit expression for the moments which has manifest symmetry for $\alpha \in \{1, 2, 3, 4\}$ has been found in \cite{MR3207468}, 
but it is even less convenient to calculate and it is still not manifestly polynomial in $q$ after multiplying by $(a^4; q)_n$.

The formula (\ref{AskeyWilson_moments_eval}) shows that 
any normalized Wilson line defect half-indices evaluates to a rational function multiplying the half index \eqref{su2_4b}. 
Since $(a^4; q)_n \mu_n$ is a polynomial we have the result for the $n$-point function of fundamental Wilson lines that
\begin{align}
\label{AskeyWilson_moments_poly}
\langle \underbrace{W_{\tiny \yng(1)}\cdots W_{\tiny \yng(1)}}_{n} \rangle^{A}_{(\mathcal{N},N)}
    & = 
    \left( (a^4; q)_n \mu_n \right) \frac{(q^{n} a^4;q)_{\infty}}
{\prod_{\alpha<\beta}^{4} (a^2 x_{\alpha} x_{\beta};q)_{\infty}} \; ,
\end{align}
where we now see the spin shift by $n$.
Here we have again set $r= 0$ for simplicity but it can be restored by replacing $a$ with $q^{r/2} a$.

We now introduce the following basis of symmetric polynomials in the four variables $X_{\alpha} = a x_{\alpha}$
\begin{align}
    \label{Basis_Sym4poly}
    P_1 = S_{(1, 0, 0, 0)}(X_{\alpha}) = & \sum_{\alpha} X_{\alpha} = a \sum_{\alpha} x_{\alpha} \equiv a p_1, \\
    P_2 = S_{(1, 1, 0, 0)}(X_{\alpha}) = & \sum_{\alpha < \beta} X_{\alpha} X_{\beta} = a^2 \sum_{\alpha < \beta} x_{\alpha} x_{\beta} \equiv a^2 p_2, \\
    P_3 = S_{(1, 1, 1, 0)}(X_{\alpha}) = & \left( \prod_{\alpha} X_{\alpha} \right) \sum_{\beta} X_{\beta}^{-1} = a^3 \sum_{\alpha} x_{\alpha}^{-1} \equiv a^3 p_3, \\
    P_4 = S_{(1, 1, 1, 1)}(X_{\alpha}) = & \prod_{\alpha} X_{\alpha} = a^4 \; ,
\end{align}
where in general the Schur polynomials in $n$ variables $w_i$ are defined as
\begin{align}
\label{SchurPoly_defn}
    S_{\underline{\lambda}}(w_i) = & \frac{\det(w_j^{\lambda_i + n - i})_{i, j = 1}^n}{\det(w_j^{n - i})_{i, j = 1}^n}, 
\end{align}
where $\underline{\lambda}$ is an integer partition with $\lambda_1 \ge \lambda_2 \ge \cdots \lambda_n \ge 0$.
Any symmetric polynomial can then be expressed as a polynomial in $a, p_1, p_2, p_3$. Note that $p_1, p_2$ and in particular $p_3$ are polynomials in $x_{\alpha}$ since $\prod_{\alpha} x_{\alpha} = 1$.

For example we have
\begin{align}
\label{AW4_moments}
    (a^4; q)_0 \mu_0 = & 1, \\
    (a^4; q)_1 \mu_1 = & a p_1 - a^3 p_3, \\
    %a \sum_{\alpha} x_{\alpha} - a^3 \sum_{\alpha} x_{\alpha}^{-1} \\
    (a^4; q)_2 \mu_2 = & \left( 1 + a^2(p_1^2 - p_2) - a^4 p_1 p_3 + a^6 p_2 - a^8 \right)
    \nonumber \\
    & + q \left( -1 + a^2 p_2 - a^4 p_1 p_3 + a^6 (-p_2 + p_3^2) + a^8 \right), \\
    %\left( 1 + a^2 \sum_{\alpha \le \beta} x_{\alpha}x_{\beta} - a^4 \sum_{\alpha, \beta} x_{\alpha}x_{\beta}^{-1} + a^6 \sum_{\alpha < \beta} x_{\alpha}x_{\beta} - a^8 \right)
    %\nonumber \\
    %& + q \left( -1 + a^2 \sum_{\alpha < \beta} x_{\alpha}x_{\beta} - a^4 \sum_{\alpha, \beta} x_{\alpha}x_{\beta}^{-1} + a^6 \sum_{\alpha \le \beta} x_{\alpha}^{-1} x_{\beta}^{-1} + a^8 \right) \\
    (a^4; q)_3 \mu_3 = & 2 a  p_1 - a^9 p_1 - a^7 ( p_3- p_1  p_2) - a^5 \left( p_1^2  p_3- p_1- p_2  p_3\right)
    \nonumber \\
    & -a^3 \left(- p_1^3+2  p_1  p_2+ p_3\right)
    +q \Bigl(a^{11} p_3 - a^9 p_2 p_3 + a^7 \left( p_1  p_3^2 + p_3\right) 
    \nonumber\\
    &- a^5 \left( p_1^2  p_3+ p_1\right) + a^3  p_1  p_2 - a  p_1\Bigr)
    \nonumber \\
    & +q^2 \left(a^{11} p_3 - a^9 p_2 p_3 + a^7 \left( p_1  p_3^2 + p_3\right) - a^5 \left( p_1^2  p_3+ p_1\right) + a^3  p_1  p_2 - a  p_1\right)
    \nonumber \\
    & +q^3 \Bigl( -2 a^{11}  p_3+a^9 \left( p_1 + 2  p_2  p_3 - p_3^3\right) - a^7 \left( p_1 \left( p_2- p_3^2\right)+ p_3\right) 
    \nonumber\\
    &- a^5 ( p_2  p_3- p_1) + a^3  p_3\Bigr). 
\end{align}

In the unflavored case $x_{\alpha} = 1$ we have the simpler expressions for the moments
\begin{align}
    (a^4; q)_0 \mu_0 = & 1, \\
    (a^4; q)_1 \mu_1 = & 4a(1 - a^2), \\
    (a^4; q)_2 \mu_2 = & (1 - a^2)\left( (1 - q) + a^2(11 + 5q) - a^4(5 + 11q) + (1 - q)a^6 \right), \\
    (a^4; q)_3 \mu_3 = & 4a(1 - a^2) \left( (2 + 5a^2 - 4a^4 + a^6) + (-1 + 5a^2 - 12a^4 + 5a^6 - a^8)q(1 + q) \right.
    \nonumber \\
    & \left. + (a^2 - 4a^4 + 5a^6 + 2a^8)q^3 \right) \; .
\end{align}
and those for the 1d defect indices 
\begin{align}
\mathcal{I}^{\textrm{1d defect}}_{\tiny\yng(1);\yng(1)}
&=\frac{1-q+11q^r-5q^{2r}+q^{3r}+5q^{1+r}-11q^{2r+1}-q^{3r+1}}{(1+q^r)(1-q^{2r+1})}, \\
\mathcal{I}^{\textrm{1d defect}}_{\tiny\yng(1);\yng(1);\yng(1)}
&=\frac{4q^{\frac{r}{2}}}{(1-q^{2r+1})(1+q^r)(1+q^{r+1})}
\Bigl(
2-q-q^2+5q^r-4q^{2r}+q^{3r}
\nonumber\\
&+7q^{r+1}+4q^{r+2}-7q^{2r+1}-5q^{2r+2}+q^{3r+1}-2q^{3r+2}
\Bigr), 
\end{align}
where we have also set $a$ to unity. 

By turning off (setting to $1$) all fugacities in the one-dimensional defect index corresponding to the normalized Wilson line defect correlator, 
one obtains the Witten index. 
The result takes the following form:
\begin{align}
\mathcal{I}^{\textrm{1d defect}}_{\underbrace{\tiny\yng(1); \cdots ;\yng(1)}_{n}}(1,1;1)= & 2^n \; .
\end{align}

%%%%%%%%%%%%%%%%%%%%%%%%%%%%%%%%%%%%%%%%%%%%%%%%%%%
\subsection{$SU(2)_{-k}$ with $N_f = 4 - 2k$ from Askey-Wilson}
%%%%%%%%%%%%%%%%%%%%%%%%%%%%%%%%%%%%%%%%%%%%%%%%%%%

%%%%%%%%%%%%%%%%%%%%%%%%%%%%%%%%%%%%%%%%%%%%%%%%%%%
\subsubsection{Boundary conditions}
%%%%%%%%%%%%%%%%%%%%%%%%%%%%%%%%%%%%%%%%%%%%%%%%%%%
We can use the results for the case of $N_f = 4$ in terms of Askey-Wilson moments to derive exact expressions for the Wilson line defect half-indices of theories with $N_f < 4$. 
This involves sending real masses of chirals to $\pm \infty$. 
In the $SU(2)$ bulk theory we can do this with either sign but with the Neumann boundary conditions we have a simple result only for the plus sign. In terms of the half-indices the derivation of the results for $N_f < 4$ is straightforward:
\begin{itemize}
    \item Define flavor fugacities $X_{\alpha} = a x_{\alpha}$.
    \item Send $k$ of the $X_{\alpha} \to 0$.
    \item Redefine $a$ and the remaining $N_f = 4 - k$ flavor fugacities as $X_{\alpha} = a x_{\alpha}$ with $\prod_{\alpha = 1}^{N_f} x_{\alpha} = 1$.
\end{itemize}
The effect in the dual theory is that we are left with $M_{\alpha \beta}$ with the reduced number of components (now an antisymmetric $N_f \times N_f$ matrix) and the chiral $V$ is removed. The latter is seen from the above limit as $a^4 \to 0$.

For the half-indices with $N_f \in \{0, 1, 2, 3\}$ we have matching of
\begin{align}
\label{su2_4mka}
\mathbb{II}^{A}_{(\mathcal{N},N)}
&=\frac{(q)_{\infty}}{2}
\oint \frac{ds}{2\pi is}
(s^{\pm2};q)_{\infty}
\prod_{\alpha=1}^{N_f}
\frac{1}{(q^{\frac{r}{2}} s^{\pm} a x_{\alpha};q)_{\infty}}
\end{align}
where $\prod_{\alpha = 1}^{N_f} x_{\alpha} = 1$,
with 
\begin{align}
\label{su2_4mkb}
\mathbb{II}^{B}_{N} & = 
\prod_{\alpha<\beta}^{N_f}
\frac{1}
{(q^{r} a^2 x_{\alpha} x_{\beta};q)_{\infty}}. 
\end{align}
This demonstrates the duality between the Neumann boundary conditions for the $SU(2)$ vector and the chiral multiplets,
and the Neumann boundary condition for the chiral multiplet $M_{\alpha \beta}$. 

%%%%%%%%%%%%%%%%%%%%%%%%%%%%%%%%%%%%%%%%%%%%%%%%%%%
\subsubsection{Line defect half-indices}
%%%%%%%%%%%%%%%%%%%%%%%%%%%%%%%%%%%%%%%%%%%%%%%%%%%
We now turn to the $n$-point fundamental Wilson line defect half-indices. 
They take the form
\begin{align}
\label{su2_Nf<4}
\langle \underbrace{W_{\tiny \yng(1)}\cdots W_{\tiny \yng(1)}}_{n} \rangle^{A}_{(\mathcal{N},N)}
&=\frac{(q)_{\infty}}{2}
\oint \frac{ds}{2\pi is}
(s^{\pm2};q)_{\infty}
\left( \prod_{\alpha=1}^{N_f}
\frac{1}{(q^{\frac{r}{2}} s^{\pm} a x_{\alpha};q)_{\infty}} \right)
(s+s^{-1})^n. 
\end{align}
These expressions for the Wilson line half-indices can similarly be evaluated from \eqref{AskeyWilson_moments} and \eqref{AskeyWilson_moments_eval} 
by taking the same limits $X_{\alpha} \to 0$ for $\alpha > N_f$. 
The associated moments are defined by
\begin{align}
\label{AskeyWilson_Nf_moments}
    \mu_n^{(N_f)}(a, x_{\alpha}; q) \mathbb{II}_{(\mathcal{N},N)}^A
    & = \langle \underbrace{W_{\tiny \yng(1)}\cdots W_{\tiny \yng(1)}}_{n} \rangle^{A}_{(\mathcal{N},N)}
\end{align}
and they involve the one-dimensional quantum mechanical degrees of freedom localized on a collection of $n$ fundamental Wilson line operators. 
Hence we refer to them as the 1d defect index
\begin{align}
 \mu_n^{(N_f)}(a, x_{\alpha}; q)&=\mathcal{I}^{\textrm{1d defect}}_{\underbrace{\tiny\yng(1); \cdots ;\yng(1)}_{n}}(a,x_{\alpha};q). 
\end{align}

Again, any normalized Wilson line defect half-index can then be written as a linear combination of $\mu_n^{(N_f)}(a, x_{\alpha}; q)$ with constant coefficients. 
For $N_f > 0$ the moments have the closed form expression
\begin{align}
\label{AskeyWilson_Nf_moments_eval}
    \mu_n^{(N_f)}(a, x_{\alpha}; q)
    = & \sum_{k = 0}^n \left( \prod_{\alpha = 2}^{N_f} (a^2 x_1 x_{\alpha}; q)_{k} \right) q^k \sum_{j = 0}^k q^{-j^2} a^{-2j} x_1^{-2j}
    \nonumber \\
    & \frac{(q^j a x_1 + q^{-j} a^{-1} x_1^{-1})^n}{(q; q)_{j} (q^{1 - 2j} a^{-2} x_1^{-2}; q)_{j} (q; q)_{k - j} (q^{1 + 2j} a^{2} x_1^{2}; q)_{k - j}}
    \; .
\end{align}
For $N_f < 4$ the moments $\mu_n$ are in fact polynomials although the expression \eqref{AskeyWilson_Nf_moments_eval} is only manifestly rational. 
This follows from the fact that for $N_f = 4$ the moments are rational functions with denominator $(a^4; q)_n$ and taking the limit to reduce the number of flavors sends $a \to 0$.

Kim and Stanton \cite{MR3207468} provided expressions for the moments which are manifestly symmetric (and polynomial) in $x_{\alpha}$. 
The result for $N_f = 4$ is rather unwieldy and, as previously noted, is anyway not manifestly rational in $q$ so we do not present it. 
However, for $N_f < 4$ we have the following expressions which are manifestly polynomial as well as symmetric in $x_{\alpha}$.
\begin{align}
\label{AskeyWilson_moments_sympoly}
    \mu_n^{(N_f = 3)} = & \sum_{k = 0}^n \left( \left( \begin{array}{c} n \\ \frac{n-k}{2} \end{array} \right) - \left( \begin{array}{c} n \\ \frac{n-k}{2} - 1 \end{array} \right) \right)
     \sum_{u+v+w+2t = k} a^{u+v+w} (-1)^t x_1^u x_2^v x_3^w q^{\frac{1}{2}t(t+1)} 
     \nonumber \\
    &
    \times \left[ \begin{array}{c} u+v+t \\ v \end{array} \right]_q
    \left[ \begin{array}{c} v+w+t \\ w \end{array} \right]_q
    \left[ \begin{array}{c} w+u+t \\ u \end{array} \right]_q, 
    \\
    \mu_n^{(N_f = 2)} = & \sum_{k = 0}^n \left( \left( \begin{array}{c} n \\ \frac{n-k}{2} \end{array} \right) - \left( \begin{array}{c} n \\ \frac{n-k}{2} - 1 \end{array} \right) \right)
    \nonumber\\
    &\times \sum_{u+v+2t = k} a^{u+v} (-1)^t x^{u-v} q^{\frac{1}{2}t(t+1)} \left[ \begin{array}{c} u+v+t \\ u, v, t \end{array} \right]_q, 
    \\
    \mu_n^{(N_f = 1)} = & \sum_{k = 0}^n \left( \left( \begin{array}{c} n \\ \frac{n-k}{2} \end{array} \right) - \left( \begin{array}{c} n \\ \frac{n-k}{2} - 1 \end{array} \right) \right)
    \sum_{u+2t = k} a^{u} (-1)^t q^{\frac{1}{2}t(t+1)} \left[ \begin{array}{c} u+t \\ u \end{array} \right]_q, 
    \\
    \mu_n^{(N_f = 0)} = & \sum_{k = 0}^n \left( \left( \begin{array}{c} n \\ \frac{n-k}{2} \end{array} \right) - \left( \begin{array}{c} n \\ \frac{n-k}{2} - 1 \end{array} \right) \right)
    (-1)^t q^{\frac{1}{2}t(t+1)} \Big\vert_{2t = k}
    \; ,
\end{align}
where the $q$-binomial coefficients are defined by
\begin{align}
\left[
\begin{array}{c}
n\\
k\\
\end{array}
\right]_{q}
&:=\frac{[n]_q [n-1]_q\cdots [n-k+1]_q}{[1]_q [2]_q \cdots [k]_q}, \\
[n]_q&:=\frac{1-q^n}{1-q}. 
\end{align}
Here in each case we have $\prod_{\alpha = 1}^{N_f} x_{\alpha} = 1$ and $u, v, w, t \in \Zb$. 
The sums over $k$ are restricted to even or odd $k$ so that $\frac{n-k}{2} \in \Zb$.  
For $N_f = 3$ the sum is over $0 \le u, v, w \le k$. 
For $N_f =2$ we have labelled $x_1 = x$ and $x_2 = x^{-1}$ and the sum is over $u, v \ge 0$ with also $t \ge 0$. 
For $N_f =1$ the sum is over $0 \le u \le k$. In all cases $k$ is restricted to integer values such that $n-k$ is even and further restrictions arise from the requirement that $t \in \Zb$. 
Let us now examine explicitly the cases with $N_f=3,2,1,0$. 

%%%%%%%%%%%%%%%%%%%%%%%%%%%%%%%%%%%%%%%%%%%%%%%%%%%
\subsubsection{$N_f=3$}
%%%%%%%%%%%%%%%%%%%%%%%%%%%%%%%%%%%%%%%%%%%%%%%%%%%
%Nf=3
For $N_f = 3$ we have the following examples
\begin{align}
\label{AW3_moments}
    \mu_0 = & 1, \\
    \label{AW3_moments1}
    \mu_1 = & a p_1 - a^3, \\
    \label{AW3_moments2}
    \mu_2 = & \left( 1 + a^2(p_1^2 - p_2) - a^4 p_1 \right)
    \nonumber \\
    & + q \left( -1 + a^2 p_2 - a^4 p_1  + a^6 \right), \\
    \label{AW3_moments3}
    \mu_3 = & 2 a  p_1 - a^5 \left( p_1^2   - p_2  \right)+a^3 \left(- p_1^3+2  p_1  p_2+ 1 \right)
    \nonumber \\
    & +q \left( a^7 p_1 - a^5 p_1^2 + a^3  p_1  p_2 - a  p_1\right)
    \nonumber \\
    & +q^2 \left( a^7 p_1 - a^5 p_1^2 + a^3  p_1  p_2 - a  p_1\right)
    \nonumber \\
    & +q^3 \left( -a^9 + a^7 p_1 - a^5 p_2 + a^3 \right), 
\end{align}
where now
\begin{align}
p_1 = & \sum_{\alpha = 1}^3 x_{\alpha} \\
p_2 = & \sum_{\alpha = 1}^3 x_{\alpha}^{-1},
\end{align}
with $\prod_{\alpha = 1}^3 x_{\alpha} = 1$. 
Consequently, we obtain the exact expressions for the line defect indices. 
In the unflavored limit they are given by
\begin{align}
\mathcal{I}^{\textrm{1d defect}}_{\tiny\yng(1)}(1,1;q)
&=3q^{\frac{r}{2}}-q^{\frac{3r}{2}}, \\
\mathcal{I}^{\textrm{1d defect}}_{\tiny\yng(1);\yng(1)}1,1;q)
&=1-q+6q^{r}-3q^{2r}+3q^{r+1}-3q^{2r+1}. 
\end{align}

%%%%%%%%%%%%%%%%%%%%%%%%%%%%%%%%%%%%%%%%%%%%%%%%%%%
\subsubsection{$N_f=2$}
%%%%%%%%%%%%%%%%%%%%%%%%%%%%%%%%%%%%%%%%%%%%%%%%%%%

%Nf=2
In the case of $N_f = 2$ the Askey-Wilson polynomials reduce to Al-Salam-Chihara polynomials \cite{MR399537} by setting $X_3 = X_4 = 0$. 
So all the results can also be expressed in terms of properties of the Al-Salam-Chihara orthogonal polynomials \cite{MR399537} and their moments \cite{MR2803800}.
For example, we find
\begin{align}
\label{AW2_moments}
    \mu_0 = & 1, \\
    \mu_1 = & a p_1, \\
    \mu_2 = & \left( 1 + a^2(p_1^2 - 1) \right)
    + q \left( -1 + a^2 \right), \\
    \mu_3 = & 2 a  p_1 + 2 a^3 p_1
    + q \left( a^3  p_1 - a  p_1\right)
    + q^2 \left( a^3  p_1 - a  p_1\right),
\end{align}
where now
\begin{align}
p_1 = & \sum_{\alpha = 1}^2 x_{\alpha}, 
\end{align}
with $\prod_{\alpha = 1}^2 x_{\alpha} = 1$. 
The unflavored defect indices are given by
\begin{align}
\mathcal{I}^{\textrm{1d defect}}_{\tiny\yng(1)}(1,1;q)
&=2q^{\frac{r}{2}}, \\
\mathcal{I}^{\textrm{1d defect}}_{\tiny\yng(1);\yng(1)}1,1;q)
&=1-q+3q^{r}+q^{r+1}. 
\end{align}

%%%%%%%%%%%%%%%%%%%%%%%%%%%%%%%%%%%%%%%%%%%%%%%%%%%
\subsubsection{$N_f=1$}
%%%%%%%%%%%%%%%%%%%%%%%%%%%%%%%%%%%%%%%%%%%%%%%%%%%
For $N_f = 1$ we have the even simpler expressions
\begin{align}
\label{AW1_moments}
    \mu_0 = & 1, \\
    \mu_1 = & a, \\
    \mu_2 = & \left( 1 + a^2 \right) - q, \\
    \mu_3 = & 2 a - q a - q^2 a \; .
\end{align}
The unflavored defect indices are 
\begin{align}
\mathcal{I}^{\textrm{1d defect}}_{\tiny\yng(1)}(1,1;q)
&=q^{\frac{r}{2}}, \\
\mathcal{I}^{\textrm{1d defect}}_{\tiny\yng(1);\yng(1)}1,1;q)
&=1-q+q^{r}. 
\end{align}

%%%%%%%%%%%%%%%%%%%%%%%%%%%%%%%%%%%%%%%%%%%%%%%%%%%
\subsubsection{$N_f=0$}
%%%%%%%%%%%%%%%%%%%%%%%%%%%%%%%%%%%%%%%%%%%%%%%%%%%

%Nf=0
For $N_f = 0$ we see that 
both $k$ and $n-k$ must be even, hence both $k$ and $n$ must be even, to have a non-zero contribution. 
Therefore $\mu_{2m+1}^{(N_f = 0)} = 0$ while
\begin{align}
\label{AskeyWilson_moments_Nf0}
    \mu_{2m}^{(N_f = 0)} = & \sum_{t = 0}^m \left( \left( \begin{array}{c} 2m \\ m - t \end{array} \right) - \left( \begin{array}{c} 2m \\ m - t - 1 \end{array} \right) \right)
    (-1)^t q^{\frac{1}{2}t(t+1)}
    \; ,
\end{align}
For example, we have 
\begin{align}
\label{AW0_moments}
    \mu_0 = & 1, \\
    \mu_1 = & 0, \\
    \mu_2 = & 1 - q, \\
    \mu_3 = & 0 \; .
\end{align}
These results are in complete agreement with those obtained in \cite{Okazaki:2024paq}. 

%%%%%%%%%%%%%%%%%%%%%%%%%%%%%%%%%%%%%%%%%%%%%%%%%%%
%%%%%%%%%%%%%%%%%%%%%%%%%%%%%%%%%%%%%%%%%%%%%%%%%%%
\section{Gustafson-type moments}
\label{sec_Gmoment}
%%%%%%%%%%%%%%%%%%%%%%%%%%%%%%%%%%%%%%%%%%%%%%%%%%%
%%%%%%%%%%%%%%%%%%%%%%%%%%%%%%%%%%%%%%%%%%%%%%%%%%%

%%%%%%%%%%%%%%%%%%%%%%%%%%%%%%%%%%%%%%%%%%%%%%%%%%%
\subsection{$SU(N)$ with $N_f=N_a=N$}
\label{sec_suNNf=Na=N}
%%%%%%%%%%%%%%%%%%%%%%%%%%%%%%%%%%%%%%%%%%%%%%%%%%%

%%%%%%%%%%%%%%%%%%%%%%%%%%%%%%%%%%%%%%%%%%%%%%%%%%%
\subsubsection{Boundary conditions}
%%%%%%%%%%%%%%%%%%%%%%%%%%%%%%%%%%%%%%%%%%%%%%%%%%%
We consider the 3d $\mathcal{N}=2$ gauge theory with gauge group $SU(N)$, 
coupled to matter fields consisting of $N_f=N$ fundamental chiral multiplets $Q_I$ 
and $N_a=N$ anti-fundamental chiral multiplets $\overline{Q}_{\alpha}$. 
We impose $\mathcal{N}=(0,2)$ half-BPS boundary conditions 
by choosing Neumann boundary conditions for the $SU(N)$ vector multiplet, 
as well as Neumann boundary conditions for all chiral multiplets. 

The resulting boundary theory, which we refer to as theory A, 
admits a dual description (theory B) in terms of gauge invariant degrees of freedom. 
The latter consists of:
\begin{itemize}

\item bifundamental chiral multiplets $M_{I\alpha}$ transforming under the non-Abelian flavor symmetry $SU(N_f)\times SU(N_a)$. 

\item two singlet chiral multiplets $B$ and $\overline{B}$

\item an additional singlet chiral multiplet $V$. 

\end{itemize}
The boundary conditions are assigned as follows: 
$M_{I\alpha}$, $B$, and $\overline{B}$ obey the Neumann boundary condition, 
whereas $V$ satisfies the Dirichlet boundary condition. 
The matter content and charge assignments of the two theories are summarized in
\begin{align}
\label{SUN_N_charges}
\begin{array}{c|c|c|c|c|c|c|c}
& \textrm{bc} & SU(N) & SU(N_f = N) & SU(N_a = N) & U(1)_a & U(1)_b & U(1)_R \\ \hline
\textrm{VM} & \mathcal{N} & {\bf Adj} & {\bf 1} & {\bf 1} & 0 & 0 & 0 \\
Q_I & \textrm{N} & {\bf N} & {\bf N_f} & {\bf 1} & 1 & 0 & 0 \\
\overline{Q}_{\alpha} & \textrm{N} & {\bf \overline{N}} & {\bf 1} & {\bf N_a} & 0 & 1 & 0 \\
 \hline
M_{I \alpha} & \textrm{N} & {\bf 1} & {\bf N_f} & {\bf N_a} & 1 & 1 & 0 \\
B & \textrm{N} & {\bf 1} & {\bf 1} & {\bf 1} & N & 0 & 0 \\
\overline{B} & \textrm{N} & {\bf 1} & {\bf 1} & {\bf 1} & 0 & N & 0 \\
V & \textrm{D} & {\bf 1} & {\bf 1} & {\bf 1} & -N & -N & 2
\end{array}
\end{align}
This construction may be regarded as a higher-rank generalization of the model analyzed in section \ref{sec_su2Nf4}. 

The theory A half-index is 
\begin{align}
\label{SUN_N_halfindex}
&
\mathbb{II}_{(\mathcal{N},N,N)}^A
=
\frac{(q)_{\infty}^{N-1}}{N!} \prod_{i=1}^N \oint \frac{ds_i}{2\pi i s_i}
\prod_{i \ne j}^N (s_i s_j^{-1}; q)_{\infty}
\nonumber\\
&\times 
\frac{1}{\prod_{i = 1}^N \left( \prod_{i = 1}^N (q^{r_a/2} a s_i x_I; q)_{\infty} \right) \left( \prod_{\alpha = 1}^{N} (q^{r_b/2} b s_i^{-1} \tilde{x}_{\alpha}; q)_{\infty} \right)}. 
\end{align}
The theory B half-index reads
\begin{align}
\label{SUN_N_halfindexD}
\mathbb{II}_{N,N,N,D}^B
&=
\frac{\left( q^\frac{Nr_a+Nr_b}{2} a^N b^N; q \right)_{\infty}}{\left( q^{Nr_a/2} a^N; q \right)_{\infty} \left( q^{Nr_b/2} b^N; q \right)_{\infty} \prod_{I = 1}^N \prod_{\alpha = 1}^{N} (q^{(r_a + r_b)/2} ab x_I \tilde{x}_{\alpha}; q)_{\infty}}. 
\end{align}
As discussed in \cite{Okazaki:2023hiv}, 
the equivalence of the two expressions (\ref{SUN_N_halfindex}) and (\ref{SUN_N_halfindexD}) follows from the Gustafson integral formula \cite{MR1139492}.

%%%%%%%%%%%%%%%%%%%%%%%%%%%%%%%%%%%%%%%%%%%%%%%%%%%
\subsubsection{One-point functions}
%%%%%%%%%%%%%%%%%%%%%%%%%%%%%%%%%%%%%%%%%%%%%%%%%%%
We now consider the insertion of a Wilson line operator in the fundamental representation of the $SU(N)$ gauge group, 
compatible with the boundary conditions specified above. 
In the presence of this line defect, the corresponding half-index is modified with the resulting expression taking the following form
\begin{align}
\label{suN_N_Wfund}
&\langle W_{\tiny \yng(1)}\rangle_{(\mathcal{N},N,N)}^A
=\frac{(q)_{\infty}^{N-1}}{N!} \prod_{i=1}^N \oint \frac{ds_i}{2\pi i s_i}
\prod_{i \ne j}^N (s_i s_j^{-1}; q)_{\infty}
\nonumber\\
&\times 
\frac{1}{\prod_{i = 1}^N \left( \prod_{i = 1}^N (q^{r_a/2} a s_i x_I; q)_{\infty} \right) \left( \prod_{\alpha = 1}^{N} (q^{r_b/2} b s_i^{-1} \tilde{x}_{\alpha}; q)_{\infty} \right)}
\left(\sum_{i=1}^{N} s_i\right),  
\end{align}
where $\prod_{i=1}^{N}s_i=1$. 

The matrix integral representation (\ref{suN_N_Wfund}) of the line defect half-index naturally suggests 
an introduction of moments of the Gustafson integral as the normalized line defect half-index. 
Indeed, the effect of inserting a Wilson line in the fundamental representation is 
to modify the integrand by a character factor, thereby promoting the Gustafson-type integral to a moment. 
The first moment is defined as
\begin{align}
\mu_1^{SU(N)}(a,x_I,\tilde{x}_{\alpha}; q) \mathbb{II}_{(\mathcal{N},N,N)}^A
&=\langle W_{\tiny \yng(1)}\rangle_{(\mathcal{N},N,N)}^A. 
\end{align}
Again it can be identified with the 1d defect index
\begin{align}
\mu_1^{SU(N)}(a,x_I,\tilde{x}_{\alpha}; q)&=\mathcal{I}^{\textrm{1d defect}}_{\tiny\yng(1)}(a,x_I,\tilde{x}_{\alpha};q), 
\end{align}
which encodes the one-dimensional quantum mechanical degrees of freedom localized on the fundamental Wilson line. 

In order to find the closed-form expression for the 1d defect index, 
we begin by taking into account the structural form of the Askey-Wilson moment (\ref{1d_su2Nf4_1ptQM}) and its physical interpretation. 
From the perspective of the dual theory B, the contribution of the chiral multiplet obeying the Dirichlet boundary condition in the $SU(2)$ half-index (\ref{su2_4b}) is shifted as
\begin{align}
q^{2r}a^4\rightarrow q^{2r+1}a^4, 
\end{align}
which is of the form of the spin shift produced by the vortex line operator. 
In complete analogy with the rank-one situation, we focus on the chiral multiplet $V$ subject to Dirichlet boundary condition 
and consider the following modification of its contribution: 
\begin{align}
q^{\frac{Nr_a+Nr_b}{2}}a^Nb^N \rightarrow  q^{\frac{Nr_a+Nr_b}{2}+1}a^N b^N. 
\end{align}
Accordingly, we are led to a conjectural higher-rank generalization of the 1d defect index associated with the Gustafson first moment
\begin{align}
\label{suN_fundWL}
&
\mathcal{I}^{\textrm{1d defect}}_{\tiny\yng(1)}(a,x_I,\tilde{x}_{\alpha};q)
\nonumber\\
&=\frac{
\sum_{\alpha=1}^N\tilde{x}_{\alpha} (q^{\frac{r_b}{2}}b-q^{\frac{Nr_a+r_b}{2}}a^Nb)
+\sum_{I=1}x_{I}^{-1} (q^{\frac{(N-1)r_a}{2}}a^{N-1}-q^{\frac{(N-1)r_a+Nr_b}{2}}a^{N-1}b^N)
}{1-q^{\frac{Nr_a+Nr_b}{2}}a^N b^N}. 
\end{align}
To arrive at this, the form of the denominator follows from the spin shift argument above while the numerator is constrained by the requirement that it reproduces (\ref{1d_su2Nf4_1ptQM}) in the case $N = 2$, after suitable identification of the flavour fugacities (since in the $SU(2)$ case there is no distinction between fundamental and antifundamental). While respecting the global flavor symmetries, this doesn't give many options for the terms appearing in the numerator. for example the term $a \sum_{\alpha = 1}^4 x_{\alpha}$ in the numerator of (\ref{1d_su2Nf4_1ptQM}) corresponds to $a \sum_{I = 1}^2 x_I + b \sum_{\alpha = 1}^2 \tilde{x}_{\alpha}$ in the notation in this section. In the $SU(2)$ case $x_! x_2 = 1$ so $a \sum_{I = 1}^2 x_I = a \sum_{I = 1}^2 x_I^{-1}$ etc., giving some ambiguity as to how this term generalizes to $N > 2$ but there are not many options for the numerator. Then looking at some examples leads us to the conjecture \eqref{suN_fundWL}.

We have performed extensive numerical checks of the above expression for $N=3$ and $N=4$ 
and verified that the proposed form reproduces the explicit expansions for a wide range of values of $r_a$ and $r_b$. 
When $N=2$, $r_a=r_b=r$, $b=a$, $\tilde{x}_1=x_3$, and $\tilde{x}_2=x_4$, 
the expression (\ref{suN_fundWL}) reduces to the 1d defect index (\ref{1d_su2Nf4_1ptQM}) with $x_1x_2=\tilde{x}_1\tilde{x}_2=1$. 
In the unflavored limit $a\rightarrow 1$, $x_{I}\rightarrow 1$, and $\tilde{x}_{\alpha}\rightarrow 1$, 
the 1d defect index (\ref{suN_fundWL}) reduces to
\begin{align}
\mathcal{I}^{\textrm{1d defect}}_{\tiny\yng(1)}(1,1,1;q) 
&=\frac{N(q^{\frac{r_b}{2}}+q^{\frac{(N-1)r_a}{2}} - q^{\frac{Nr_a+r_b}{2}} -q^{\frac{(N-1)r_a+Nr_b}{2}})}
{1-q^{\frac{Nr_a+Nr_b}{2}}}. 
\end{align}
Furthermore, when we take the limit $q\rightarrow1$, 
we obtain the Witten index
\begin{align}
\mathcal{I}^{\textrm{1d defect}}_{\tiny\yng(1)}(1,1,1;1)
&=N.  
\end{align}

%%%%%%%%%%%%%%%%%%%%%%%%%%%%%%%%%%%%%%%%%%%%%%%%%%%
\subsubsection{Multi-point functions}
%%%%%%%%%%%%%%%%%%%%%%%%%%%%%%%%%%%%%%%%%%%%%%%%%%%
The multi-point functions of the fundamental Wilson lines associated with the higher moments exhibit a structurally richer behavior compared to the first moment. 
Nevertheless, by analogy with the $SU(2)$ case, i.e.\ the Askey-Wilson moment, 
we propose that their full expressions can be obtained 
by implementing a spin shift of the singlet $V$ satisfying Dirichlet boundary conditions, 
by the amount appearing in the one-point function, together with a finite set of additional quantum mechanical contributions. 
In particular, for $N=3$, we find that in the unflavored limit the 1d defect index corresponding to the second moment can be written explicitly in the following form: 
\begin{align}
&
\mu_2^{SU(3)}(a,b;q)
=\mathcal{I}^{\textrm{1d defect}}_{\tiny\yng(1);\yng(1)}(a,b;q)
=\frac{\langle W_{\tiny \yng(1)} W_{\tiny \yng(1)}\rangle_{(\mathcal{N},N,N)}^A(a,b;q)}
{\mathbb{II}_{(\mathcal{N},N,N)}^A(a,b;q)}
\nonumber\\
&=\frac{1}{(1-q^{\frac{3r_a+3r_b}{2}}a^3b^3)(1-q^{\frac{3r_a+3r_b}{2}+1}a^3b^3)}
\Bigl(
3q^{\frac{r_a}{2}}a+9q^{r_b}b^2+9q^{r_a+\frac{r_b}{2}}a^2b+6q^{2r_a}a^4
\nonumber\\
&-3q^{\frac{r_a+3r_b}{2}}ab^3-12q^{\frac{3r_a}{2}+r_b}a^3b^2-9q^{\frac{5r_a+r_b}{2}}a^5b-3q^{1+\frac{r_a}{2}}a
-9q^{r_a+2r_b}a^2b^4-6q^{2r_a+\frac{3r_b}{2}}a^4b^3
\nonumber\\
&+3q^{3r_a+r_b}a^6b^2+9q^{1+r_a+\frac{r_b}{2}}a^2b+3q^{1+2r_a}a^4
+3q^{\frac{3r_a+5r_b}{2}}a^3b^5+9q^{\frac{5r_a}{2}+2r_b}a^5b^4+3q^{1+\frac{r_a+3r_b}{2}}ab^3
\nonumber\\
&-6q^{1+\frac{3r_a}{2}+r_b}a^3b^2-9q^{1+\frac{5r_a+r_b}{2}}a^5b
-3q^{3r_a+\frac{5r_b}{2}}a^6b^5-9q^{1+r_a+2r_b}a^2b^4-12q^{1+2r_a+\frac{3r_b}{2}}a^4b^3
\nonumber\\
&-3q^{1+\frac{3r_a+5r_b}{2}}a^3b^5+6q^{1+3r_a+r_b}a^6b^2+9q^{1+\frac{5r_a}{2}+2r_b}a^5b^4
+9q^{1+2r_a+3r_b}a^4b^6+3q^{1+3r_a+\frac{5r_b}{2}}a^6b^5
\Bigr). 
\end{align}

Also, the unflavored 1d defect index corresponding to a pair of the fundamental and antifundamental Wilson lines for the $SU(3)$ gauge theory is given by
\begin{align}
&
\mathcal{I}^{\textrm{1d defect}}_{\tiny\yng(1);\overline{\yng(1)}}(a,b;q)
=\frac{\langle W_{\tiny \yng(1)} W_{\tiny \overline{\yng(1)}}\rangle_{(\mathcal{N},N,N)}^A(a,b;q)}
{\mathbb{II}_{(\mathcal{N},N,N)}^A(a,b;q)}
\nonumber\\
&=\frac{1}{(1-q^{\frac{3r_a+3r_b}{2}}a^3b^3)(1-q^{\frac{3r_a+3r_b}{2}+1}a^3b^3)}
\Bigl(
1+9q^{\frac{r_a+r_b}{2}}ab+8q^{\frac{3r_a}{2}}a^3+8q^{\frac{3r_b}{2}}b^3
\nonumber\\
&-9q^{2r_a+\frac{r_b}{2}}a^4b-9q^{\frac{r_a}{2}+2r_b}ab^4
-18q^{\frac{3r_a+3r_b}{2}}a^3b^3+9q^{2r_a+2r_b}a^4b^4+q^{3r_a+\frac{3r_b}{2}}a^6b^3
\nonumber\\
&+q^{\frac{3r_a}{2}+3r_b}a^3b^6+q^{3r_a+3r_b}a^6b^6
-q+q^{1+\frac{3r_a}{2}}a^3+q^{1+\frac{3r_b}{2}}b^3+q^{1+r_a+r_b}a^2b^2
\nonumber\\
&-18q^{1+\frac{3r_a+3r_b}{2}}a^3b^3-9q^{1+\frac{5r_a+2r_b}{2}}a^5b^2-9q^{1+r_a+\frac{5r_b}{2}}a^2b^5
+8q^{1+3r_a+\frac{3r_b}{2}}a^6b^3
\nonumber\\
&+8q^{1+\frac{3r_a}{2}+3r_b}a^3b^6+9^{1+\frac{5r_a+5r_b}{2}}a^5b^5
+q^{1+3r_a+3r_b}a^6b^6
\Bigr). 
\end{align}

We find that, upon introducing flavor fugacities, 
all contributions beyond the shift of the chiral multiplet $V$ can be organized 
in terms of the $SU(3)$ characters $\chi^{\mathfrak{su}(3)}_{\mathcal{R}}(x_I)$ and $\chi^{\mathfrak{su}(3)}_{\mathcal{R}}(\tilde{x}_{\alpha})$. 
Although a general closed-form expression for higher moments of this Gustafson-type integral is presently unavailable, 
we conjecture - by analogy with the $SU(2)$ Askey-Wilson moments - 
that these additional contributions admit an interpretation as Higgs-branch sectors, 
and can be expressed in terms of flavor symmetry characters as finite terms. 

On the other hand, 
for the Witten index corresponding to special values of the general moments whose fugacities are turned off, 
explicit computations across several examples lead us to conjecture the following general structure: 
\begin{align}
\mu_k^{SU(N)}=\mathcal{I}^{\textrm{1d defect}}_{\underbrace{\tiny\yng(1);\cdots;\yng(1)}_{k}}=N^k. 
\end{align}

%%%%%%%%%%%%%%%%%%%%%%%%%%%%%%%%%%%%%%%%%%%%%%%%%%%
\subsubsection{Reduced number of flavors}
%%%%%%%%%%%%%%%%%%%%%%%%%%%%%%%%%%%%%%%%%%%%%%%%%%%
We can also reduce the number of fundamental and antifundamental flavors by sending masses to infinity which generates a Chern-Simons level $-(N - (N_f + N_a)/2))$. This is implemented in the half-index expressions by defining $X_I = q^{r_a/2} a x_I$, $\tilde{X}_{\alpha} = q^{r_b/2} b \tilde{x}_{\alpha}$ then setting $X_I = 0$ for $I > N_f$ and $\tilde{X}_{\alpha} = 0$ for $\alpha > N_a$ and then rescaling $a$ and the remaining $x_I$, as well as $b$ and $\tilde{x}_{\alpha}$, so that $\prod_{I = 1}^{N_f} x_I = 1$ and $\prod_{\alpha = 1}^{N_a} \tilde{x}_{\alpha} = 1$. This gives the general result
\begin{align}
\label{suN_moment1_gen}
\langle W_{\tiny \yng(1)}\rangle_{(\mathcal{N},N,N)}^{A; N_f, N_a}
= & \mathcal{I}^{\textrm{1d defect}; N_f, N_a}_{\tiny\yng(1)} \mathbb{II}_{(\mathcal{N},N,N)}^{A; N_f. N_a}
\end{align}
where
\begin{align}
&
    \mathbb{II}_{(\mathcal{N},N,N)}^{A; N_f, N_a} 
    \nonumber\\
    & = \left\{ \begin{array}{rcl}
\frac{1}
{(q^{\frac{Nr_b}{2}} b^N;q)_{\infty} \prod_{I, \alpha = 1}^N (q^{\frac{r_a+r_b}{2}} ab x_I \tilde{x}_{\alpha};q)_{\infty}} & , & N_f < N, N_a = N \\
\frac{1}
{(q^{\frac{Nr_a}{2}} a^N;q)_{\infty} \prod_{I, \alpha = 1}^N (q^{\frac{r_a+r_b}{2}} ab x_I \tilde{x}_{\alpha};q)_{\infty}} & , & N_f = N, N_a < N \\
\frac{1}
{\prod_{I, \alpha = 1}^N (q^{\frac{r_a+r_b}{2}} ab x_I \tilde{x}_{\alpha};q)_{\infty}} & , & N_f < N, N_a < N
\end{array}
\right.
\end{align}
and
\begin{align}
&
   \mathcal{I}^{\textrm{1d defect}; N_f, N_a}_{\tiny\yng(1)} 
   \nonumber\\
  & =  \left\{ \begin{array}{rcl}
    \sum_{\alpha=1}^N\tilde{x}_{\alpha} (q^{\frac{r_b}{2}}b-q^{\frac{Nr_a+r_b}{2}}a^Nb)
+\sum_{I=1}^N x_{I}^{-1} q^{\frac{(N-1)r_a}{2}}a^{N-1} & , & N_f = N, N_a < N \\
    \sum_{\alpha=1}^N\tilde{x}_{\alpha} q^{\frac{r_b}{2}}b
+ a^{N-1}(q^{\frac{(N-1)r_a}{2}} - q^{\frac{(N-1)r_a+Nr_b}{2}}b^N) & , & N_f = N-1, N_a = N \\
    \sum_{\alpha=1}^{N_a}\tilde{x}_{\alpha} q^{\frac{r_b}{2}}b
+ a^{N-1} q^{\frac{(N-1)r_a}{2}} & , & N_f = N-1, N_a < N \\
    \sum_{\alpha=1}^{N_a} \tilde{x}_{\alpha} q^{\frac{r_b}{2}}b & , & N_f < N-1
\end{array}
\right.
\end{align}

%%%%%%%%%%%%%%%%%%%%%%%%%%%%%%%%%%%%%%%%%%%%%%%%%%%
\subsection{$USp(2n)$ with $2n+2$ fundamental chirals}
\label{sec_usp2n_2n+2}
%%%%%%%%%%%%%%%%%%%%%%%%%%%%%%%%%%%%%%%%%%%%%%%%%%%

%%%%%%%%%%%%%%%%%%%%%%%%%%%%%%%%%%%%%%%%%%%%%%%%%%%
\subsubsection{Boundary conditions}
%%%%%%%%%%%%%%%%%%%%%%%%%%%%%%%%%%%%%%%%%%%%%%%%%%%
Next consider another higher-rank generalization of the $SU(2)$ gauge theory discussed in section \ref{sec_su2Nf4}, by choosing a symplectic gauge group. 
The theory A has gauge group $USp(2n)$ and $2n+2$ fundamental chiral multiplets $Q_{\alpha}$
We impose Neumann boundary conditions for both the vector multiplet and all chiral multiplets. 
The theory B as the Seiberg-like dual description contains no dynamical gauge fields. 
The matter content consists of a chiral multiplet $M_{\alpha\beta}$ transforming  in the rank-two antisymmetric representation of the flavor symmetry group $SU(2n+2)$ 
and a singlet chiral multiplet $V$ \cite{Karch:1997ux}. 
As discussed in \cite{Okazaki:2023hiv}, 
the dual boundary conditions are realized as the Neumann boundary condition for $M_{\alpha\beta}$ and the Dirichlet boundary condition for $V$. 
The field content and their transformation properties under the global symmetries, together with the corresponding boundary conditions, are summarized as
\begin{align}
\label{USp2n_2np2_charges}
\begin{array}{c|c|c|c|c|c}
& \textrm{bc} & USp(2n) & SU(N_f = 2n + 2) & U(1)_a & U(1)_R \\ \hline
\textrm{VM} & \mathcal{N} & {\bf Adj} & {\bf 1} & 0 & 0 \\
Q_{\alpha} & \textrm{N} & {\bf 2n} & {\bf N_f} & 1 & 0 \\
 \hline
M_{\alpha \beta} & \textrm{N} & {\bf 1} & {\bf N_f(N_f - 1)/2} & 2 & 0 \\
V & \textrm{D} & {\bf 1} & {\bf 1} & -N_f & 2
\end{array}
\end{align}

The corresponding half-index for theory A is given by
\begin{align}
\label{USp2n_2n+2_hindexA}
\mathbb{II}_{(\mathcal{N},N)}^A
&=\frac{(q)_{\infty}^n}{n! 2^n} \prod_{i=1}^n \oint \frac{ds_i}{2\pi i s_i}
\prod_{i \ne j}^n (s_i s_j^{-1}; q)_{\infty} \prod_{i \le j}^n (s_i^{\pm} s_j^{\pm}; q)_{\infty}
\nonumber\\
&\times 
\frac{1}{\prod_{\alpha = 1}^{2n + 2} \prod_{i = 1}^n (q^{r/2} a s_i^{\pm} x_{\alpha}; q)_{\infty}}, 
\end{align}
with $\prod_{\alpha=1}^{2n+2}x_{\alpha}$ $=$ $1$. 
The half-index for theory B is
\begin{align}
\label{USp2n_2n+2_hindexB}
\mathbb{II}_{N,D}^B
&=\frac{\left( q^{(n+1)r} a^{2n+2}; q \right)_{\infty}}{\prod_{\alpha < \beta}^{2n + 2} (q^r a^2 x_{\alpha} x_{\beta}; q)_{\infty}} \; .
\end{align}
The equality of the two expressions (\ref{USp2n_2n+2_hindexA}) and (\ref{USp2n_2n+2_hindexB})
follows from Theorem 7.1 in \cite{MR1139492}. 

%%%%%%%%%%%%%%%%%%%%%%%%%%%%%%%%%%%%%%%%%%%%%%%%%%%
\subsubsection{One-point functions}
%%%%%%%%%%%%%%%%%%%%%%%%%%%%%%%%%%%%%%%%%%%%%%%%%%%
Let us consider an insertion of the Wilson line operator in the fundamental representation of the $USp(2n)$ gauge group, 
compatible with the above boundary conditions specified in Theory A. 
In the presence of such a line defect, the half-index is modified by the insertion of the corresponding character. 
The line defect half-index takes the form
\begin{align}
\label{USp2n_2n+2_W}
\langle W_{\tiny \yng(1)}\rangle_{(\mathcal{N},N)}^A
&=\frac{(q)_{\infty}^n}{n! 2^n} \prod_{i=1}^n \oint \frac{ds_i}{2\pi i s_i}
\prod_{i \ne j}^n (s_i s_j^{-1}; q)_{\infty} \prod_{i \le j}^n (s_i^{\pm} s_j^{\pm}; q)_{\infty}
\nonumber\\
&\times 
\frac{1}{\prod_{\alpha = 1}^{2n + 2} \prod_{i = 1}^n (q^{r/2} a s_i^{\pm} x_{\alpha}; q)_{\infty}} 
\left(\sum_{i=1}s_i+s_i^{-1}\right). 
\end{align}
The matrix integral representation (\ref{USp2n_2n+2_W}) naturally admits an interpretation 
in terms of the first moments of a $USp(2n)$ analogue of the Gustafson integral defined by
\begin{align}
\mu_1^{USp(2n)}(a,x_{\alpha};q) \mathbb{II}_{(\mathcal{N},N)}^A
&=\langle W_{\tiny \yng(1)}\rangle_{(\mathcal{N},N)}^A, 
\end{align}
which is interpreted as the 1d defect index
\begin{align}
\mu_1^{USp(2n)}(a,x_{\alpha};q)&=
\mathcal{I}^{\textrm{1d defect}}_{\tiny\yng(1)}(a,x_{\alpha};q). 
\end{align}

To find a closed-form expression for the defect index, 
we again take into account the dual (Theory B) description. 
The general mechanism closely parallels that of the $SU(N)$ case.  
The insertion of a Wilson line is encoded, on the dual side, 
as a relevant deformation of the contribution from boundary chiral multiplet $V$ subject to Dirichlet boundary condition as the vortex line.
This contribution is modified by a shift in the power of $q$ as
\begin{align}
q^{(n+1)r}a^{2n+2} \rightarrow  q^{(n+1)r+1}a^{2n+2}. 
\end{align}
From a physical perspective, this modification can again be naturally interpreted as a spin shift induced by the vortex line operator. 
Having identified the contribution associated with the spin shift, 
we observe that the remaining contributions are entirely captured by finite terms. 
This structural simplification leads us to propose the following conjectural formula:
\begin{align}
\label{usp2n_fundWL}
&
\mu_1^{USp(2n)}(a,x_{\alpha};q)=
\mathcal{I}^{\textrm{1d defect}}_{\tiny\yng(1)}(a,x_{\alpha};q)
\nonumber\\
&=\frac{\sum_{\alpha=1}^{2n+2} x_{\alpha} q^{\frac{r}{2}}a -\sum_{\alpha=1}^{2n+2}x_{\alpha}^{-1} q^{\frac{(2n+1)r}{2}} a^{2n+1}}{1-q^{(n+1)r}a^{2(n+1)}}. 
\end{align}
We have checked that, for $n=2$ and $n=3$, 
the proposed expression (\ref{usp2n_fundWL}) correctly reproduces the expected expansion for various values of the parameter $r$. 
In addition, for $n=1$, it reduces to the Askey-Wilson first moment (\ref{1d_su2Nf4_1ptQM}). 
In particular, in the unflavored limit, it becomes
\begin{align}
\mathcal{I}^{\textrm{1d defect}}_{\tiny\yng(1)}(1,1;q)
&=(2n+2)\frac{q^{\frac{r}{2}}-q^{\frac{(2n+1)r}{2}}}{1-q^{(n+1)r}}. 
\end{align}
Moreover, in the limit $q\rightarrow 1$, the expression yields a non-trivial Witten index
\begin{align}
\mathcal{I}^{\textrm{1d defect}}_{\tiny\yng(1)}(1,1;1)&=2n. 
\end{align}

%%%%%%%%%%%%%%%%%%%%%%%%%%%%%%%%%%%%%%%%%%%%%%%%%%%
\subsubsection{Multi-point functions}
%%%%%%%%%%%%%%%%%%%%%%%%%%%%%%%%%%%%%%%%%%%%%%%%%%%

We now turn to the multi-point functions of fundamental Wilson line operators.
Motivated by the structure observed in the Askey-Wilson moments for the $SU(2)$ $\cong$ $USp(2)$ case, 
we conjecture that an analogous mechanism persists in the present setting. 
Namely, we propose that the dependence on the number of insertions can be systematically incorporated 
through an appropriate spin shift acting on the chiral multiplet $V$. 
This shift isolates a universal defect index contribution, while the remaining data are captured by an additional polynomial contribution. 
In particular, the unflavored two-point function corresponding to the second moment for the $USp(4)$ gauge theory takes the form 
\begin{align}
\label{usp2n_fundWLWL}
&
\mu_2^{USp(4)}
:=
\mathcal{I}^{\textrm{1d defect}}_{\tiny\yng(1);\yng(1)}(a,1;q)
=\frac{\langle W_{\tiny \yng(1)} W_{\tiny \yng(1)}\rangle_{(\mathcal{N},N)}^A}
{ \mathbb{II}_{(\mathcal{N},N)}^A}
\nonumber\\
&=\frac{1}{(1-q^{3r}a^6)(1-q^{3r+1}a^6)}
\Bigl(
1+36q^r a^2-15q^{2r}a^4-36q^{3r}a^6+15q^{4r}a^8-q^{6r}a^{12}
\nonumber\\
&-q+15q^{1+2r}a^4-36q^{1+3r}a^6-15q^{1+4r}a^8+36q^{1+5r}a^{10}+q^{1+6r}a^{12}
\Bigr), 
\end{align}
where the denominator factors arise as a consequence of the corresponding shift
\begin{align}
q^{3r} a^6\rightarrow q^{3r+2}a^6. 
\end{align}
which is twice that appearing in the one-point function. 
Similarly, the third moment corresponding to the three-point function is given by
\begin{align}
\label{usp2n_fundWLWLWL}
&
\mu_3^{USp(4)}(a,1;q)
:=
\mathcal{I}^{\textrm{1d defect}}_{\tiny\yng(1);\yng(1);\yng(1)}(a,1;q)
=\frac{\langle W_{\tiny \yng(1)} W_{\tiny \yng(1)} W_{\tiny \yng(1)}\rangle_{(\mathcal{N},N)}^A}
{ \mathbb{II}_{(\mathcal{N},N)}^A}
\nonumber\\
&=\frac{1}{(1-q^{3r}a^6)(1-q^{3r+1}a^6)(1-q^{3r+2}a^6)}
\nonumber\\
\Bigl(
&18 q^{\frac{r}{2}} a + 196 q^{\frac{3r}{2}} a^3 - 186 q^{\frac{5r}{2}} a^5 - 120 q^{\frac{7r}{2}} a^7 + 110 q^{\frac{9r}{2}} a^9 \nonumber\\
&- 18 q^{\frac{r}{2}+1} a - 12 q^{\frac{11r}{2}} a^{11} + 40 q^{\frac{3r}{2}+1} a^3 - 6 q^{\frac{13r}{2}} a^{13} \nonumber\\
&+ 90 q^{\frac{5r}{2}+1} a^5 - 396 q^{\frac{7r}{2}+1} a^7 + 210 q^{\frac{9r}{2}+1} a^9 + 144 q^{\frac{11r}{2}+1} a^{11} \nonumber\\
&- 20 q^{\frac{3r}{2}+2} a^3 - 90 q^{\frac{13r}{2}+1} a^{13} + 90 q^{\frac{5r}{2}+2} a^5 + 20 q^{\frac{15r}{2}+1} a^{15} \nonumber\\
&- 144 q^{\frac{7r}{2}+2} a^7 - 210 q^{\frac{9r}{2}+2} a^9 + 396 q^{\frac{11r}{2}+2} a^{11} - 90 q^{\frac{13r}{2}+2} a^{13} \nonumber\\
&+ 6 q^{\frac{5r}{2}+3} a^5 - 40 q^{\frac{15r}{2}+2} a^{15} + 12 q^{\frac{7r}{2}+3} a^7 + 18 q^{\frac{17r}{2}+2} a^{17} \nonumber\\
&- 110 q^{\frac{9r}{2}+3} a^9 + 120 q^{\frac{11r}{2}+3} a^{11} + 186 q^{\frac{13r}{2}+3} a^{13} \nonumber\\
&- 196 q^{\frac{15r}{2}+3} a^{15} - 18 q^{\frac{17r}{2}+3} a^{17}
\Bigr).
\end{align}
This is again obtained from a shift of the effective spin, now by three units. 
With all fugacities turned off, 
the second moment (\ref{usp2n_fundWLWL}) and third moment (\ref{usp2n_fundWLWLWL}) for the $USp(4)$ theory become 16 and 64, respectively. 
We conjecture that
\begin{align}
\mu_k^{USp(2n)}(1,1;1)
&=(2n)^k. 
\end{align}

%%%%%%%%%%%%%%%%%%%%%%%%%%%%%%%%%%%%%%%%%%%%%%%%%%%
\subsection{$USp(4)$ with $2$ fundamental and $2$ antisymmetric chirals}
\label{sec_GNR_USp4_2_2AS}
%%%%%%%%%%%%%%%%%%%%%%%%%%%%%%%%%%%%%%%%%%%%%%%%%%%

%%%%%%%%%%%%%%%%%%%%%%%%%%%%%%%%%%%%%%%%%%%%%%%%%%%
\subsubsection{Boundary conditions}
%%%%%%%%%%%%%%%%%%%%%%%%%%%%%%%%%%%%%%%%%%%%%%%%%%%
Next consider Theory A with gauge group $USp(4)$, 
coupled to two fundamental and two rank-two antisymmetric chiral multiplets, all subject to Neumann boundary conditions. 
The rank-two antisymmetric representation of $USp(2n)$ is reducible, decomposing into an irreducible component of dimension $n(2n-1)-1$ together with a singlet.
As discussed in \cite{Okazaki:2023hiv}, the dual boundary condition in Theory B is described by a collection of chiral multiplets with Neumann boundary conditions: 
a singlet $M$, a pair of $SU(2)_A$ fundamentals $\phi_I$ and $B_I$, an $SU(2)_A$ symmetric chiral $\phi_{IJ}$, and an $SU(2)_a$ symmetric chiral $B_{\alpha\beta}$ 
as well as a singlet chiral $V$ with Dirichlet boundary condition. 
The content and boundary conditions of both theories are summarized as
\begin{align}
\label{USp4_2_2AS_charges}
\begin{array}{c|c|c|c|c|c|c|c}
& \textrm{bc} & USp(4) & SU(2)_a & SU(2)_A & U(1)_a & U(1)_A & U(1)_R \\ \hline
\textrm{VM} & \mathcal{N} & {\bf Adj} & {\bf 1} & {\bf 1} & 0 & 0 & 0 \\
Q_{\alpha} & \textrm{N} & {\bf 4} & {\bf 2} & {\bf 1} & 1 & 0 & 0 \\
\Phi_I & \textrm{N} & {\bf 6} & {\bf 1} & {\bf 2} & 0 & 1 & 0 \\
 \hline
M & \textrm{N} & {\bf 1} & {\bf 1} & {\bf 1} & 2 & 0 & 0 \\
\phi_I & \textrm{N} & {\bf 1} & {\bf 1} & {\bf 2} & 0 & 1 & 0 \\
\phi_{IJ} & \textrm{N} & {\bf 1} & {\bf 1} & {\bf 3} & 0 & 2 & 0 \\
B_{\alpha \beta} & \textrm{N} & {\bf 1} & {\bf 3} & {\bf 1} & 2 & 2 & 0 \\
B_I & \textrm{N} & {\bf 1} & {\bf 1} & {\bf 2} & 2 & 1 & 0 \\
V & \textrm{D} & {\bf 1} & {\bf 1} & {\bf 1} & -4 & -4 & 2
\end{array}
\end{align}

The half-index of theory A reads
\begin{align}
\label{USp4_2_2AS_hindexA}
\mathbb{II}_{(\mathcal{N},N,N)}^{A}
&=\frac{(q)_{\infty}^2}{8} \prod_{i=1}^2 \oint \frac{ds_i}{2\pi i s_i}
(s_1^{\pm} s_2^{\mp}; q)_{\infty} \prod_{i \le j}^2 (s_i^{\pm} s_j^{\pm}; q)_{\infty}
\frac{1}{\prod_{\alpha = 1}^2 \prod_{i = 1}^2 (q^{r_a/2} a s_i^{\pm} x_{\alpha}; q)_{\infty}}
\nonumber \\
& \times
\frac{1}{\prod_{I = 1}^2 (q^{r_A/2} A s_1^{\pm} s_2^{\mp} \tilde{x}_I; q)_{\infty} (q^{r_A/2} A s_1^{\pm} s_2^{\pm} \tilde{x}_I; q)_{\infty} (q^{r_A/2} A \tilde{x}_I; q)_{\infty}^2}. 
\end{align}
The half-index of theory B is 
\begin{align}
\label{USp4_2_2AS_hindexB}
&
\mathbb{II}_{N,N,N,N,N,D}^{B}
\nonumber\\
&=
\frac{(q^{2r_a + 2r_A} a^4 A^4; q)_{\infty}}
{(q^{r_a} a^2; q)_{\infty} \left( \prod_{I \le J}^{2} (q^{r_A} A^2 \tilde{x}_I \tilde{x}_J; q)_{\infty} \right) \left( \prod_{I=1}^{2} (q^{r_A/2} A \tilde{x}_I; q)_{\infty} (q^{r_a + r_A/2} a^2 A \tilde{x}_I; q)_{\infty} \right)}
\nonumber\\
&\times 
\frac{1}{
\prod_{\alpha \le \beta}^{2} (q^{r_a + r_A} a^2 A^2 x_{\alpha} x_{\beta}; q)_{\infty}
}
\end{align}
where $\prod_{\alpha = 1}^2 x_{\alpha} = 1 = \prod_{I = 1}^2 \tilde{x}_I$. 

It follows from Theorem 4.1 in \cite{MR1266569} 
that the half-indices (\ref{USp4_2_2AS_hindexA}) and (\ref{USp4_2_2AS_hindexB}) are equivalent. 

%%%%%%%%%%%%%%%%%%%%%%%%%%%%%%%%%%%%%%%%%%%%%%%%%%%
\subsubsection{One-point function}
%%%%%%%%%%%%%%%%%%%%%%%%%%%%%%%%%%%%%%%%%%%%%%%%%%%
We now introduce the Wilson line operator in the fundamental representation of the $USp(4)$ gauge group, 
chosen so as to preserve the boundary conditions of theory A described above. 
In the presence of this line defect, the half-index is accordingly deformed by the insertion of the associated character. 
The resulting line defect half-index is given by
\begin{align}
\label{USp4_2_2AS_W}
&
\langle W_{\tiny \yng(1)}\rangle_{(\mathcal{N},N,N)}^{A}
\nonumber\\
&=\frac{(q)_{\infty}^2}{8} \prod_{i=1}^2 \oint \frac{ds_i}{2\pi i s_i}
(s_1^{\pm} s_2^{\mp}; q)_{\infty} \prod_{i \le j}^2 (s_i^{\pm} s_j^{\pm}; q)_{\infty}
\frac{1}{\prod_{\alpha = 1}^2 \prod_{i = 1}^2 (q^{r_a/2} a s_i^{\pm} x_{\alpha}; q)_{\infty}}
\nonumber \\
& \times
\frac{1}{\prod_{I = 1}^2 (q^{r_A/2} A s_1^{\pm} s_2^{\mp} \tilde{x}_I; q)_{\infty} (q^{r_A/2} A s_1^{\pm} s_2^{\pm} \tilde{x}_I; q)_{\infty} (q^{r_A/2} A \tilde{x}_I; q)_{\infty}^2}
(s_1+s_2+s_1^{-1}+s_2^{-1}). 
\end{align}
The matrix integral (\ref{USp4_2_2AS_W}) can be naturally reinterpreted as 
encoding the first moment defined by
\begin{align}
\mu_1^{USp(4)}(a,x_{\alpha},\tilde{x}_{I};q) \mathbb{II}_{(\mathcal{N},N,N)}^{A}
&=\langle W_{\tiny \yng(1)}\rangle_{(\mathcal{N},N,N)}^{A}, 
\end{align}
which is physically identified with the 1d defect index
\begin{align}
\mu_1^{USp(4)}(a,x_{\alpha},\tilde{x}_{I};q)&=
\mathcal{I}^{\textrm{1d defect}}_{\tiny\yng(1)}(a,x_{\alpha},\tilde{x}_I;q). 
\end{align}
To obtain a closed-form expression for the defect index, 
we again appeal to the dual description in Theory B. 
On the dual side, one has a relevant deformation of the boundary contribution from the chiral multiplet $V$, which obeys Dirichlet boundary conditions. 
Concretely, this deformation manifests as a shift 
\begin{align}
q^{2r_a+2r_A}a^4A^4\rightarrow q^{2r_a+2r_A+1}a^4A^4. 
\end{align}
Isolating the spin-shift contribution, 
we find that all residual terms reduce to finite contributions. 
This simplification in structure leads us to the following closed-form expression:
\begin{align}
\label{USp4_2_2AS_WL}
&
\mathcal{I}^{\textrm{1d defect}}_{\tiny\yng(1)}(a,x_{\alpha},\tilde{x}_{I};q)
\nonumber\\
&=\frac{1}{1-q^{2r_a+2r_A}a^4A^4}
\Biggl(
\sum_{\alpha=1}^2x_{\alpha} q^{\frac{r_a}{2}}a
+\sum_{\alpha=1}^2\sum_{I=1}^2 x_{\alpha}\tilde{x}_I q^{\frac{r_a+r_A}{2}}aA
+\sum_{\alpha=1}^2 q^{\frac{r_a+2r_A}{2}} x_{\alpha} aA^2
\nonumber\\
&-\sum_{\alpha=1}^2 x_{\alpha}q^{\frac{3r_a+2r_A}{2}}a^3 A^2
-\sum_{\alpha=1}^2\sum_{I=1}^2 x_{\alpha} \tilde{x}_I q^{\frac{3r_a+3r_A}{2}}a^3 A^3
-\sum_{\alpha=1}^2x_{\alpha}q^{\frac{3r_a+4r_A}{2}}a^3A^4
\Biggr). 
\end{align}

Taking $q\to 1$ first, and then setting all other fugacities to one, yields the Witten index
\begin{align}
\mathcal{I}^{\textrm{1d defect}}_{\tiny\yng(1)}(1,1,1;1)
&=4. 
\end{align}

%%%%%%%%%%%%%%%%%%%%%%%%%%%%%%%%%%%%%%%%%%%%%%%%%%%
\subsection{$USp(4)$ with $3$ antisymmetric chirals}
\label{sec_GNR_USp4_3AS}
%%%%%%%%%%%%%%%%%%%%%%%%%%%%%%%%%%%%%%%%%%%%%%%%%%%

%%%%%%%%%%%%%%%%%%%%%%%%%%%%%%%%%%%%%%%%%%%%%%%%%%%
\subsubsection{Boundary conditions}
%%%%%%%%%%%%%%%%%%%%%%%%%%%%%%%%%%%%%%%%%%%%%%%%%%%
For $USp(4)$ gauge theories with a boundary, there exists another distinct and intriguing duality of boundary conditions. 
In theory A, we consider a model with three rank-two antisymmetric chiral multiplets, 
all of which are taken to obey the Neumann boundary conditions. 
It is dual to a collection of boundary conditions for theory B, 
involving $\phi_I$ in the fundamental representation of $SU(3)$, 
$\phi_{IJ}$ in the symmetric representation of $SU(3)$, both with the Neumann boundary conditions, 
and a singlet $V$ with the Dirichlet boundary condition (see \cite{Okazaki:2023hiv} for the details). 
The field content and boundary conditions of both theories are shown in the following table: 
\begin{align}
\label{USp4_3AS_charges}
\begin{array}{c|c|c|c|c|c}
& \textrm{bc} & USp(4) & SU(3) & U(1)_A & U(1)_R \\ \hline
\textrm{VM} & \mathcal{N} & {\bf Adj} & {\bf 1} & 0 & 0 \\
\Phi_I & \textrm{N} & {\bf 6} & {\bf 3} & 1 & 0 \\
 \hline
\phi_I & \textrm{N} & {\bf 1} & {\bf 3} & 1 & 0 \\
\phi_{IJ} & \textrm{N} & {\bf 1} & {\bf 6} & 2 & 0 \\
V & \textrm{D} & {\bf 1} & {\bf 1} & -3 & 0
\end{array}
\end{align}

The Neumann half-index of theory A takes the form 
\begin{align}
\label{USp4_3AS_hindexA}
&
\mathbb{II}_{(\mathcal{N},N)}^A
\nonumber\\
&=\frac{(q)_{\infty}^2}{8} \prod_{i=1}^2 \oint \frac{ds_i}{2\pi i s_i}
\frac{(s_1^{\pm} s_2^{\mp}; q)_{\infty} \prod_{i \le j}^2 (s_i^{\pm} s_j^{\pm}; q)_{\infty}}
{\prod_{I = 1}^3 (q^{r_A/2} A s_1^{\pm} s_2^{\mp} \tilde{x}_I; q)_{\infty} (q^{r_A/2} A s_1^{\pm} s_2^{\pm} \tilde{x}_I; q)_{\infty} (q^{r_A/2} A \tilde{x}_I; q)_{\infty}^2}. 
\end{align}
The half-index of theory B is
\begin{align}
\label{USp4_3AS_hindexB}
\mathbb{II}_{N,N,D}^B
&=\frac{(q^{1 + 3r_A/2} A^3; q)_{\infty}}{\prod_{I=1}^{3} (q^{r_A/2} A \tilde{x}_I; q)_{\infty} \prod_{J \ge I}^{3} (q^{r_A} A^2 \tilde{x}_I \tilde{x}_J; q)_{\infty}}. 
\end{align}
It can be shown that 
the half-index (\ref{USp4_3AS_hindexA}) precisely agrees with the half-index (\ref{USp4_3AS_hindexB}) according to Theorem 4.4 in \cite{MR1266569}. 

%%%%%%%%%%%%%%%%%%%%%%%%%%%%%%%%%%%%%%%%%%%%%%%%%%%
\subsubsection{Two-point function}
%%%%%%%%%%%%%%%%%%%%%%%%%%%%%%%%%%%%%%%%%%%%%%%%%%%

%1pt
The Wilson line in the fundamental representation of $USp(4)$ carries an unsaturated fundamental index. 
In order to obtain a nonzero one-point function, this index must be absorbed by the BPS local operator inserted at the endpoint, 
thereby forming a gauge invariant composite. 
However, in the present theory all matter fields transform in the rank-two antisymmetric representation, and there are no chiral multiplets in the fundamental representation. Consequently, the one-point function of the fundamental Wilson line vanishes identically in this case. 

%2pt
We are thus led to consider the two-point function, 
which provides the simplest non-trivial observable in the absence of a gauge invariant one-point function. 
It can be evaluated as 
\begin{align}
\label{USp4_3AS_W}
&
\langle W_{\tiny \yng(1)}W_{\tiny \yng(1)} \rangle_{(\mathcal{N},N)}^A
\nonumber\\
&=\frac{(q)_{\infty}^2}{8} \prod_{i=1}^2 \oint \frac{ds_i}{2\pi i s_i}
\frac{(s_1^{\pm} s_2^{\mp}; q)_{\infty} \prod_{i \le j}^2 (s_i^{\pm} s_j^{\pm}; q)_{\infty}}
{\prod_{I = 1}^3 (q^{r_A/2} A s_1^{\pm} s_2^{\mp} \tilde{x}_I; q)_{\infty} (q^{r_A/2} A s_1^{\pm} s_2^{\pm} \tilde{x}_I; q)_{\infty} (q^{r_A/2} A \tilde{x}_I; q)_{\infty}^2}
\nonumber\\
&\times (s_1+s_2+s_1^{-1}+s_2^{-1})^2. 
\end{align}
In this setup, it is natural to introduce the corresponding second moment defined by
\begin{align}
\mu_2^{USp(4)}(a,\tilde{x}_{I};q) \mathbb{II}_{(\mathcal{N},N)}^A
&=\langle W_{\tiny \yng(1)}W_{\tiny \yng(1)} \rangle_{(\mathcal{N},N)}^A. 
\end{align} 
This quantity admits a direct physical interpretation as the index of a one-dimensional defect operator
\begin{align}
\mu_2^{USp(4)}(a,\tilde{x}_{I};q)&=\mathcal{I}^{\textrm{1d defect}}_{\tiny\yng(1);\yng(1)}(a,\tilde{x}_I;q). 
\end{align}

Proceeding as in the previous analysis, 
we incorporate the spin shift associated with the chiral multiplet obeying Dirichlet boundary condition. 
Apart from this contribution, we find that all remaining terms are exhausted by finite contributions. 
As a result, we find the full expression for the 1d defect index takes the following closed form:
\begin{align}
\label{USp4_3AS_WL}
\mathcal{I}^{\textrm{1d defect}}_{\tiny\yng(1);\yng(1)}(a,\tilde{x}_{I};q)
&=\frac{
(1-q)(1+\sum_{I=1}^3\tilde{x}_I q^{\frac{r_A}{2}}A
+\sum_{I=1}^3 \tilde{x}_I^{-1} q^{r_A} A^2
+q^{\frac{3r_A}{2}}A^3)
}
{1-q^{1+\frac{3r_A}{2}}A^3}. 
\end{align}

Taking the limit $q\to 1$, and subsequently setting the remaining fugacities to unity, we obtain the vanishing Witten index
\begin{align}
\mathcal{I}^{\textrm{1d defect}}_{\tiny\yng(1);\yng(1)}(1,1;1)&=0. 
\end{align}

%%%%%%%%%%%%%%%%%%%%%%%%%%%%%%%%%%%%%%%%%%%%%%%%%%%
\subsection{$USp(6)$ with $2$ antisymmetric chirals}
\label{sec_GNR_USp6_2AS}
%%%%%%%%%%%%%%%%%%%%%%%%%%%%%%%%%%%%%%%%%%%%%%%%%%%

%%%%%%%%%%%%%%%%%%%%%%%%%%%%%%%%%%%%%%%%%%%%%%%%%%%
\subsubsection{Boundary conditions}
%%%%%%%%%%%%%%%%%%%%%%%%%%%%%%%%%%%%%%%%%%%%%%%%%%%
Consider theory A given by a $USp(6)$ gauge theory with two rank-$2$ antisymmetric chiral multiplets. 
Under Neumann boundary conditions, this system exhibits an $s$-confining-type dual description \cite{Okazaki:2023hiv}. 
The dual boundary condition in theory B is formulated in terms of $SU(2)$ matter fields: 
a fundamental chiral $\phi_I$, an adjoint chiral $\phi_{IJ}$, and a symmetric chiral $\phi_{IJK}$, 
together with two singlet chirals $\phi$ and $V$. 
These fields are subject to boundary conditions $(N,N,N,N,D)$, respectively. 
\begin{align}
\label{USp6_2AS_charges}
\begin{array}{c|c|c|c|c|c}
& \textrm{bc} & USp(6) & SU(2) & U(1)_A & U(1)_R \\ \hline
\textrm{VM} & \mathcal{N} & {\bf Adj} & {\bf 1} & 0 & 0 \\
\Phi_I & \textrm{N} & {\bf 15} & {\bf 2} & 1 & 0 \\
 \hline
\phi_{I} & \textrm{N} & {\bf 1} & {\bf 2} & 1 & 0 \\
\phi_{IJ} & \textrm{N} & {\bf 1} & {\bf 3} & 2 & 0 \\
\phi_{IJK} & \textrm{N} & {\bf 1} & {\bf 4} & 3 & 0 \\
\phi & \textrm{N} & {\bf 1} & {\bf 1} & 4 & 0 \\
V & \textrm{D} & {\bf 1} & {\bf 1} & -6 & 0
\end{array}
\end{align}

The half-index of theory A is 
\begin{align}
\label{bdy_USp6_2AS_hindexA}
&\mathbb{II}_{(\mathcal{N},N)}^A
\nonumber\\
&=\frac{(q)_{\infty}^3}{48} \prod_{i=1}^3 \oint \frac{ds_i}{2\pi i s_i}
\frac{\left( \prod_{i < j}^3 (s_i^{\pm} s_j^{\mp}; q)_{\infty} \right) \prod_{i \le j}^3 (s_i^{\pm} s_j^{\pm}; q)_{\infty}}
{\prod_{I = 1}^2 (q^{r_A/2} A \tilde{x}_I; q)_{\infty}^3 \prod_{i < j}^3 (q^{r_A/2} A s_i^{\pm} s_j^{\mp} \tilde{x}_I; q)_{\infty} (q^{r_A/2} A s_i^{\pm} s_j^{\pm} \tilde{x}_I; q)_{\infty}}. 
\end{align}
The half-index of theory B is
\begin{align}
\label{bdy_USp6_2AS_hindexB}
&\mathbb{II}_{N,N,N,N,D}^B
\nonumber\\
&=
 \frac{(q^{1 + 3r_A} A^6; q)_{\infty}}{(q^{2r_A} A^4; q)_{\infty} \prod_{I=1}^{2} (q^{r_A/2} A \tilde{x}_I; q)_{\infty} \prod_{J \ge I}^{2} (q^{r_A} A^2 \tilde{x}_I \tilde{x}_J; q)_{\infty} \prod_{K \ge J \ge I}^2 (q^{3r_A/2} A^3 \tilde{x}_I \tilde{x}_J \tilde{x}_K; q)_{\infty}}. 
\end{align}
A direct consequence of Theorem 4.6 in \cite{MR1266569} is that the integral \eqref{bdy_USp6_2AS_hindexA} coincides with \eqref{bdy_USp6_2AS_hindexB}.

%%%%%%%%%%%%%%%%%%%%%%%%%%%%%%%%%%%%%%%%%%%%%%%%%%%
\subsubsection{Two-point function}
%%%%%%%%%%%%%%%%%%%%%%%%%%%%%%%%%%%%%%%%%%%%%%%%%%%

%1pt
As in the $USp(4)$ theory with $3$ antisymmetric chirals in section \ref{sec_GNR_USp4_3AS}, 
the Wilson line in the fundamental representation of $USp(6)$ carries an unsaturated fundamental index. 
To obtain a nonzero one-point function, this index must be absorbed by a BPS local operator inserted at the endpoint. 
However, since all matter fields transform in the rank-2 antisymmetric representation 
and no chiral multiplet in the fundamental representation is present, such a contraction is impossible. 
It follows that the one-point function of the fundamental Wilson line vanishes identically also in this case.

%2pt
As in the $USp(4)$ case, 
we are naturally led to consider instead the two-point function as the simplest non-trivial observable, which can be evaluated as follows: 
\begin{align}
\label{bdy_USp6_2AS_W}
&
\langle W_{\tiny \yng(1)}W_{\tiny \yng(1)}\rangle_{(\mathcal{N},N)}^A
\nonumber\\
&=\frac{(q)_{\infty}^3}{48} \prod_{i=1}^3 \oint \frac{ds_i}{2\pi i s_i}
\frac{\left( \prod_{i < j}^3 (s_i^{\pm} s_j^{\mp}; q)_{\infty} \right) \prod_{i \le j}^3 (s_i^{\pm} s_j^{\pm}; q)_{\infty}}
{\prod_{I = 1}^2 (q^{r_A/2} A \tilde{x}_I; q)_{\infty}^3 \prod_{i < j}^3 (q^{r_A/2} A s_i^{\pm} s_j^{\mp} \tilde{x}_I; q)_{\infty} (q^{r_A/2} A s_i^{\pm} s_j^{\pm} \tilde{x}_I; q)_{\infty}}
\nonumber\\
&\times \sum_{i=1}^3 (s_i+s_i^{-1})^2. 
\end{align}
We introduce the second moment for the $USp(6)$ theory as the two-point function of fundamental Wilson lines,
\begin{align}
\mu_2^{USp(6)}(a,\tilde{x}_{I};q) \mathbb{II}_{(\mathcal{N},N)}^A
= \langle W_{\tiny \yng(1)} W_{\tiny \yng(1)} \rangle_{(\mathcal{N},N)}^A,
\end{align}
which is naturally identified with the index of the corresponding one-dimensional defect 
\begin{align}
\mu_2^{USp(6)}(a,\tilde{x}_{I};q)
= \mathcal{I}^{\mathrm{1d\ defect}}{\tiny \yng(1);\yng(1)}(a,\tilde{x}_I;q).
\end{align}
Incorporating the spin shift induced by the Dirichlet boundary condition, 
we find that all remaining contributions are finite, and the resulting 1d defect index admits the closed-form expression
\begin{align}
&
\mathcal{I}^{\mathrm{1d\ defect}}{\tiny \yng(1);\yng(1)}(a,\tilde{x}_I;q)
\nonumber\\
&=\frac{1-q}{1-q^{1+3r_A}A^6}
\Biggl(
1+\sum_{I=1}^2\tilde{x}_I q^{\frac{r_A}{2}}A
+\sum_{I,J=1}^2\tilde{x}_I\tilde{x}_J q^{r_A}A^2
+\sum_{I,J=1}^2\tilde{x}_I\tilde{x}_J q^{\frac{3r_A}{2}}A^3
\nonumber\\
&+\sum_{I,J=1}^2\tilde{x}_I\tilde{x}_J q^{2r_A}A^4
+\sum_{I=1}^2\tilde{x}_I q^{\frac{5r_A}{2}}A^5
+q^{3r_A}A^6
\Biggr). 
\end{align}

In the limit $q\to 1$, followed by setting the remaining fugacities to unity, the index vanishes
\begin{align}
\mathcal{I}^{\mathrm{1d\ defect}}{\tiny \yng(1);\yng(1)}(1,1;1)&=0. 
\end{align}

%%%%%%%%%%%%%%%%%%%%%%%%%%%%%%%%%%%%%%%%%%%%%%%%%%%
\subsection{$SO(N)$ with $N-2$ fundamental chirals}
\label{sec_soN_N-2}
%%%%%%%%%%%%%%%%%%%%%%%%%%%%%%%%%%%%%%%%%%%%%%%%%%%

%%%%%%%%%%%%%%%%%%%%%%%%%%%%%%%%%%%%%%%%%%%%%%%%%%%
\subsubsection{Boundary conditions}
%%%%%%%%%%%%%%%%%%%%%%%%%%%%%%%%%%%%%%%%%%%%%%%%%%%
We now turn to the case of an orthogonal gauge group. 
In particular, we consider theory A as the 3d $\mathcal{N}=2$ $SO(N)$ gauge theory with $N-2$ fundamental chiral multiplets $Q_{\alpha}$. 
We impose $\mathcal{N}=(0,2)$ boundary conditions such that all fields obey Neumann boundary conditions, which are free from the gauge anomaly. 
As discussed in \cite{Okazaki:2023hiv}, these boundary conditions are dual to a set of boundary conditions in theory B 
that has a symmetric rank-$2$ $SU(N_f = N - 2)$ chiral $M_{\alpha\beta}$ obeying the Neumann boundary condition 
and a singlet chiral $V$ with $U(1)_a$ charge $-N_f$ obeying the Dirichlet boundary condition. 
We summarize the field content, their representations under the global symmetries, and the corresponding boundary conditions as follows:
\begin{align}
\label{SON_Nm2_charges}
\begin{array}{c|c|c|c|c|c}
& \textrm{bc} & SO(N) & SU(N_f = N - 2) & U(1)_a & U(1)_R \\ \hline
\textrm{VM} & \mathcal{N} & {\bf Adj} & {\bf 1} & 0 & 0 \\
Q_{\alpha} & \textrm{N} & {\bf N} & {\bf N_f} & 1 & 0 \\
 \hline
M_{\alpha \beta} & \textrm{N} & {\bf 1} & {\bf N_f(N_f + 1)/2} & 2 & 0 \\
V & \textrm{D} & {\bf 1} & {\bf 1} & -N_f & 0
\end{array}
\end{align}

We can write the half-index for theory A as
\begin{align}
\label{bdy_SON_Nm2_hindexA}
\mathbb{II}_{(\mathcal{N},N)}^{A}
&=
\frac{(q)_{\infty}^n}{n! 2^{n - 1 + \epsilon}} \prod_{i=1}^n \oint \frac{ds_i}{2\pi i s_i}
(s_i^{\pm}; q)_{\infty}^{\epsilon}
\prod_{i < j}^n (s_i^{\pm} s_j^{\mp}; q)_{\infty} (s_i^{\pm} s_j^{\pm}; q)_{\infty}
 \nonumber \\
& \times
\frac{1}{\prod_{\alpha = 1}^{N - 2} \left( (q^{r/2} a x_\alpha; q)_{\infty}^{\epsilon} \prod_{i = 1}^n (q^{r/2} s_i^{\pm} a x_\alpha; q)_{\infty} \right)}, 
\end{align}
where $N = 2n + \epsilon$ with $n \in \Zb$ and $\epsilon \in \{0, 1\}$. 
The half-index for theory B reads
\begin{align}
\label{bdy_SON_Nm2_hindexB}
\mathbb{II}_{N,D}^B
&=\frac{\left( q^{1 + (N/2 - 1)r} a^{N-2}; q \right)_{\infty}}{\prod_{\alpha \le \beta}^{N-2} (q^r a^2 x_\alpha x_\beta; q)_{\infty}}. 
\end{align}
It follows from Theorems 7.14 and 7.16 in \cite{MR1139492} 
that the half-indices (\ref{bdy_SON_Nm2_hindexA}) and (\ref{bdy_SON_Nm2_hindexB}) are equivalent. 

We may further refine the above identities by introducing discrete fugacities $\chi$ and $\zeta$ 
associated with the global $\mathbb{Z}_2$ symmetries, 
$\mathcal{C}$ (charge conjugation symmetry) and $\mathcal{M}$ (magnetic symmetry) respectively \cite{Aharony:2013kma}. 
Consistency of the dual description imposes a constraint on these fugacities. 
In particular, since the dual theory has trivial gauge group $SO(0)$, the corresponding dual magnetic fugacity must satisfy
\begin{align}
\chi\zeta&=+1. 
\end{align}
As a result, there are two admissible choices: 
$(\chi,\zeta)$ $=$ $(+1,+1)$ and $(-1,-1)$. 
For the case $N=2n+1$, the half-index corresponding to $(\chi,\zeta)$ $=$ $(-1,-1)$ is related to that of $(\chi,\zeta)$ $=$ $(+1,+1)$ by a simple change of variables, 
$s_i\rightarrow -s_i$, $a\rightarrow -a$, 
and therefore does not lead to a genuinely new identity. 
In contrast, for $N=2n$, this equivalence no longer holds, and one finds a non-trivial matching of half-indices, leading to a distinct identity. 
As discussed in \cite{Okazaki:2023hiv}, for the choice $(\chi,\zeta)$ $=$ $(-1,-1)$, 
the boundary duality states that the half-index
\begin{align}
\mathbb{II}_{(\mathcal{N},N)}^A
&=
\frac{(q)_{\infty}^{n-1} (-q; q)_{\infty}}{(n-1)! 2^{n-1}} \prod_{i=1}^{n-1} \oint \frac{ds_i}{2\pi i s_i}
(s_i^{\pm}; q)_{\infty} (-s_i^{\pm}; q)_{\infty}
\prod_{i < j}^{n-1} (s_i^{\pm} s_j^{\mp}; q)_{\infty} (s_i^{\pm} s_j^{\pm}; q)_{\infty}
 \nonumber \\
 &\times
\frac{1}{\prod_{\alpha = 1}^{N - 2} \left( (\pm q^{r/2} a x_{\alpha}; q)_{\infty} \prod_{i = 1}^{n-1} (q^{r/2} s_i^{\pm} a x_{\alpha}; q)_{\infty} \right)} 
\end{align}
of theory A coincides with the half-index 
\begin{align}
\mathbb{II}_{N,D}^B= & \frac{\left( -q^{1 + (N/2 - 1)r} a^{N-2}; q \right)_{\infty}}{\prod_{\alpha \le \beta}^{N-2} (q^r a^2 x_{\alpha} x_{\beta}; q)_{\infty}}
\end{align}
of theory B. 

%%%%%%%%%%%%%%%%%%%%%%%%%%%%%%%%%%%%%%%%%%%%%%%%%%%
\subsubsection{One-point functions}
%%%%%%%%%%%%%%%%%%%%%%%%%%%%%%%%%%%%%%%%%%%%%%%%%%%
We introduce the Wilson line operator in the fundamental representation of the $SO(N)$ gauge group, compatible with the boundary conditions of Theory A. 
In the presence of this line defect, the half-index is modified by the insertion of the corresponding character. 
The resulting line defect half-index takes the form
\begin{align}
\label{bdy_SON_Nm2_W}
\langle W_{\tiny \yng(1)}\rangle_{(\mathcal{N},N)}^{A}
&=
\frac{(q)_{\infty}^n}{n! 2^{n - 1 + \epsilon}} \prod_{i=1}^n \oint \frac{ds_i}{2\pi i s_i}
(s_i^{\pm}; q)_{\infty}^{\epsilon}
\prod_{i < j}^n (s_i^{\pm} s_j^{\mp}; q)_{\infty} (s_i^{\pm} s_j^{\pm}; q)_{\infty}
 \nonumber \\
& \times
\frac{1}{\prod_{\alpha = 1}^{N - 2} \left( (q^{r/2} a x_\alpha; q)_{\infty}^{\epsilon} \prod_{i = 1}^n (q^{r/2} s_i^{\pm} a x_\alpha; q)_{\infty} \right)}
\chi_{\tiny \yng(1)}^{\mathfrak{so}(N)}(s), 
\end{align}
where
\begin{align}
\chi_{\tiny \yng(1)}^{\mathfrak{so}(N)}(s)
&=\begin{cases}
\sum_{i=1}^n (s_i+s_i^{-1})&\textrm{for $N=2n$}\cr
1+\sum_{i=1}^n (s_i+s_i^{-1})&\textrm{for $N=2n+1$}\cr
\end{cases}
\end{align}
is the $\mathfrak{so}(N)$ character of the fundamental representation \cite{MR1153249}. 
The expression \eqref{bdy_SON_Nm2_W} is naturally interpreted in terms of the first moment of an $SO(N)$-type Gustafson integral defined by
\begin{align}
\mu_1^{SO(N)}(a,x_{\alpha};q) \mathbb{II}_{(\mathcal{N},N)}^A
&=\langle W_{\tiny \yng(1)}\rangle_{(\mathcal{N},N)}^A. 
\end{align}
The moment is again viewed as the 1d defect index
\begin{align}
\mu_1^{SO(N)}(a,x_{\alpha};q)&=
\mathcal{I}^{\textrm{1d defect}}_{\tiny\yng(1)}(a,x_{\alpha};q), 
\end{align}
which enumerates the extra 1d degrees of freedom supported on the line operator. 

To derive a closed-form expression for the line defect index in the $SO(N)$ gauge theory, 
we again exploit the dual (theory B) description. 
The underlying mechanism closely parallels that of the $SU(N)$ and $USp(2n)$ cases. 
On the dual side, the vortex line would deform the contribution from the chiral multiplet $V$ with the Dirichlet boundary condition. 
This induces a shift in the effective spin, so that the contribution is organized in the powers of $q$ as
\begin{align}
q^{1+\left(\frac{N}{2}-1\right)r}a^{N-2}
\rightarrow q^{2+\left(\frac{N}{2}-1\right)r}a^{N-2}. 
\end{align}
After isolating this contribution, the remaining terms are finite, leading to the following conjectural formula:
\begin{align}
\label{soN_fundWL}
&
\mu_1^{SO(N)}(a,x_{\alpha};q)
=\mathcal{I}^{\textrm{1d defect}}_{\tiny\yng(1)}(a,x_{\alpha};q)
\nonumber\\
&=\frac{\sum_{\alpha=1}^{N-2}x_{\alpha}q^{\frac{r}{2}}a-\sum_{\alpha=1}^{N-2}x_{\alpha}^{-1}q^{1+\frac{N-3}{2}r}a^{N-3}
}{1-q^{1+\left(\frac{N}{2}-1\right)r}a^{N-2}}. 
\end{align}
We have checked numerically that, for $N=3,4,5$, 
the expression (\ref{soN_fundWL}) correctly reproduces the expansion coefficients for various choices of R-charge $r$. 

%SO(2n+1)-
We turn to the $SO(2n+1)$ theory with discrete fugacities $(\chi,\zeta)$ $=$ $(-1,-1)$. 
For the Wilson line in the fundamental representation, the corresponding line defect half-index is evaluated as
\begin{align}
\langle W_{\tiny \yng(1)}\rangle_{(\mathcal{N},N)}^{A}
&=
\frac{(q)_{\infty}^n}{n! 2^{n}} \prod_{i=1}^n \oint \frac{ds_i}{2\pi i s_i}
(-s_i^{\pm}; q)_{\infty}
\prod_{i < j}^n (s_i^{\pm} s_j^{\mp}; q)_{\infty} (s_i^{\pm} s_j^{\pm}; q)_{\infty}
 \nonumber \\
& \times
\frac{1}{\prod_{\alpha = 1}^{2n - 1} \left( (-q^{r/2} a x_\alpha; q)_{\infty} \prod_{i = 1}^n (q^{r/2} s_i^{\pm} a x_\alpha; q)_{\infty} \right)}
\left(-1+\sum_{i=1}^n (s_i+s_i^{-1})\right). 
\end{align}

Note that the charge conjugation $\mathbb{Z}_2$ symmetry 
acts by exchanging the weights $\pm e_i$, leaving their contributions $s_i$ and $s_i^{-1}$ invariant, 
while it acts with a definite parity on the zero-weight component. 
In the odd sector, the latter acquires a minus sign, resulting in the above character.
Taking the effective spin shift into account and identifying the polynomial contribution, we obtain a closed-form expression
\begin{align}
\label{soNodd-_fundWL}
&
\mu_1^{SO(2n+1)_{\chi=-,\zeta=-}}(a,x_{\alpha};q)
=\mathcal{I}^{\textrm{1d defect}}_{\tiny\yng(1)}(a,x_{\alpha};q)
\nonumber\\
&=\frac{\sum_{\alpha=1}^{2n-1}x_{\alpha} q^{\frac{r}{2}}a
+\sum_{\alpha=1}^{2n-1}x_{\alpha}^{-1}q^{1+(n-1)r}a^{2n-2}
}{1+q^{1+(n-\frac12)r}a^{2n-1}}. 
\end{align}
However, note that as for the half-indices the equality of these expressions is related to the $(\chi,\zeta)$ $=$ $(1,1)$ case by taking $s_i \to -s_i$ and $a \to -a$, hence this does not provide an independent identity. 

%SO(2n)--
We next consider the case of the $SO(2n)$ gauge theory with discrete fugacities $(\chi,\zeta)$ $=$ $(-1,-1)$. 
In this setting, the one-point function of the Wilson line in the fundamental representation takes the following form:
\begin{align}
\langle W_{\tiny \yng(1)}\rangle_{(\mathcal{N},N)}^A
&=
\frac{(q)_{\infty}^{n-1} (-q; q)_{\infty}}{(n-1)! 2^{n-1}} \prod_{i=1}^{n-1} \oint \frac{ds_i}{2\pi i s_i}
(s_i^{\pm}; q)_{\infty} (-s_i^{\pm}; q)_{\infty}
\prod_{i < j}^{n-1} (s_i^{\pm} s_j^{\mp}; q)_{\infty} (s_i^{\pm} s_j^{\pm}; q)_{\infty}
 \nonumber \\
 &\times
\frac{1}{\prod_{\alpha = 1}^{N - 2} \left( (\pm q^{r/2} a x_{\alpha}; q)_{\infty} \prod_{i = 1}^{n-1} (q^{r/2} s_i^{\pm} a x_{\alpha}; q)_{\infty} \right)} 
\left(\sum_{i=1}^{n-1} s_i+s_i^{-1}\right). 
\end{align}
By an analogous argument, we arrive at the following expression for the 1d defect index:
\begin{align}
\label{soNeven-_fundWL}
&
\mu_1^{SO(2n)_{\chi=-,\zeta=-}}(a,x_{\alpha};q)
=
\mathcal{I}^{\textrm{1d defect}}_{\tiny\yng(1)}(a,x_{\alpha};q)
\nonumber\\
&=\frac{\sum_{\alpha=1}^{2n-2}x_{\alpha} q^{\frac{r}{2}}a+\sum_{\alpha=1}^{2n-2}x_{\alpha}^{-1}q^{1+(n-\frac{3}{2})r}a^{2n-3}
}{1+q^{1+(n-1)r}a^{2(n-1)}}. 
\end{align}

It is worth noting that the expression (\ref{soNodd-_fundWL}) for $N=2n+1$ 
and the expression (\ref{soNeven-_fundWL}) for $N=2n$ with $(\chi,\zeta)=(-1,-1)$ admit a unified simple expression
\begin{align}
\label{soN-_fundWL}
&
\mu_1^{SO(N)_{\chi=-,\zeta=-}}(a,x_{\alpha};q)
=
\mathcal{I}^{\textrm{1d defect}}_{\tiny\yng(1)}(a,x_{\alpha};q)
\nonumber\\
&=\frac{\sum_{\alpha=1}^{N-2}x_{\alpha} q^{\frac{r}{2}}a+\sum_{\alpha=1}^{N-2}x_{\alpha}^{-1}q^{1+(\frac{N-3}{2})r}a^{N-3}
}{1+q^{1+(\frac{N}{2}-1)r}a^{N-2}}. 
\end{align}
In comparison with the $(\chi,\zeta)$ $=$ $(+1,+1)$ cases, 
the two terms in each of the numerator and denominator carry opposite relative signs. 
Turning off the flavor fugacities $a$ and $x_{\alpha}$, the expression (\ref{soN-_fundWL}) reduces to
\begin{align}
&
\mu_1^{SO(N)_{\chi=-,\zeta=-}}(1,1;q)
=\mathcal{I}^{\textrm{1d defect}}_{\tiny\yng(1)}(1,1;q)
\nonumber\\
&=
(N-2)\frac{q^{\frac{r}{2}}+q^{1+\frac{(N-3)r}{2}}}
{1+q^{1+(\frac{N}{2}-1)r}}. 
\end{align}

Let us consider the Witten indices. 
In the sector with $\chi=+$ and $\zeta=+$, the order of limits becomes essential, as discussed in section \ref{sec_VortexLG}.
To avoid an unphysical dependence on the trial R-charge, one must first take the limit $q\to 1$.
Subsequently setting the remaining fugacities to unity, we obtain from (\ref{soN_fundWL}) and (\ref{soN-_fundWL}) the Witten indices
\begin{align}
\mu_1^{SO(N)_{\chi=+,\zeta=+}}(1,1;1)
&=N-4, \\
\mu_1^{SO(N)_{\chi=-,\zeta=-}}(1,1;1)
&=N-2. 
\end{align}

%%%%%%%%%%%%%%%%%%%%%%%%%%%%%%%%%%%%%%%%%%%%%%%%%%%
\subsubsection{Multi-point functions}
%%%%%%%%%%%%%%%%%%%%%%%%%%%%%%%%%%%%%%%%%%%%%%%%%%%

Next consider higher-point functions of the fundamental Wilson lines in the orthogonal gauge theories. 
In analogy with the previous analysis, we propose that the exact $k$-point function is obtained 
by applying a spin shift of order $k$ to the singlet chiral multiplet $V$, 
supplemented by polynomial contributions arising from the Higgs branch operators. 
As illustrative examples, the unflavored multi-point functions for the $SO(3)$, $SO(4)$ and $SO(5)$ gauge theories  
that are even under both the charge conjugation and magnetic $\mathbb{Z}_2$ symmetries, are given by
\begin{align}
&
\mu_2^{SO(3)_{\chi=+,\zeta=+}}
:=
\mathcal{I}^{\textrm{1d defect}}_{\tiny\yng(1);\yng(1)}(a,1;q)
=\frac{\langle W_{\tiny \yng(1)} W_{\tiny \yng(1)}\rangle_{(\mathcal{N},N)}^A}
{ \mathbb{II}_{(\mathcal{N},N)}^A}
\nonumber\\
&=\frac{1}{(1-q^{\frac{r}{2}+1}a) (1-q^{\frac{r}{2}+2}a)}
\Bigl(
1+q^{\frac{r}{2}}a+q^r a^2
-q-2q^{1+\frac{r}{2}}a
\nonumber\\
&-2q^{2+\frac{r}{2}}a-q^{2+r}a^2+q^3+q^{3+\frac{r}{2}}a+q^{3+r}a^2
\Bigr), 
\end{align}
\begin{align}
&
\mu_3^{SO(3)_{\chi=+,\zeta=+}}
:=
\mathcal{I}^{\textrm{1d defect}}_{\tiny\yng(1);\yng(1);\yng(1)}(a,1;q)
=\frac{\langle W_{\tiny \yng(1)} W_{\tiny \yng(1)} W_{\tiny \yng(1)}\rangle_{(\mathcal{N},N)}^A}
{ \mathbb{II}_{(\mathcal{N},N)}^A}
\nonumber\\
&=\frac{1}{(1-q^{\frac{r}{2}+1}a) (1-q^{\frac{r}{2}+2}a) (1-q^{\frac{r}{2}+3}a)}
\Bigl(
1+3q^{\frac{r}{2}}a+2q^r a^2+q^{\frac{3r}{2}}a^3-3q-3q^{\frac{r}{2}+1}a-q^{1+r}a^2
\nonumber\\
&-3q^{2+\frac{r}{2}}a-4q^{2+r}a^2+2q^3+3q^{3+\frac{r}{2}}a-3q^{3+r}a^2
-2q^{3+\frac{3r}{2}}a^3+4q^{4+\frac{r}{2}}a+3q^{4+r}a^2+q^{5+\frac{r}{2}}a
\nonumber\\
&+3q^{5+r}a^2+3q^{5+\frac{3r}{2}}a^3-q^6-2q^{6+\frac{r}{2}}a-3q^{6+r}a^2-q^{6+\frac{3r}{2}}a^3
\Bigr), 
\end{align}
\begin{align}
&
\mu_2^{SO(4)_{\chi=+,\zeta=+}}
:=
\mathcal{I}^{\textrm{1d defect}}_{\tiny\yng(1);\yng(1)}(a,1;q)
=\frac{\langle W_{\tiny \yng(1)} W_{\tiny \yng(1)}\rangle_{(\mathcal{N},N)}^A}
{ \mathbb{II}_{(\mathcal{N},N)}^A}
\nonumber\\
&=\frac{1}{(1-q^{r+1}a^2) (1-q^{r+2}a^2)}
\Bigl(
1+5q^r a^2-2q-5q^{1+r}a^2+q^{1+2r}a^4
\nonumber\\
&+q^2-5q^{2+r}a^2-2q^{2+2r}a^4+5q^{3+r}a^2+q^{3+2r}a^4
\Bigr), 
\end{align}
and 
\begin{align}
&
\mu_2^{SO(5)_{\chi=+,\zeta=+}}
:=
\mathcal{I}^{\textrm{1d defect}}_{\tiny\yng(1);\yng(1)}(a,1;q)
=\frac{\langle W_{\tiny \yng(1)} W_{\tiny \yng(1)}\rangle_{(\mathcal{N},N)}^A}
{ \mathbb{II}_{(\mathcal{N},N)}^A}
\nonumber\\
&=\frac{1}{(1-q^{\frac{3r}{2}+1}a^2) (1-q^{\frac{3r}{2}+2}a^2)}
\Bigl(
1+9q^{r}a^2+q^{\frac{3r}{2}}a^3-q-3q^{1+\frac{r}{2}}a-10q^{1+\frac{3r}{2}}a^3+3q^{1+\frac{5r}{2}}a^5
\nonumber\\
&+3q^{2+\frac{r}{2}}a-10q^{2+\frac{3r}{2}}a^3-3q^{2+\frac{5r}{2}}a^5
-q^{2+3r}a^6+q^{3+\frac{3r}{2}}a^3+9q^{3+2r}a^4+q^{3+3r}a^6
\Bigr). 
\end{align}
For $(\chi,\zeta)$ $=$ $(-1,-1)$ 
the expressions have the same structure as in the $(\chi,\zeta)$ $=$ $(1,1)$ cases, but with different signs. 
For example, we have
\begin{align}
&
\mu_2^{SO(3)_{\chi=-,\zeta=-}}
:=
\mathcal{I}^{\textrm{1d defect}}_{\tiny\yng(1);\yng(1)}(a,1;q)
=\frac{\langle W_{\tiny \yng(1)} W_{\tiny \yng(1)}\rangle_{(\mathcal{N},N)}^A}
{ \mathbb{II}_{(\mathcal{N},N)}^A}
\nonumber\\
&=\frac{1}{(1+q^{\frac{r}{2}+1}a) (1+q^{\frac{r}{2}+2}a)}
\Bigl(
1-q^{\frac{r}{2}}a+q^r a^2
-q+2q^{1+\frac{r}{2}}a
\nonumber\\
&+2q^{2+\frac{r}{2}}a-q^{2+r}a^2+q^3-q^{3+\frac{r}{2}}a+q^{3+r}a^2
\Bigr), 
\end{align}
\begin{align}
&
\mu_2^{SO(5)_{\chi=-,\zeta=-}}
:=
\mathcal{I}^{\textrm{1d defect}}_{\tiny\yng(1);\yng(1)}(a,1;q)
=\frac{\langle W_{\tiny \yng(1)} W_{\tiny \yng(1)}\rangle_{(\mathcal{N},N)}^A}
{ \mathbb{II}_{(\mathcal{N},N)}^A}
\nonumber\\
&=\frac{1}{(1+q^{\frac{3r}{2}+1}a^2) (1+q^{\frac{3r}{2}+2}a^2)}
\Bigl(
1+9q^{r}a^2-q^{\frac{3r}{2}}a^3-q+3q^{1+\frac{r}{2}}a+10q^{1+\frac{3r}{2}}a^3-3q^{1+\frac{5r}{2}}a^5
\nonumber\\
&-3q^{2+\frac{r}{2}}a+10q^{2+\frac{3r}{2}}a^3+3q^{2+\frac{5r}{2}}a^5
-q^{2+3r}a^6-q^{3+\frac{3r}{2}}a^3+9q^{3+2r}a^4+q^{3+3r}a^6
\Bigr). 
\end{align}

We now consider more general Witten indices associated with the multi-point functions. 
As in the one-point function the order of limits is crucial for the $(\chi,\zeta)=(+,+)$ sector. 
In order to avoid an unphysical dependence on the trial R-charge, one must first take the limit $q\to 1$.
We then propose the following general conjecture: 
\begin{align}
\mu_k^{SO(N)_{\chi=+,\zeta=+}}(1,1;1)&
=(N-4)^k,\\
\mu_k^{SO(N)_{\chi=-,\zeta=-}}(1,1;1)&
=(N-2)^k. 
\end{align}

%%%%%%%%%%%%%%%%%%%%%%%%%%%%%%%%%%%%%%%%%%%%%%%%%%%
\subsection{$G_2$ with $4$ fundamental chirals}
\label{sec_G2_4}
%%%%%%%%%%%%%%%%%%%%%%%%%%%%%%%%%%%%%%%%%%%%%%%%%%%

%%%%%%%%%%%%%%%%%%%%%%%%%%%%%%%%%%%%%%%%%%%%%%%%%%%
\subsubsection{Boundary conditions}
%%%%%%%%%%%%%%%%%%%%%%%%%%%%%%%%%%%%%%%%%%%%%%%%%%%
Consider a 3d $\mathcal{N}=2$ $G_2$ gauge theory with four fundamental chirals $Q_\alpha$, $\alpha=1,\dots,4$, 
which is known to be $s$-confining \cite{Nii:2017npz} (see also \cite{Pesando:1995bq,Giddings:1995ns} for the 4d analogue). 
We choose the Neumann boundary conditions for the vector and chiral multiplets at the boundary. 
As discussed in \cite{Okazaki:2023kpq}, we have the dual boundary condition of theory B with 
an $SU(4)$ rank-$2$ symmetric chiral $M_{\alpha\beta}$, an $SU(4)$ antifundamental chiral $B^{\alpha}$, a singlet $B$ all obeying Neumann boundary conditions, 
as well as a singlet $V$ with Dirichlet boundary condition.  
The field content is summarized as follows: 
\begin{align}
\label{G2_4_charges}
\begin{array}{c|c|c|c|c|c}
& \textrm{bc} & G_2 & SU(N_f = 4) & U(1)_a & U(1)_R \\ \hline
\textrm{VM} & \mathcal{N} & {\bf Adj} & {\bf 1} & 0 & 0 \\
Q_{\alpha} & \textrm{N} & {\bf 7} & {\bf 4} & 1 & 0 \\
 \hline
M_{\alpha \beta} & \textrm{N} & {\bf 1} & {\bf 10} & 2 & 0 \\
B^{\alpha} & \textrm{N} & {\bf 1} & {\bf \overline{4}} & 3 & 0 \\
B & \textrm{N} & {\bf 1} & {\bf 1} & 4 & 0 \\
V & \textrm{D} & {\bf 1} & {\bf 1} & -8 & 2
\end{array}
\end{align}

The half-index for theory A reads
\begin{align}
\label{bdy_G2_4_hindexA}
\II^A_{\Ncal, N} = & \frac{(q)_{\infty}^2}{2^2 3} \prod_{i=1}^2 \oint \frac{ds_i}{2\pi i s_i}
\frac{\left( \prod_{i \ne j}^3 (s_i s_j^{-1}; q)_{\infty} \right) \prod_{i = 1}^3 (s_i^{\pm}; q)_{\infty} }{\prod_{\alpha = 1}^4 (q^{r/2} a x_{\alpha}; q)_{\infty} \prod_{i = 1}^3 (q^{r/2} s_i^{\pm} a x_{\alpha}; q)_{\infty}}, 
\end{align}
where $\prod_{i = 1}^3 s_i = \prod_{\alpha = 1}^4 x_{\alpha} = 1$. 
The half-index for theory B is given by
\begin{align}
\label{bdy_G2_4_hindexB}
\II^B_{N, N, N, D} = & \frac{(q^{4r}a^8;q)_{\infty}}{(q^{2r}a^4;q)_{\infty} \left( \prod_{\alpha \le \beta}^4 (q^r a^2 x_{\alpha} x_{\beta}; q)_{\infty} \right) \prod_{\alpha = 1}^4 (q^{3r/2} a^3 x_{\alpha}^{-1};q)_{\infty}}. 
\end{align}
The two expressions (\ref{bdy_G2_4_hindexA}) and (\ref{bdy_G2_4_hindexB}) are shown to be equivalent \cite{MR1139492}. 

%%%%%%%%%%%%%%%%%%%%%%%%%%%%%%%%%%%%%%%%%%%%%%%%%%%
\subsubsection{One-point function}
%%%%%%%%%%%%%%%%%%%%%%%%%%%%%%%%%%%%%%%%%%%%%%%%%%%
Let us introduce a Wilson line operator in the fundamental representation of the $G_2$ gauge group, 
compatible with the boundary conditions of Theory A. 
The insertion of this line defect modifies the half-index by the corresponding $G_2$ character, 
and the resulting expression is naturally interpreted as the first moment of the associated $G_2$-type integral, 
\begin{align}
\mu_1^{G_2}(a,x_{\alpha};q) \mathbb{II}_{(\mathcal{N},N)}^A
&=\langle W_{\tiny \yng(1)}\rangle_{(\mathcal{N},N)}^A, 
\end{align}
where 
\begin{align}
\label{bdy_G2_4_W}
\langle W_{\tiny \yng(1)}\rangle^A_{\Ncal, N} = & \frac{(q)_{\infty}^2}{2^2 3} \prod_{i=1}^2 \oint \frac{ds_i}{2\pi i s_i}
\frac{\left( \prod_{i \ne j}^3 (s_i s_j^{-1}; q)_{\infty} \right) \prod_{i = 1}^3 (s_i^{\pm}; q)_{\infty} }{\prod_{\alpha = 1}^4 (q^{r/2} a x_{\alpha}; q)_{\infty} \prod_{i = 1}^3 (q^{r/2} s_i^{\pm} a x_{\alpha}; q)_{\infty}}
\nonumber\\
&\times 
(1+s_1+s_2+s_1^{-1}+s_2^{-1}+s_1s_2+s_1^{-1}s_2^{-1}). 
\end{align}
The first moment that captures the additional degrees of freedom localized on the line operator can be regarded as the 1d defect index 
\begin{align}
\mu_1^{G_2}(a,x_{\alpha};q)
&=\mathcal{I}^{\textrm{1d defect}}_{\tiny\yng(1)}(a,x_{\alpha};q). 
\end{align}

Similarly, on the dual theory B side, the effect of the vortex line can be captured by a deformation of the Dirichlet chiral multiplet $V$, which results in a shift of its effective spin. 
Once this contribution is extracted, the remaining terms organize into a finite expression. 
We find 
\begin{align}
&
\mu_1^{G_2}(a,x_{\alpha};q)=
\mathcal{I}^{\textrm{1d defect}}_{\tiny\yng(1)}(a,x_{\alpha};q)
\nonumber\\
&=\frac{
\sum_{\alpha=1}^4x_{\alpha} q^{\frac{r}{2}}a 
+\sum_{\alpha<\beta}x_{\alpha}^{-1}x_{\beta}^{-1}q^{r}a^2
+\sum_{\alpha=1}^{4}x_{\alpha}^{-1}q^{\frac{3r}{2}}a^3
}
{1+q^{2r}a^4}. 
\end{align}
We have confirmed that 
this closed-form expression correctly reproduces the terms in the expansions for various values of the R-charge. 

In the unflavored limit the defect index reduces to
\begin{align}
\mu_1^{G_2}(1,1;q)=
\mathcal{I}^{\textrm{1d defect}}_{\tiny\yng(1)}(1,1;q)
&=\frac{4q^{\frac{r}{2}}+6q^{r}+4q^{\frac{3r}{2}}}{1+q^{2r}}. 
\end{align}
The limit $q\rightarrow 1$ gives the Witten index
\begin{align}
\mu_1^{G_2}(1,1;1)=
\mathcal{I}^{\textrm{1d defect}}_{\tiny\yng(1)}(1,1;1)
&=7. 
\end{align}

%%%%%%%%%%%%%%%%%%%%%%%%%%%%%%%%%%%%%%%%%%%%%%%%%%%
\subsubsection{Multi-point functions}
%%%%%%%%%%%%%%%%%%%%%%%%%%%%%%%%%%%%%%%%%%%%%%%%%%%
We proceed to analyze higher-point functions of fundamental Wilson line operators. 
As in the preceding cases, the effect of multiple insertions is accounted for by a spin shift of the monopole-dual singlet chiral multiplet $V$, increasing by one unit per insertion. 
This produces a universal prefactor, while the remaining contributions assemble into a polynomial term. 
Determining this polynomial leads to an exact expression. 
In particular, we find that the two-point function is given by
\begin{align}
&
\mu_2^{G_{2}}(a,1;q)
:=
\mathcal{I}^{\textrm{1d defect}}_{\tiny\yng(1);\yng(1)}(a,1;q)
=\frac{\langle W_{\tiny \yng(1)} W_{\tiny \yng(1)}\rangle_{(\mathcal{N},N)}^A}
{ \mathbb{II}_{(\mathcal{N},N)}^A}
\nonumber\\
&=\frac{1}{(1-q^{4r}a^8) (1-q^{4r+1}a^8)}
\Bigl(
1 - q + 4 q^{\frac{r}{2}} a + 22 q^{r} a^{2} + 48 q^{\frac{3r}{2}} a^{3} + 51 q^{2r} a^{4} + 16 q^{\frac{5r}{2}} a^{5} \nonumber\\
&- 38 q^{3r} a^{6} - 76 q^{\frac{7r}{2}} a^{7} - 68 q^{4r} a^{8} - 20 q^{\frac{9r}{2}} a^{9}
+ 22 q^{5r} a^{10} + 32 q^{\frac{11r}{2}} a^{11} + 17 q^{6r} a^{12}  \nonumber\\
&- 6 q^{7r} a^{14} - 4 q^{\frac{15r}{2}} a^{15} - q^{8r} a^{16} 
- 4 q^{\frac{r}{2}+1} a - 6 q^{r+1} a^{2} + 17 q^{2r+1} a^{4} + 32 q^{\frac{5r}{2}+1} a^{5}  \nonumber\\
&+ 22 q^{3r+1} a^{6} - 20 q^{\frac{7r}{2}+1} a^{7} - 68 q^{4r+1} a^{8}
- 76 q^{\frac{9r}{2}+1} a^{9} - 38 q^{5r+1} a^{10} + 16 q^{\frac{11r}{2}+1} a^{11} \nonumber\\
&+ 51 q^{6r+1} a^{12} + 48 q^{\frac{13r}{2}+1} a^{13} + 22 q^{7r+1} a^{14} + 4 q^{\frac{15r}{2}+1} a^{15} + q^{8r+1} a^{16}
\Bigr). 
\end{align}
As $q\rightarrow 1$, the expression reduces to the Witten index
\begin{align}
\mu_2^{G_{2}}(1,1;1)=49. 
\end{align}
We thus conjecture the following values for the $k$-th moments corresponding to the $k$-point functions of the fundamental Wilson lines
\begin{align}
\mu_k^{G_{2}}(1,1;1)=7^k. 
\end{align}

%%%%%%%%%%%%%%%%%%%%%%%%%%%%%%%%%%%%%%%%%%%%%%%%%
%%%%%%%%%%%%%%%%%%%%%%%%%%%%%%%%%%%%%%%%%%%%%%%%%
\subsection{$U(N)$ with $N_f=N_a=N$}
%%%%%%%%%%%%%%%%%%%%%%%%%%%%%%%%%%%%%%%%%%%%%%%%%
%%%%%%%%%%%%%%%%%%%%%%%%%%%%%%%%%%%%%%%%%%%%%%%%%

%%%%%%%%%%%%%%%%%%%%%%%%%%%%%%%%%%%%%%%%%%%%%%%%%
\subsubsection{Boundary conditions}
%%%%%%%%%%%%%%%%%%%%%%%%%%%%%%%%%%%%%%%%%%%%%%%%%
Consider theory A as 3d $\mathcal{N}=2$ $U(N)$ gauge theory 
with $N_f=N$ fundamental chiral multiplets $Q_{I}$, $I=1,\cdots, N$ 
and $N_a=N$ antifundamental chiral multiplets $\overline{Q}_{\alpha}$, $\alpha=1,\cdots, N$ 
each assigned R-charge $r$. 
Choosing the $\mathcal{N}=(0,2)$ Neumann boundary conditions for both gauge and matter multiplets, 
the residual gauge anomaly can be cancelled by introducing the Fermi multiplet $\Lambda$ transforming in the determinant representation. 

In the dual description (theory B), 
one has the chiral multiplets $Y$ and $Z$ which match the monopole operators of theory A 
and the chiral multiplet $M_{I \alpha}$ transforming in the bifundamental representation of $SU(N)\times SU(N)$, 
which maps to the mesons of theory A. 
It has the superpotential 
\begin{align}
\mathcal{W}&=\det (M_{I \alpha}) YZ. 
\end{align}
The dual boundary condition consists of the $\mathcal{N}=(0,2)$ Dirichlet boundary condition for $Y,Z$ 
and the Neumann boundary condition for $M_{I \alpha}$ \cite{Dimofte:2017tpi}. 
The field content and their boundary conditions are summarized as
\begin{align}
\label{uN_NN_charges}
\begin{array}{c|c|c|c|c|c|c|c}
& \textrm{bc} & U(N) & SU(N_f = N) & SU(N_a = N) & U(1)_z& U(1)_a & U(1)_R \\ \hline
\textrm{VM} & \mathcal{N} & {\bf Adj} &{\bf 1} & {\bf 1} & 0 & 0 & 0 \\
Q_{I} & \textrm{N} & {\bf N} & {\bf N} & {\bf 1}& 0& 1 & 0 \\
\overline{Q}_{\alpha} & \textrm{N} & \overline{\bf N} & {\bf 1} & {\bf N} & 0 & 1 & 0 \\
\Lambda&  & {\bf det} & {\bf 1} & {\bf 1}& 1& 0 & 0 \\
 \hline
M_{I \alpha} & \textrm{N} & {\bf 1} & {\bf N}& {\bf N} & 0& 2 & 0 \\
Y & \textrm{D} & {\bf 1}& {\bf 1} & {\bf 1} & 1& -N & 1\\
Z & \textrm{D} & {\bf 1}& {\bf 1} & {\bf 1} & -1& -N & 1\\
\end{array}
\end{align}

The half-index for theory A is 
\begin{align}
\label{uN_halfindA}
\mathbb{II}_{(\mathcal{N},N,N)+\Lambda}^A
&=\frac{(q)_{\infty}^N}{N!}
\prod_{i=1}^N \oint \frac{ds_i}{2\pi is_i}
\prod_{i\neq j}(s_is_j^{-1};q)_{\infty}
\nonumber\\
&\times \frac{(q^{1/2} \prod_{i=1}^N s_i^{\pm}z^{\pm};q)_{\infty}}
{\prod_{i=1}^N (\prod_{I=1}^N (q^{r/2}as_ix_I;q)_{\infty}) (\prod_{\alpha=1}^N (q^{r/2}as_i^{-1}x_{\alpha};q)_{\infty}) }
\end{align}
with $\prod_{I}x_{I}=\prod_{\alpha}x_{\alpha}=1$. 
The half-index for theory B is
\begin{align}
\label{uN_halfindB}
\mathbb{II}_{N,D,D}^B
&=\frac{(q^{\frac{1+Nr}{2}}a^Nz^{\pm};q)_{\infty}}
{\prod_{I=1}^N \prod_{\alpha=1}^N (q^ra^2x_Ix_{\alpha};q)_{\infty}}. 
\end{align}
According to the duality of the boundary conditions, 
the equality of the two expressions (\ref{uN_halfindA}) and (\ref{uN_halfindB}) is conjectured. 

%%%%%%%%%%%%%%%%%%%%%%%%%%%%%%%%%%%%%%%%%%%%%%%%%
\subsubsection{One-point functions}
%%%%%%%%%%%%%%%%%%%%%%%%%%%%%%%%%%%%%%%%%%%%%%%%%
We introduce a Wilson line in the fundamental representation of the $U(N)$ gauge group, 
chosen to be compatible with the above boundary conditions. 
In the presence of this line operator, the line defect half-index takes the form
\begin{align}
\label{uN_halfindA_W1}
\langle W_{\tiny \yng(1)}\rangle_{(\mathcal{N},N,N)+\Lambda}^A
&=\frac{(q)_{\infty}^N}{N!}
\prod_{i=1}^N \oint \frac{ds_i}{2\pi is_i}
\prod_{i\neq j}(s_is_j^{-1};q)_{\infty}
\nonumber\\
&\times \frac{(q^{1/2} \prod_{i=1}^N s_i^{\pm}z^{\pm};q)_{\infty}}
{\prod_{i=1}^N (\prod_{I=1}^N (q^{r/2}as_ix_I;q)_{\infty}) (\prod_{\alpha=1}^N (q^{r/2}as_i^{-1}x_{\alpha};q)_{\infty}) }
\left(\sum_{i=1}^N s_i\right). 
\end{align}
Let us define the first moment as the normalized line defect half-index as
\begin{align}
\mu_1^{U(N)}(a,x_I,\tilde{x}_{\alpha};q) \mathbb{II}_{(\mathcal{N},N,N)+\Lambda}^A
&=\langle W_{\tiny \yng(1)}\rangle_{(\mathcal{N},N,N)+\Lambda}^A. 
\end{align}
Physically it can be identified with the 1d defect index
\begin{align}
\mu_1^{U(N)}(a,x_I,\tilde{x}_{\alpha};q)
&=\mathcal{I}_{\tiny \yng(1)}^{\textrm{1d defect}}(a,x_I,\tilde{x}_{\alpha};q). 
\end{align}

Again, the closed-form expression for the 1d defect index can be addressed from the viewpoint of the dual theory B. 
In theory B, the chiral multiplet $Y$ and $Z$ map to the monopole operator. 
A key distinction of the $U(N)$ gauge theory, as opposed to theories with a simple gauge group such as $SU(N)$ is the presence of an Abelian $U(1)$ factor.
It allows the Wilson lines to carry an additional $U(1)$ electric charge. 
In the dual description, this charge maps to the topological $U(1)$ symmetry rotating $Y$ and $Z$, 
so that the line operator induces a background topological charge. 
This background selects chiral multiplets according to their $U(1)$ charges, 
effectively reducing the effect to a choice between monopole operators of charge $\pm1$. 
We find that adopting the following spin shift allows for an exact closed-form expression: 
\begin{align}
\label{uN_shift}
(q^{\frac{1+Nr}{2}}a^Nz^{\pm})\rightarrow
\begin{cases}
(q^{\frac{1+Nr}{2}}a^Nz)(q^{\frac{1+Nr}{2}+1}a^Nz^{-1})&\textrm{for $W_{\tiny \yng(1)}$}\cr
(q^{\frac{1+Nr}{2}+1}a^Nz)(q^{\frac{1+Nr}{2}}a^Nz^{-1})&\textrm{for $W_{\tiny \overline{\yng(1)}}$}\cr
\end{cases}
\end{align}
In other words, the shift for the positively (resp. negatively) charged Wilson line in the fundamental (resp. antifundamental) representation 
corresponds to selecting the negatively charged chiral $Z$ (resp. positively charged chiral $Y$). 
We thus arrive at the closed-form expression 
\begin{align}
\label{uN_W1}
&
\mu_1^{U(N)}(a,x_I,\tilde{x}_{\alpha};q)
=\mathcal{I}_{\tiny \yng(1)}^{\textrm{1d defect}}(a,x_I,\tilde{x}_{\alpha};q)
\nonumber\\
&=\frac{q^{\frac{r}{2}}a\sum_{\alpha=1}^N x_{\alpha}-q^{\frac12+\frac{(N-1)r}{2}}a^{N-1}z^{-1}\sum_{I}^N x_I^{-1}}
{1-q^{\frac12+\frac{Nr}{2}}a^N z^{-1}}. 
\end{align}

We now consider the Witten index.
In this case, one must be careful about the order of limits, as discussed in section \ref{sec_VortexLG}.
To avoid an unphysical dependence on the trial R-charge, we first take the limit $q\to 1$, 
from which the following expression is obtained from (\ref{uN_W1}): 
\begin{align}
\mu_1^{U(N)}(1,1,1;1)
&=N-2. 
\end{align}

We further consider the Wilson line operators in higher-rank representations. 
For the rank-$k$ antisymmetric representation, 
it can be simply computed by replacing the character of the fundamental representation in the integrand (\ref{uN_halfindA_W1}) 
with the $k$-th elementary symmetric function $e_k(s)$ 
\begin{align}
\langle W_{(1^k)}\rangle_{(\mathcal{N},N,N)+\Lambda}^A
&=\frac{(q)_{\infty}^N}{N!}
\prod_{i=1}^N \oint \frac{ds_i}{2\pi is_i}
\prod_{i\neq j}(s_is_j^{-1};q)_{\infty}
\nonumber\\
&\times \frac{(q^{1/2} \prod_{i=1}^N s_i^{\pm}z^{\pm};q)_{\infty}}
{\prod_{i=1}^N (\prod_{I=1}^N (q^{r/2}as_ix_I;q)_{\infty}) (\prod_{\alpha=1}^N (q^{r/2}as_i^{-1}x_{\alpha};q)_{\infty}) }
e_k(s). 
\end{align}
It is found to be evaluated by a shift of the same type as in the fundamental representation given by (\ref{uN_shift}). 
We propose the following closed-form formula for the unflavored one-point function:
\begin{align}
\label{uN_Wasymk}
&
\mathcal{I}_{(1^k)}^{\textrm{1d defect}}(a,1,1;q)
=\frac{\langle W_{(1^k)}\rangle_{(\mathcal{N},N,N)+\Lambda}^A}{\mathbb{II}_{(\mathcal{N},N,N)+\Lambda}^A}
\nonumber\\
&=
\left(
\begin{matrix}
N\\
k\\
\end{matrix}
\right)
\frac{q^{\frac{kr}{2}}a^k-q^{\frac{1}{2}+\frac{(N-k)r}{2}}a^{N-k}}
{1-q^{\frac12+\frac{Nr}{2}}a^N}. 
\end{align}
Applying the same procedure, we take the limit $q\rightarrow 1$ and then setting the remaining fugacity $a$ to unity, 
we obtain the following Witten index:
\begin{align}
\mathcal{I}_{(1^k)}^{\textrm{1d defect}}(1,1,1;1)
&=\left(
\begin{matrix}
N-1\\
k\\
\end{matrix}
\right)
-\left(
\begin{matrix}
N-1\\
k-1\\
\end{matrix}
\right). 
\end{align}

%%%%%%%%%%%%%%%%%%%%%%%%%%%%%%%%%%%%%%%%%%%%%%%%%
\subsubsection{Multi-point functions}
%%%%%%%%%%%%%%%%%%%%%%%%%%%%%%%%%%%%%%%%%%%%%%%%%
Consider higher-point functions of the (anti)fundamental Wilson lines.
The insertion of a single fundamental Wilson line (resp. antifundamental Wilson line) acts  
by shifting the spin of the monopole-dual singlet chiral multiplet $Z$ (resp. $Y$) by one unit per insertion. 
This results in a universal prefactor, with all remaining contributions captured by a polynomial contribution.
Once this polynomial is fixed, an exact expression follows. 
In particular, the unflavored two-point function of the fundamental Wilson line and the antifundamental Wilson line is given by
\begin{align}
&
\mathcal{I}_{\tiny \yng(1);\overline{\yng(1)}}^{\textrm{1d defect}}(a,1,1;q)
=\frac{\langle W_{\tiny \yng(1)} W_{\tiny \overline{\yng(1)}}\rangle_{(\mathcal{N},N,N)+\Lambda}^A}
{\mathbb{II}_{(\mathcal{N},N,N)+\Lambda}^A}
\nonumber\\
&=\frac{
1+N^2q^ra^2-q-2N^2q^{\frac{Nr+1}{2}}a^N-q^{Nr}a^{2N}+N^2q^{(N-1)r+1}a^{2(N-1)}+q^{Nr+1}a^{2N}
}
{(1-q^{\frac{1+Nr}{2}} a^{N})^2}. 
\end{align}

We conjecture that, for correlation functions involving a total of $k$ fundamental and antifundamental Wilson lines, 
the associated Witten indices are given by the $k$-th power of the one-point function
\begin{align}
\mathcal{I}_{\underbrace{\tiny \yng(1);\cdots;\yng(1);\overline{\yng(1)},\cdots, \overline{\yng(1)}}_{k}}
^{\textrm{1d defect}}(1,1,1;1)
&=(N-2)^k. 
\end{align}

%%%%%%%%%%%%%%%%%%%%%%%%%%%%%%%%%%
%%%%%%%%%%%%%%%%%%%%%%%%%%%%%%%%%%
\section{Macdonald-Koornwinder moments}
\label{sec_MK_moments}
%%%%%%%%%%%%%%%%%%%%%%%%%%%%%%%%%%
%%%%%%%%%%%%%%%%%%%%%%%%%%%%%%%%%%
Here we focus on the case where theory A has general $USp(2n)$ gauge group 
with a rank-$2$ antisymmetric chiral as well as fundamental chirals. 
For the case of $5(+1)$ chirals (the `$+1$' chiral having different charges and boundary condition) 
the Neumann half-indices can be identified with the Gustafson-Rakha integrals \cite{MR1266569} as noted in \cite{Okazaki:2023hiv} and reviewed below. 
If we integrate out $1(+1)$ chirals by sending masses to infinity, 
the half-index integrand reduces to the weight for Macdonald-Koornwinder polynomials. 
We then note some results in the literature for Macdonald-Koornwinder moments, 
albeit with a specialization of the flavor fugaticty for the antisymmetric chiral, which correspond to exact results for line operator half-indices.

%%%%%%%%%%%%%%%%%%%%%%%%%%%%%%%%%%
\subsection{$USp(2n)$ with rank-$2$ antisymmetric and $4$ fundamental chirals}
\label{sec_MacKoor_USp_AS_integrals}
%%%%%%%%%%%%%%%%%%%%%%%%%%%%%%%%%%

%%%%%%%%%%%%%%%%%%%%%%%%%%%%%%%%%%
\subsubsection{$USp(2n)$ with rank-$2$ antisymmetric and $5(+1)$ fundamental chirals}
\label{sec_GNR_USp_AS_integrals}
%%%%%%%%%%%%%%%%%%%%%%%%%%%%%%%%%%
We begin by reviewing the $USp(2n)$ gauge theory with a rank-2 antisymmetric chiral multiplet 
and $5(+1)$ fundamental chirals, together with its boundary conditions and associated dualities, as discussed in \cite{Okazaki:2023hiv}.
Theory A has a $USp(2n)$ vector multiplet obeying Neumann boundary condition, 
a $USp(2n)$ rank-2 antisymmetric chiral $\Phi$ with Neumann boundary condition, 
5 fundamental chirals $Q_{\alpha}$ with Neumann boundary conditions, 
and one fundamental chiral $\widetilde{Q}$ with Dirichlet boundary condition. 
The dual theory B has $n$ copiesi (labelled by $\lambda$ and differing only in the $U(1)_A$ charge) of the following chirals
in representations of the $SU(5)$ global flavor symmetry group:
rank-2 antisymmetric chirals $M_{\alpha\beta}^{(\lambda)}$ with Neumann boundary conditions, 
singlets $\phi^{(\lambda)}$ with Neumann boundary conditions and fundamentals $\widetilde{M}_{\alpha}^{(\lambda)}$ with Dirichlet boundary conditions. 
The field content and the charges are given by
\begin{align}
\label{USp2n_AS_6_charges}
\begin{array}{c|c|c|c|c|c|c|c|c}
& \textrm{bc} & USp(2n) & SU(N_f = 5) & U(1)_A & U(1)_a & U(1)_R \\ \hline
\textrm{VM} & \mathcal{N} & {\bf Adj} & {\bf 1} & 0 & 0 & 0 \\
\Phi &\textrm{N}& {\bf n(2n-1)} & {\bf 1} & 1 & 0 & 0 \\
Q_{\alpha} & \textrm{N} & {\bf 2n} & {\bf 5} & 0 & 1 & 0 \\
\widetilde{Q} & \textrm{D} & {\bf 2n} & {\bf 1} & 2 - 2n & -5 & 2 \\
 \hline
M_{\alpha \beta}^{(\lambda)} & \textrm{N} & {\bf 1} & {\bf 10} & \lambda-1 & 2 & 0 \\
\phi^{(\lambda)} & \textrm{N} & {\bf 1} & {\bf 1} & \lambda & 0 & 0 \\
\widetilde{M}_{\alpha}^{(\lambda)} & \textrm{D} & {\bf 1} & {\bf 5} & \lambda + 1 - 2n & -4 & 2
\end{array}
\end{align}
where $\lambda \in \{1, 2, \ldots , n\}$.

The Neumann half-index for theory A takes the form: 
\begin{align}
\label{bdy_USp2n_AS_6_hindexA}
&
\mathbb{II}_{(\mathcal{N,N,N,D})}^{A}
\nonumber\\
&=\frac{(q)_{\infty}^n}{n! 2^n} \prod_{i=1}^n \oint \frac{ds_i}{2\pi i s_i}
\prod_{i \ne j}^n (s_i s_j^{-1}; q)_{\infty} \prod_{i \le j}^n (s_i^{\pm} s_j^{\pm}; q)_{\infty}
\frac{\prod_{i = 1}^n (q^{(n-1)r_A+5r_a/2} A^{2n - 2} a^5 s_i^{\pm}; q)_{\infty}}{\prod_{\alpha = 1}^{5} \prod_{i = 1}^n (q^{r_a/2} s_i^{\pm} a x_{\alpha}; q)_{\infty}}
\nonumber \\
& \times \frac{1}{(q^{r_A/2} A; q)_{\infty}^n \prod_{i < j}^n (q^{r_A/2} A s_i^{\pm} s_j^{\mp}; q)_{\infty} (q^{r_A/2} A s_i^{\pm} s_j^{\pm}; q)_{\infty}}
\end{align}
where $\prod_{\alpha = 1}^{5} x_{\alpha} = 1$.
The half-index of theory B reads
\begin{align}
\label{bdy_USp2n_AS_6_hindexB}
\mathbb{II}_{N,N,D}^{B}
&=\prod_{\lambda = 1}^n \frac{\prod_{\alpha=1}^{5} (q^{(2n - 1 - \lambda)r_A/2 + 2r_a} A^{2n - 1 - \lambda} a^4 x_{\alpha}^{-1}; q)_{\infty}}{(q^{\lambda r_A/2} A^\lambda; q)_{\infty} \prod_{1 \le \alpha < \beta \le 5} (q^{(\lambda-1)r_A/2 + r_a} A^{\lambda-1} a^2 x_{\alpha} x_{\beta}; q)_{\infty}}. 
\end{align}
By making use of Theorem 2.1 in \cite{MR1266569}, 
we can show that (\ref{bdy_USp2n_AS_6_hindexA}) and (\ref{bdy_USp2n_AS_6_hindexB}) are identical.

%%%%%%%%%%%%%%%%%%%%%%%%%%%%%%%%%%
\subsubsection{$USp(2n)$ with rank-$2$ antisymmetric and $4$ fundamental chirals}
%%%%%%%%%%%%%%%%%%%%%%%%%%%%%%%%%%
Starting from the $USp(2n)$ gauge theory with a rank-2 antisymmetric chiral multiplet and $5(+1)$ fundamental chirals, 
we integrate out $1(+1)$ fundamental chiral multiplets by taking their masses to infinity. 
The resulting theory A is the $USp(2n)$ gauge theory with a rank-2 antisymmetric chiral multiplet and four fundamental chiral multiplets. 
The dual theory B consists of $SU(4)$ rank-2 antisymmetric chiral multiplets $M_{\alpha\beta}^{(\lambda)}$, 
singlets $\phi^{(\lambda)}$ and singlets $\widetilde{M}^{(\lambda)}$. 
The field content and the charges of theories A and B are given by
\begin{align}
\label{USp2n_AS_4_charges}
\begin{array}{c|c|c|c|c|c|c|c|c}
& USp(2n) & SU(N_f = 4) & U(1)_A & U(1)_a & U(1)_R \\ \hline
\textrm{VM} & {\bf Adj} & {\bf 1} & 0 & 0 & 0 \\
\Phi & {\bf n(2n-1)} & {\bf 1} & 1 & 0 & 0 \\
Q_{\alpha} & {\bf 2n} & {\bf 4} & 0 & 1 & 0 \\
 \hline
M_{\alpha \beta}^{(\lambda)} & {\bf 1} & {\bf 6} & \lambda-1 & 2 & 0 \\
\phi^{(\lambda)} & {\bf 1} & {\bf 1} & \lambda & 0 & 0 \\
\widetilde{M}^{(\lambda)} & {\bf 1} & {\bf 1} & \lambda + 1 - 2n & -4 & 2
\end{array}
\end{align}
where $\lambda \in \{1, 2, \ldots , n\}$. 
The mapping of operators between theory A and theory B is given by
$M_{\alpha \beta}^{(\lambda)} \sim Q_{\alpha} \Phi^{\lambda-1} Q_{\beta}$ and $\phi^{(\lambda)} \sim \Phi^\lambda$ where the gauge indices are contracted with the $USp(2n)$-invariant rank-$2$ antisymmetric tensor. While, in terms of charges, $\widetilde{M}^{(\lambda)}$ here corresponds to $\widetilde{M}_{\alpha = 5}^{(\lambda)}$ from the case with $5(+1)$ fundamentals, the interpretation is different. In the case with $5(+1)$ fundamentals, the mapping is $\widetilde{M}_{\alpha}^{(\lambda)} \sim Q_{\alpha} \Phi^{\lambda-1} \widetilde{Q}$ but in the reduction to $4$ fundamentals both $Q_5$ and $\widetilde{Q}$ are removed. Instead the interpretation of $\widetilde{M}^{(\lambda)}$ is the dual of the theory A minimal monopole dressed with $\Phi^{\lambda - 1}$.

Since, to our knowledge, this duality has not been established in the literature, we take this opportunity to verify its consistency by computing the supersymmetric full-indices. 
The full-index for theory A can be evaluated as
\begin{align}
&
I^A=
\frac{1}{n! 2^n}\sum_{m_1,\cdots, m_n}
\prod_{i=1}^n 
\oint \frac{ds_i}{2\pi is_i}
\prod_{i\neq j}^{n}
(1-q^{\frac{|m_i-m_j|}{2}}s_i s_j^{-1})
\prod_{i\le j}^n
(1-q^{\frac{|m_i+m_j|}{2}}s_i^{\pm} s_j^{\pm})
\nonumber\\
&\times 
\prod_{i=1}^{n}
\prod_{\alpha=1}^4
\frac{
(q^{1-\frac{r_a}{2}+\frac{|m_i|}{2}} s_i^{\mp}a^{-1}x_{\alpha}^{-1};q)_{\infty}
}
{
(q^{\frac{r_a+|m_i|}{2}} s_i^{\pm}ax_{\alpha};q)_{\infty}
}
\nonumber\\
&\times 
\prod_{i<j}^n 
\frac{
(q^{1-\frac{r_A}{2}+\frac{|m_i-m_j|}{2}} s_i^{\pm}s_j^{\mp}A^{-1};q)_{\infty}
}
{
(q^{\frac{r_A+|m_i-m_j|}{2}} s_i^{\pm}s_j^{\mp}A;q)_{\infty}
}
\frac{
(q^{1-\frac{r_A}{2}+\frac{|m_i+m_j|}{2}} s_i^{\mp}s_j^{\mp}A^{-1};q)_{\infty}
}
{
(q^{\frac{r_A+|m_i+m_j|}{2}} s_i^{\pm}s_j^{\pm}A;q)_{\infty}
}
\nonumber\\
&\times 
\frac{
(q^{1-\frac{r_A}{2}}A^{-1};q)_{\infty}^n
}
{
(q^{\frac{r_A}{2}}A;q)_{\infty}^n
}
\nonumber\\
&\times 
q^{2(1-r_a)\sum_i|m_i|+\frac{1}{2}(1-r_A)\sum_{i<j}|m_i\pm m_j| 
-\frac12 \sum_{i<j}|m_i\pm m_j|
-\sum_i|m_i|}
\nonumber\\
&\times 
A^{-\sum_{i<j}|m_i\pm m_j|}
a^{-4\sum_{i}|m_i|}. 
\end{align}
The full-index of theory B is 
\begin{align}
I^B&=
\prod_{\lambda=1}^n 
\prod_{\alpha<\beta}
\frac{
(q^{1-(\lambda-1)\frac{r_A}{2}-r_a}A^{-\lambda+1}a^{-2}x_{\alpha}^{-1}x_{\beta}^{-1};q)_{\infty}
}
{
(q^{(\lambda-1)\frac{r_A}{2}+r_a}A^{\lambda-1}a^{2}x_{\alpha}x_{\beta};q)_{\infty}
}
\nonumber\\
&\times 
\frac{
(q^{(2n-1-\lambda)\frac{r_A}{2}+2r_a}A^{2n-1-\lambda}a^4;q)_{\infty}
}
{
(q^{1-(2n-1-\lambda)\frac{r_A}{2}-2r_a}A^{-2n+1+\lambda}a^{-4};q)_{\infty}
}
\frac{
(q^{1-\lambda\frac{r_A}{2}}A^{-\lambda};q)_{\infty}
}
{
(q^{\lambda\frac{r_A}{2}}A^{\lambda};q)_{\infty}
}. 
\end{align}
We have checked that the indices coincide for several choices of R-charge assignments for $n=2, 3$. 
As an illustrative example, for $n=2$, $r_a=1/6$ ,and $r_A=1/4$, the unflavored indices expand as follows: 
\begin{align}
&I^A=I^B
\nonumber\\
&=1+Aq^{\frac{1}{8}}+6a^2q^{\frac{1}{6}}+2A^2q^{\frac{1}{4}}+12a^2Aq^{\frac{7}{24}}+21a^4q^{\frac{1}{3}}+2A^3q^{\frac{3}{8}}
+\left(\frac{1}{a^4A^2}+18a^2A^2\right)q^{\frac{5}{12}}
\nonumber\\
&+56a^4Aq^{\frac{11}{24}}
+\left(56a^6+3A^4\right)q^{\frac{1}{2}}+\frac{2+24a^6A^4}{a^4A}q^{\frac{13}{24}}
+\frac{6+97a^6A^4}{a^2A^2}q^{\frac{7}{12}}+\cdots, 
\end{align}
exhibiting precise agreement between the two sides. 

%%%%%%%%%%%%%%%%%%%%%%%%%%%%%%%%%%
\subsubsection{Boundary conditions}
%%%%%%%%%%%%%%%%%%%%%%%%%%%%%%%%%%
Given the dual pair of theories A and B, we proceed to extend this correspondence to a duality between their $\mathcal{N}=(0,2)$ supersymmetric boundary conditions.
We consider theory A with a $USp(2n)$ vector multiplet obeying Neumann boundary condition, 
a $USp(2n)$ rank-2 antisymmetric chiral $\Phi$ with Neumann boundary condition and $4$ fundamental chirals $Q_{\alpha}$ with Neumann boundary conditions. 
This can be arrived at from the case of $5(+1)$ fundamental chirals by sending the mass of $1(+1)$ fundamental chirals to infinity. 
They are summarized as
\begin{align}
\label{USp2n_AS_4_bdy_charges}
\begin{array}{c|c|c|c|c|c|c|c|c}
& \textrm{bc} & USp(2n) & SU(N_f = 4) & U(1)_A & U(1)_a & U(1)_R \\ \hline
\textrm{VM} & \mathcal{N} & {\bf Adj} & {\bf 1} & 0 & 0 & 0 \\
\Phi &\textrm{N}& {\bf n(2n-1)} & {\bf 1} & 1 & 0 & 0 \\
Q_{\alpha} & \textrm{N} & {\bf 2n} & {\bf 4} & 0 & 1 & 0 \\
 \hline
M_{\alpha \beta}^{(\lambda)} & \textrm{N} & {\bf 1} & {\bf 6} & \lambda-1 & 2 & 0 \\
\phi^{(\lambda)} & \textrm{N} & {\bf 1} & {\bf 1} & \lambda & 0 & 0 \\
\widetilde{M}^{(\lambda)} & \textrm{D} & {\bf 1} & {\bf 1} & \lambda + 1 - 2n & -4 & 2
\end{array}
\end{align}

The Neumann half-index can be derived from the half-index for the theory with $5(+1)$ fundemntal chirals \eqref{bdy_USp2n_AS_6_hindexA} by defining $X_{\alpha} = a x_{\alpha}$ and taking $X_5 \to 0$ while keeping $X_{\alpha}$ fixed for $\alpha \in \{ 1, 2, 3, 4 \}$. 
After then rescaling $a$ and $x_{\alpha}$ for $\alpha \in \{ 1, 2, 3, 4 \}$ so that $\prod_{\alpha = 1}^4 x_{\alpha} = 1$ the half-index of theory A takes the form
\begin{align}
\label{bdy_USp2n_AS_4_hindexA}
&
\mathbb{II}_{(\mathcal{N},N,N)}^{A}
\nonumber\\
&=\frac{(q)_{\infty}^n}{n! 2^n} \prod_{i=1}^n \oint \frac{ds_i}{2\pi i s_i}
\prod_{i \ne j}^n (s_i s_j^{-1}; q)_{\infty} \prod_{i \le j}^n (s_i^{\pm} s_j^{\pm}; q)_{\infty}
\frac{1}{\prod_{\alpha = 1}^{4} \prod_{i = 1}^n (q^{r_a/2} s_i^{\pm} a x_{\alpha}; q)_{\infty}}
\nonumber \\
& \times \frac{1}{(q^{r_A/2} A; q)_{\infty}^n \prod_{i < j}^n (q^{r_A/2} A s_i^{\pm} s_j^{\mp}; q)_{\infty} (q^{r_A/2} A s_i^{\pm} s_j^{\pm}; q)_{\infty}}, 
\end{align}
while the theory B half-index \eqref{bdy_USp2n_AS_6_hindexB} becomes
\begin{align}
\label{bdy_USp2n_AS_4_hindexB}
\mathbb{II}_{(N, N, D)}^B = & \prod_{\lambda = 1}^n \frac{(q^{(2n - 1 - \lambda)r_A/2 + 2r_a} A^{2n - 1 - \lambda} a^4; q)_{\infty}}{(q^{\lambda r_A/2} A^\lambda; q)_{\infty} \prod_{1 \le \alpha < \beta \le 4} (q^{(\lambda-1)r_A/2 + r_a} A^{\lambda-1} a^2 x_{\alpha} x_{\beta}; q)_{\infty}}. 
\end{align}
The agreement of the half-indices (\ref{bdy_USp2n_AS_4_hindexA}) and (\ref{bdy_USp2n_AS_4_hindexB}) 
follows from the equality of the half-indices (\ref{bdy_USp2n_AS_6_hindexA}) and (\ref{bdy_USp2n_AS_6_hindexB}) of the parent theory.

%%%%%%%%%%%%%%%%%%%%%%%%%%%%%%%%%%
\subsubsection{Line defect half-indices}
%%%%%%%%%%%%%%%%%%%%%%%%%%%%%%%%%%
We proceed by inserting, at the boundary of theory A, the Wilson line operator transforming in a representation $\lambda$ of the gauge group $USp(2n)$. 
The associated line defect half-index is given by the following matrix integral: 
\begin{align}
\label{bdy_USp2n_AS_4_W}
&
\langle W_{\lambda}\rangle_{(\mathcal{N},N,N)}^{A}
\nonumber\\
&=\frac{(q)_{\infty}^n}{n! 2^n} \prod_{i=1}^n \oint \frac{ds_i}{2\pi i s_i}
\prod_{i \ne j}^n (s_i s_j^{-1}; q)_{\infty} \prod_{i \le j}^n (s_i^{\pm} s_j^{\pm}; q)_{\infty}
\frac{1}{\prod_{\alpha = 1}^{4} \prod_{i = 1}^n (q^{r_a/2} s_i^{\pm} a x_{\alpha}; q)_{\infty}}
\nonumber \\
& \times \frac{1}{(q^{r_A/2} A; q)_{\infty}^n \prod_{i < j}^n (q^{r_A/2} A s_i^{\pm} s_j^{\mp}; q)_{\infty} (q^{r_A/2} A s_i^{\pm} s_j^{\pm}; q)_{\infty}}
\chi_{\lambda}^{\mathfrak{usp}(2n)}(s), 
\end{align}
where 
\begin{align}
\label{ch_usp2N_irrep}
\chi_{\lambda}^{\mathfrak{usp}(2n)}(s)
&=\frac{\det(s_j^{\lambda_i+N-i+1}-s_j^{-\lambda_i-N+i-1})}{\det(s_j^{N-i+1}-s_j^{-N+i-1})}
\end{align}
is the $\mathfrak{usp}(2n)$ character of the irreducible representation labeled by the Young diagram $\lambda$ in \cite{MR1153249}. 

Now, in the case of $n=1$, we have $USp(2) \simeq SU(2)$ 
and the integrand in the theory A half-index \eqref{bdy_USp2n_AS_4_hindexA} is the weight with respect to which Askey-Wilson polynomials are orthogonal. 
We saw that exact results for line operators in theory A could then be derived using known results for Askey-Wilson moments. 
For general $n$ we have a similar story. 
The half-index integrand is the weight for Macdonald-Koornwinder polynomials and line operator half-indices would be given in terms of Macdonald-Koornwinder moments, $\widehat{M}_{\underline{\lambda}}^{USp(2n)}$. Here we define these moments, 
where $\underline{\lambda}$ is an integer partition with $\lambda_1 \ge \lambda_2 \ge \cdots \lambda_n \ge 0$, in terms of Schur polynomials as
\begin{align}
\label{KoornwinderMoments_A_defn}
& \widehat{M}_{\underline{\lambda}}^{USp(2n)} 
\prod_{\rho = 1}^n \frac{(A^{2n - 1 - \rho} a^4; q)_{\infty}}{(A^{\rho}; q)_{\infty} \prod_{1 \le \alpha < \beta \le 4} (A^{\rho - 1} a^2 x_{\alpha} x_{\beta}; q)_{\infty}}
\nonumber \\
= & \frac{(q)_{\infty}^n}{n! 2^n} \prod_{i=1}^n \oint \frac{ds_i}{2\pi i s_i}
\prod_{i \ne j}^n (s_i s_j^{-1}; q)_{\infty} \prod_{i \le j}^n (s_i^{\pm} s_j^{\pm}; q)_{\infty}
\frac{S_{\underline{\lambda}}(s_i + s_i^{-1})}{\prod_{\alpha = 1}^{4} \prod_{i = 1}^n (q^{r_a/2} s_i^{\pm} a x_{\alpha}; q)_{\infty}}
\nonumber \\
& \times \frac{1}{(q^{r_A/2} A; q)_{\infty}^n \prod_{i < j}^n (q^{r_A/2} A s_i^{\pm} s_j^{\mp}; q)_{\infty} (q^{r_A/2} A s_i^{\pm} s_j^{\pm}; q)_{\infty}} \; .
\end{align}
Note that to simplify the presentation we have set $r_a = r_A = 0$ but these parameters can easily be restored by scaling $a \to q^{r_a/2} a$ and $A \to q^{r_A/2} A$.

%%%%%%%%%%%%%%%%%%%%%%%%%%%%%%%%%%
\subsubsection{One-point functions}
%%%%%%%%%%%%%%%%%%%%%%%%%%%%%%%%%%
The motivation for defining the moments in terms of Schur polynomial half-indices is to connect to exact results in the mathematical literature, specifically the Koornwinder moments described below. However, Wilson line half-indices in arbitrary representations of $USp(2n)$ can be expressed 
as linear combinations of Schur polynomial half-indices as indeed can multi-point Wilson line half-indices with arbitrary representations. 
The simplest case is the choice $\underline{\lambda} = (1, 0, \ldots , 0)$ which directly gives the fundamental Wilson line half-index. 

Unfortunately we are not aware of exact results for such moments in general, but we will conjecture a result below in the case of 
$\underline{\lambda} = (1, 0, \ldots , 0)$. 
However, with the specialization of the fugacity corresponding to the $U(1)$ flavor symmetry for the antisymmetric chiral, 
$A = q$ (or in general $Aq^{r_A/2} = q$), exact result for these moments, called Koornwinder moments, were given in \cite{MR3816505}.

Specifically, the Koornwinder moments \cite{MR3816505} $M_{\underline{\lambda}}$ are then defined by Schur polynomial half-indices as
\begin{align}
\label{KoornwinderMoments_defn}
& M_{\underline{\lambda}}^{USp(2n)} 
\prod_{\rho = 1}^n \frac{(q^{2n - 1 - \rho + 2r_a} a^4; q)_{\infty}}{(q^{\rho}; q)_{\infty} \prod_{1 \le \alpha < \beta \le 4} (q^{\rho - 1 + r_a} a^2 x_{\alpha} x_{\beta}; q)_{\infty}}
\nonumber \\
= &
\frac{1}{n! 2^n} \prod_{i=1}^n \oint \frac{ds_i}{2\pi i s_i}
\prod_{i \ne j}^n (1 - s_i s_j^{-1}) \prod_{i \le j}^n (1 - s_i^{\pm} s_j^{\pm})
\frac{S_{\underline{\lambda}}(s_i + s_i^{-1})}{\prod_{\alpha = 1}^{4} \prod_{i = 1}^n (q^{r_a/2} s_i^{\pm} a x_{\alpha}; q)_{\infty}} \; .
\end{align}
As shown in \cite{MR3816505} these moments can be evaluated in terms of the Askey-Wilson moments $\mu_k$ in \eqref{AskeyWilson_moments} and \eqref{AskeyWilson_moments_eval} as
\begin{align}
    \label{KoornwinderMoments_eval}
    M_{\underline{\lambda}}^{USp(2n)}
    = & \frac{\det(\mu_{\lambda_i + 2n - i - j})_{i, j = 1}^n}{\det(\mu_{2n - i - j})_{i, j = 1}^n} \; .
\end{align}

For example, if we take $\underline{\lambda} = (1, 0, \ldots , 0)$ the Schur polynomial is simply $\sum_{i = 1}^n (s_i + s_i^{-1})$ 
so the moment $\widehat{M}_{(1,\vec{0})}^{USp(2n)}$ and the Koornwinder moment $M_{(1,\vec{0})}^{USp(2n)}$ are related to the 1d defect index and the fundamental Wilson line half-index as follows: 
\begin{align}
\mathcal{I}_{\tiny \yng(1)}^{\textrm{1d defect}}
&=
\frac{\langle W_{\tiny \yng(1)}\rangle_{(\mathcal{N},N,N)}^{A}}
{\mathbb{II}_{(\mathcal{N},N,N)}^{A}}
=\widehat{M}_{(1,\vec{0})}^{USp(2n)}\Bigl|_{a\rightarrow q^{r_a/2}a,A\rightarrow q^{r_A/2}A}, \\
M_{(1,\vec{0})}^{USp(2n)}
&=\widehat{M}_{(1,\vec{0})}^{USp(2n)}\Bigl|_{A\rightarrow q}. 
\end{align} 
For $n = 1$ this just reproduces the Askey-Wilson result while for $n = 2$ we have
\begin{align}
    \label{KoornwinderMoments_eval_n2}
    M_{(1, 0)}^{USp(4)}
    = & \frac{\mu_3 - \mu_1 \mu_2}{\mu_2 - \mu_1^2}
    \nonumber \\
    = & \frac{a (1 + q) \left( \sum_{\alpha} x_{\alpha} - a^2 q \sum_{\alpha} x_{\alpha}^{-1} \right)}{1 - a^4 q^2} \; .
\end{align}
In the unflavored case $x_{\alpha} = 1$ this simplifies to
\begin{align}
    \label{KoornwinderMoments_eval_UF_n2}
    M_{(1, 0)}^{USp(4)}(x_{\alpha} = 1)
    = & \frac{4a(1 + q)}{1 + a^2q} \; .
\end{align}

We find that reintroducing the fugacity $A$ (for the rank-$2$ antisymmetric tensor) 
we have in the unflavored (in terms of the fundamental chirals, i.e.\ setting $x_{\alpha} = 1$) case
\begin{align}
    \label{KoornwinderMoments_eval_UF_A_n2}
    \widehat{M}_{(1, 0)}^{USp(4)}(x_{\alpha} = 1)
    = & \frac{4a(1 + A)}{1 + a^2A} \; .
\end{align}
We expect that in the flavored case
\begin{align}
    \label{KoornwinderMoments_eval_A_n2}
    \widehat{M}_{(1, 0)}^{USp(4)}
    = & \frac{a (1 + A) \left( \sum_{\alpha} x_{\alpha} - a^2 A \sum_{\alpha} x_{\alpha}^{-1} \right)}{1 - a^4 A^2} \; .
\end{align}

For the $USp(6)$ case (with $n = 3$) we have
\begin{align}
    \label{KoornwinderMoments_eval_n3}
    M_{(1, 0, 0)}^{USp(6)}
    = & \frac{a(1 + q + q^2)\left( \sum_{\alpha} x_{\alpha} - a^2 q^2 \sum_{\alpha} x_{\alpha}^{-1} \right)}{1 - a^4q^4} \; .
\end{align}
In the unflavored case $x_{\alpha} = 1$ this simplifies to
\begin{align}
    \label{KoornwinderMoments_eval_UF_n3}
    M_{(1, 0, 0)}^{USp(6)}(x_{\alpha} = 1)
    = & \frac{4a(1 + q + q^2)}{1 + a^2q^2} \; .
\end{align}

For the $USp(8)$ case (with $n = 4$) we have
\begin{align}
    \label{KoornwinderMoments_eval_n4}
    M_{(1, 0, 0, 0)}^{USp(8)}
    = & \frac{a(1 + q + q^2 + q^3)\left( \sum_{\alpha} x_{\alpha} - a^2 q^3 \sum_{\alpha} x_{\alpha}^{-1} \right)}{1 - a^4q^6} \; .
\end{align}
and in the unflavored case
\begin{align}
    \label{KoornwinderMoments_eval_UF_n4}
    M_{(1, 0, 0, 0)}^{USp(8)}(x_{\alpha} = 1)
    = & \frac{4a(1 + q + q^2 + q^3)}{1 + a^2q^3} \; .
\end{align}

We therefore have the obvious conjecture
\begin{align}
    \label{KoornwinderMoments_eval_n}
    M_{(1, \vec{0})}^{USp(2n)}
    = & \frac{a(1 - q^n)\left( \sum_{\alpha} x_{\alpha} - a^2 q^{n-1} \sum_{\alpha} x_{\alpha}^{-1} \right)}{(1 - a^4q^{2n-2})(1 - q)} \; 
\end{align}
and for the unflavored case
\begin{align}
    \label{KoornwinderMoments_eval_UF_n}
    M_{(1, \vec{0})}^{USp(2n)}(x_{\alpha} = 1)
    = & \frac{4a(1 - q^n)}{(1 + a^2q^{n-1})(1 - q)} \; .
\end{align}
It is interesting to note that while the expression for the Koornwinder moments 
giving the fundamental Wilson line half-indices for $USp(2n)$, in the case of specialization of fugacity $A = q$,
 is given in terms of the Askey-Wilson moments 
up to $\mu_{2n - 1}$, the result is much simpler than the expression for $\mu_{2n - 1}$. 
This hints at an alternative, simpler formulation for these Koornwinder moments.

We further conjecture that the results without specializing the fugacity $A$, 
the Macdonald-Koornwinder moments defined by (\ref{KoornwinderMoments_A_defn}), are
\begin{align}
    \label{KoornwinderMoments_eval_A_n}
    \widehat{M}_{(1, \vec{0})}^{USp(2n)}
    = & \frac{a(1 - A^n)\left( \sum_{\alpha} x_{\alpha} - a^2 A^{n-1} \sum_{\alpha} x_{\alpha}^{-1} \right)}{(1 - a^4 A^{2n-2})(1 - A)} \; .
\end{align}
and for the unflavored case
\begin{align}
    \label{KoornwinderMoments_eval_UF_A_n}
    \widehat{M}_{(1, \vec{0})}^{USp(2n)}(x_{\alpha} = 1)
    = & \frac{4a(1 - A^n)}{(1 + a^2 A^{n-1})(1 - A)} \; .
\end{align}
Accordingly, the 1d defect index is
\begin{align}
&
\mathcal{I}_{\tiny \yng(1)}^{\textrm{1d defect}}(a,A,x_{\alpha};q)
\nonumber\\
&=\frac{q^{\frac{r_a}{2}}a (1+q^{\frac{r_A}{2}}A+\cdots+q^{\frac{(n-1)r_A}{2}}A^{n-1})
(\sum_{\alpha}x_{\alpha}-q^{r_a+\frac{(n-1)r_A}{2}}a^2 A^{n-1}\sum_{\alpha}x_{\alpha}^{-1})}
{1-q^{2r_a+(n-1)r_A}a^4A^{2n-2}}. 
\end{align}
The factor $1-q^{2r_a+(n-1)r_A}a^4A^{2n-2}$ in the denominator appears from 
the effective shift of the spin of the singlet chiral multiplet $V$ $=$ $\widetilde{M}^{(\lambda=1)}$, 
upon shifting one of the $q$-Pochhammer terms by the factor $q$ of the Dirichlet half-index of 
$V$ $=$ $\widetilde{M}^{(\lambda=1)}$ in the half-index (\ref{bdy_USp2n_AS_4_hindexB}). 
Hence, the fundamental Wilson line half index can be seen as a polynomial multiplying this shifted half-index as we saw for the Askey Wilson case.
The unflavored 1d defect index is
\begin{align}
\mathcal{I}_{\tiny \yng(1)}^{\textrm{1d defect}}(1,1,1;q)
&=\frac{4q^{\frac{r_a}{2}}(1-q^{\frac{nr_A}{2}})}{(1+q^{r_a+\frac{(n-1)r_A}{2}})(1-q^{\frac{r_A}{2}})}. 
\end{align}
Turning off the fugacities, we obtain the Witten index
\begin{align}
\mathcal{I}_{\tiny \yng(1)}^{\textrm{1d defect}}(1,1,1;1)
&=2n. 
\end{align}

Results with $N_f < 4$ fundamental chirals will follow as in the Askey-Wilson case 
by specializing fugacities (in the flavored cases) which will simply mean restricting the values of indices $\alpha, \beta$ 
and using the $N_f < 4$ version of the Askey-Wilson moments in the same expressions. 
The correspondence for the unflavored cases must be derived from the flavored cases.

%%%%%%%%%%%%%%%%%%%%%%%%%%%%%%%%%%%
\subsection*{Acknowledgements}
The work of TO was supported by the Startup Funding no.\ 4007012317 of the Southeast University. 
The work of DJS was supported in part by the STFC Consolidated grant ST/X000591/1.
%%%%%%

\bibliographystyle{utphys}
\bibliography{ref}

\end{document}